\documentclass[nonblindrev]{informs-pom}

\OneAndAHalfSpacedXI
\usepackage[]{endnotes}
\usepackage{amsmath,amssymb,amsfonts}
 
\usepackage{rotating}
\usepackage{fancyvrb}
\usepackage[hidelinks=True]{hyperref}
\usepackage{soul}
\usepackage{mdframed}
\usepackage{ifthen}
\usepackage{multirow}

 \usepackage{natbib}
 \bibpunct[, ]{(}{)}{,}{a}{}{,}%
 \def\bibfont{\small}%
\TheoremsNumberedThrough     
\ECRepeatTheorems

\EquationsNumberedThrough    

\MANUSCRIPTNO{OPRE-2024-05-1050}

\usepackage{cleveref}
\usepackage{bbm}
\usepackage{amsmath}
\usepackage{float}
\usepackage[ruled]{algorithm2e}
\usepackage{wrapfig}
\usepackage{comment}
\usepackage[]{todonotes}
\usepackage{dcolumn}
\newcolumntype{d}[1]{D..{#1}} 

\newboolean{truncate}
\setboolean{truncate}{true} 

\newcommand\RR{\mathbb{R}}
\newcommand\Ind{\mathbbm{1}}
\newcommand\MM{\mathcal{M}}

\newcommand{\instance}{(v, \delta_C, \delta_F, \kappa_C, \kappa_F, p, \lambda)}

\newcommand{\val}[3][]{
	\def\temp{#1}\ifx\temp\empty
	v_{#2}({#3})
	\else
	v^{#1}_{#2}(#3)
	\fi
}

\newcommand{\strat}[3][]{
	\def\temp{#3}\ifx\temp\empty
	s^{#1}_{#2}
	\else
	s^{#1}_{#2}(#3)
	\fi
}

\newcommand{\ut}[3][]{
	u^{#1}_{#2}(#3)
}

\newcommand{\utt}[3]{
	\bar{u}^{#1}_{#2}(#3)
}

\newcommand{\bp}[4][]{
	\def\temp{#4}\ifx\temp\empty
	\beta^{#1}_{#2#3}
	\else
	\beta^{#1}_{#2#3}(#4)
	\fi
}

\newcommand{\se}[1]{
	s^{#1}
}

\definecolor{Quotes}{rgb}{0,0.3,0.9}
\definecolor{Rev}{rgb}{0.8,0,0}

\begin{document}

\RUNTITLE{Search and Matching for Adoption from Foster Care}
\TITLE{Search and Matching for Adoption from Foster Care}

\RUNAUTHOR{Dierks, Olberg, Seuken, Slaugh and \"{U}nver}

\ARTICLEAUTHORS{
	\AUTHOR{Ludwig Dierks}
	\AFF{Department of Information and Decision Sciences, University of Illinois at Chicago, USA, \EMAIL{ldierks@uic.edu}}
    \AUTHOR{Nils Olberg}
	\AFF{Department of Informatics, University of Zurich, Switzerland, \EMAIL{olberg@ifi.uzh.ch}}
	\AUTHOR{Sven Seuken}
	\AFF{Department of Informatics, University of Zurich, Switzerland, \EMAIL{seuken@ifi.uzh.ch}}
	\AUTHOR{Vincent W. Slaugh}
	\AFF{SC Johnson College of Business, Cornell University, USA, \EMAIL{vslaugh@cornell.edu}}
	\AUTHOR{M. Utku \"{U}nver}
	\AFF{Department of Economics, Boston College, USA, \EMAIL{unver@bc.edu}}
}

\ABSTRACT{To find families for the more than 70,000 children in need of adoptive placements, most United States child welfare agencies have employed a \emph{family-driven search} approach in which prospective families respond to announcements made by the agency. However, some agencies have switched to a \emph{caseworker-driven search} approach in which the caseworker directly contacts families recommended for a child. We introduce a novel search-and-matching model that captures the key features of the adoption process and compare family-driven with caseworker-driven search in a game-theoretical framework. Under either approach, the equilibria are generated by threshold strategies and form a lattice structure. Our main theoretical finding then shows that no family-driven equilibrium can Pareto dominate any caseworker-driven outcome, whereas it is possible that each caseworker-driven equilibrium Pareto dominates every equilibrium attainable under family-driven search.  We also find that, within our model, when families are sufficiently impatient, caseworker-driven search is better for all children. We numerically illustrate that most agents are better off under caseworker-driven search across a wide range of parameter values. Finally, we present an empirical study of an agency that switched to caseworker-driven search, finding a three-year adoption probability that outperformed a statewide benchmark by $44.9\%$, along with a statistically significant $54\%$ higher adoption hazard rate. 
}

\SUBJECTCLASS{Government/Services: Child-Welfare System; Government/Programs: Children, Youth, and Families; Population/Family planning: Adoption}


\maketitle

\section{Introduction} \label{sec:introduction}

Child welfare systems worldwide face the challenge of finding families for children in need of adoption. For example, the United States foster care system serves over 525,000 children annually. While the goal for most of these children is to reunite them with their parents or relatives, 70,421 out of 328,947 children in foster care on September 30, 2024, were waiting for permanent adoptive placements \citep{acf_afcars_dashboard}. Finding an adoptive family for these children has become a public policy priority due to high levels of incarceration, homelessness, unemployment, and teen pregnancy observed in the population of children ``aging out'' of the child welfare system without a permanent family relationship \citep{triseliotis_long-term_2002, courtney_midwest_2010, gypen2017outcomes}. In this paper, we study the search and matching process for children waiting in the child welfare system for adoptive placement and compare two prominent search methods that differ in whether families or the children's caseworkers drive the search process. We provide structural insights by analytically and numerically analyzing a game-theoretic model. Furthermore, we conduct an empirical study of a Florida child welfare agency that switched its search strategy in 2018, and we find outcomes consistent with our analytical results.

The search for an adoptive placement officially begins once a judge issues a termination of parental rights (TPR) order. A \emph{caseworker} represents the child's interests throughout this process to find an adoptive family if adoption by a relative or foster care family is not possible. Identifying a family willing to adopt the child and capable of caring for the child can be a difficult task, and the challenge varies greatly according to the child's demographic characteristics and special needs. While relatively little research has studied best practices for search operations --- either through empirical studies of how caseworkers find families or prescriptive studies for how search should be conducted --- states invest significant resources in trying to help the most vulnerable children find permanency; for example, the \citet{fl_dcf_2022,fl2024} reports spending over \$20 million annually to promote and support searches for approximately 3,800 children with a goal of adoption at any point in time. The statewide case data analyzed in Section~\ref{sec:empirical_casestudy} highlights the difficulty of securing adoptive placements for a sub-population of children requiring adoptive search services: only 16.1\% of children finalize an adoption within one year of TPR, 38.0\% within two years, and 49.0\% within three years.

Different states, counties, and agencies adhere to different paradigms for the practice of search and matching. Families first register with an adoption agency and provide a home study evaluation.  In the predominant approach, caseworkers then announce children via email to a set of registered families, each of whom has the opportunity to express interest in a child. The caseworker receives these inquiries and works to identify the family that best fits the child's needs. We label this approach \emph{family-driven search}, as families direct the search process by expressing interest in available children. \citet{hanna_innovative_2011} demonstrates how the child welfare literature has almost exclusively focused on what we view as family-driven search processes. One important downside of this search process is that some children may attract hundreds of interested families, all of which the caseworker has to consider simultaneously. 

These interactions can be time-consuming and emotionally stressful for all parties involved, and some states require agencies to engage with every family that has expressed interest. For example, Florida Administrative Code Rule 65C-16.003 states: 
\begin{quote}
\emph{Once the potential adoptive families have been identified, the staffing team will rate each family based on the family's ability to meet the identified needs of the child based on information documented in [the Florida Safe Families Network information system], the Child Study and the adoptive parent's home study. The documentation must include a key of the rating scale used by the team.}
\end{quote}
An adoptions manager for a multi-county agency in Florida directed us to this rule to emphasize the imperative that the agency must respond to every family that expresses interest in a child. Failure to respond to families can result in complaints to the governor's office or negative comments on social media. The manager also commented that the agency dreaded announcing the availability of a ``cute'' young child, who would attract dozens or even hundreds of responses from families that required thorough consideration and individual responses.  This is especially problematic because caseworkers often handle twice as many cases as they should ideally handle \citep{yamatani2009child}, an issue that has intensified since the COVID-19 pandemic \citep{lushin2023burdened}.  On the other hand, other children may attract very few interested families, and nearly 20,000 children age out of foster care each year \citep{childrens_bureau_afcars_2020}. 

We have also heard families share their frustration with the emotional and time costs that they incurred while participating in a family-driven search system. In a video shared on social media, one adoptive father described his journey that began as a prospective parent engaged with an adoption agency that announced available children and had families respond to express interest in particular children:\footnote{\href{https://www.linkedin.com/posts/adoption-share_by-inverting-the-outreach-for-adoption-matching-activity-7131265613947138048-YCM1}{https://www.linkedin.com/posts/adoption-share\_by-inverting-the-outreach-for-adoption-matching-activity-7131265613947138048-YCM1} (Retrieved 03/31/2026)}
\begin{quote}
     \emph{Basically, you put your life on hold, and you have your hopes set on this one particular child that you've fallen in love with their profile and their picture. You dream every day and night and go to sleep hoping that this kid is the one, and every time that happens for me, 30-45 days later I would find out that I wasn't chosen. So, there's this tremendous sense of disappointment, rejection, and ``why didn't they choose me?'' I went through that process over and over and over again for about 2+ years until I was finally connected with [a caseworker-driven search platform]. What I loved about the concept was that I wouldn't have to go through that process. This time it was my profile, and I'd just have to wait for the caseworker or my forever match to find me. I'm happy to report that a couple weeks later I did find my forever match, and I have my son now.}
\end{quote}

In response to these challenges, some child welfare agencies have recently sought to improve outcomes by adopting \emph{caseworker-driven search} \citep{riley2019}. In this approach, caseworkers sequentially contact specific prospective families to share details about the child. Caseworkers use their expertise to decide which families to contact based on the child's and the family's characteristics. Optionally, technological tools may also aid their decisions. 
This approach removes the burden of engaging a large number of families simultaneously and allows caseworkers to target compatible families for children with very specific needs.  

However, even though these are clear advantages of caseworker-driven search, it is important to note that changing the search paradigm may alter families' incentives, influencing their interest in different types of children. For a match to occur under either approach, both the caseworker and the prospective family need to agree. This can, for example, mean that if caseworker-driven search reduces search costs for families, they may stop being interested in certain children, leaving those children unmatched. As such, it is unclear whether caseworker-driven search leads to more desirable outcomes for the entire population of children requiring adoptive placements.  

To assess how different search disciplines affect outcomes, we 
pursue two complementary approaches in this paper. First, we analyze the strategic behavior of agents under both search disciplines in a game-theoretic search-and-matching model. Second, we provide empirical evidence for the performance of caseworker-driven search compared to classic search approaches by studying outcomes of a real-world child welfare agency that switched its approach in 2018. 

Our analytical contributions begin with the introduction of a new search-and-matching model (\Cref{sec:model}), which captures critical features. First, as both approaches to adoption matching are inherently dynamic, we assume that children and families (hereafter referred to as \emph{agents}) may enter and depart the system at any time. However, to keep the model tractable, we assume the distribution of agent types in the system remains stable over time. Second, we allow for uncertainty regarding whether a child-family pair is compatible. Third, our model captures the heterogeneous preferences of agents, which is a key distinction between our paper and most earlier literature. While we focus specifically on analyzing adoption systems, our work is also relevant to more general search-and-matching theory. 

We perform a game-theoretic comparison of the two approaches within our model.
 In \Cref{sec:equilibrium_existence}, we establish that (pure strategy) equilibria are guaranteed to exist under both \emph{search technologies} (i.e., family-driven search and caseworker-driven search). Furthermore, we find that equilibria form a lattice reminiscent of the structure of the set of stable matchings in standard two-sided matching markets.
 We then present our main theoretical result: Family-driven search equilibrium outcomes can never be Pareto improvements over caseworker-driven search equilibrium outcomes, but there are instances where each caseworker-driven search equilibrium outcome is a Pareto improvement over all family-driven search equilibrium outcomes (\Cref{sec:theory_comparison_pareto}). This holds because caseworker-driven search can reduce wasted search efforts. Thus, agents can worry less about accumulating search costs when they express interest in a child. However, because of multiplicity of equilibria and the lattice structure over these equilibria, all children can be strictly better off in either system. The same holds for families. Even when both family-driven search and caseworker-driven search only admit a unique equilibrium each, an agent can be better off in either setting (\Cref{sec:theory_comparison_no_domination}). This may be surprising, given that caseworker-driven search reduces wasteful search efforts on both sides of the market. We therefore explore the conditions under which caseworker-driven search usually leads to more agents being better off.
 In \Cref{sec:parameter_effects_discounting}, we show that all children will be better off in caseworker-driven search if families are sufficiently impatient. Furthermore, increasing family supply in family-driven search can have a negative effect on children's utilities, but not in caseworker-driven search (\Cref{sec:parameter_effects_market_thickness}).
 We find numerically that caseworker-driven search leads to more desirable outcomes for a wide variety of model parameter choices (\Cref{sec:num_eval}). 
 
 To supplement these analytical and numerical insights, we empirically analyze case-level data from a technology platform that a multi-county child welfare agency in Florida has been using for the majority of its search efforts since 2018 (\Cref{sec:empirical_casestudy}). We compare outcomes for over 300 children in need of adoptive placements to a statewide case-level benchmark, covering nearly 10,000 children, that adjusts for each child’s demographics and disability status. To measure how child- and case-level factors influence the timing of adoptions in Florida, we fit a Cox proportional hazards model \citep{cox1972regression} to statewide records from the federal Adoption and Foster Care Analysis and Reporting System (AFCARS). We then use this model to predict the adoption outcomes of children registered on the platform if traditional search disciplines were used instead of the agency's caseworker-driven search approach. These predictions can then be compared to the actual outcome data from the focal agency. For the agency's children, the actual number of adoptions within three years was 44.9\% higher than the expected number under the benchmark model. We further extend our statistical analysis to directly measure the focal agency effect using a time-varying Cox model in a conservatively combined dataset. We find a statistically significant 54\% improvement in adoption hazard rates compared to statewide case data.
These results are consistent with our analytical results, although operational differences between agencies and data limitations prevent us from attributing the differences solely to the search discipline.   

\section{Related Literature}\label{sec:rel_lit}

Recently, market designers and operations researchers have shown increased interest in how to best serve historically disadvantaged communities. Researchers have, for example, studied refugee resettlement (\citealp{andersson_dynamic_2018,bansak_improving_2018, delacretaz_matching_2020}), the improvement of teacher quality at disadvantaged schools (\citealp{combe_mechanism_2021, combe_market_2025}),  the management of volunteer workforces for non-profit organizations (\citealp{berenguer2023managing}), and allocation of public housing (\citealp{arnosti_design_2020, kawasaki_mechanism_2021, you_strategy_2022}).

Our paper contributes to the limited literature that studies child welfare systems from operations or market design perspectives --- a challenge first articulated by \citet{spindler1970social} and renewed by \citet{slaugh2025child}. \citet{slaugh_pennsylvania_2016} investigated how the Pennsylvania Statewide Adoption \& Permanency Network could utilize a match recommendation tool to improve their process of matching children to prospective parents. Their spreadsheet-based tool can be seen as a simple example of the previously mentioned technological tools that caseworkers may use. \citet{robinson-cortes_who_2019} worked with a foster care data set to analyze placements of children in foster homes. His model predicts that allowing placements across administrative regions would be beneficial for children. \citet{macdonald_foster_2019} studied a dynamic matching problem where children and families can either form reversible matches (foster placements) or irreversible matches (adoptions). In her model, children are heterogeneous in the sense that there are children with disabilities and children without disabilities, while families are homogeneous.  To the best of our knowledge, we are the first to formally analyze the economic effects of search and matching in the child welfare domain, while taking into account the full heterogeneity of preferences.
\citet{baccara_child-adoption_2014} estimated families' preferences over children available for adoption from a data set documenting the operations of adoption agencies.
 
Within the child welfare literature, relatively little research has investigated the effectiveness of search disciplines for children in need of adoptive placements. Some research has reported positive impacts from intensive multi-faceted search efforts by caseworkers: \citet{vandivere2015experimental} show that children served by skilled recruiters from the Wendy's Wonderful Kids organization were 1.7 times as likely to have an adoptive placement than a control group in an experiment with over 1,000 children. In a similar context focusing on hard-to-place youth in New York, \citet{feldman2016not} show that a program of enhanced casework improved outcomes for children. The program utilized a variety of channels to promote 88 children and conduct searches. The search methods in both experiments require extensive work from skilled caseworkers funded by grants, while the platform we study provides an example of technology assisting caseworkers. \citet{avery_adoptuskids_2009} study national photolisting service AdoptUSKids and use a hazards model to show better outcomes for children based on activity on AdoptUSKids. However, photolistings have drawn increased scrutiny since the early 2000s; \citet{roby2010adoption} describe risks for exploitation and bullying for children publicly listed online.

Even though child adoption matching markets bear similarities to other two-sided matching markets such as centralized labor markets \citep{roth_evolution_1984, roth_natural_1991} or ride-sharing platforms \citep{ma_spatio-temporal_2020}, there are important differences that necessitate new models and analyses. Adoption matching is inherently dynamic, and there is no centralized clearinghouse that determines final matches. Matches are only ever proposed, and both sides of the market have to invest \emph{search efforts} to identify a match candidate.

One matching market similar to adoption from foster care is online dating \citep{hitsch_matching_2010,hitsch_what_2010,lee_propose_2015,HalaburdaPiskorskiYildirim2018,kanoria_facilitating_2021}.
 However, most online dating markets are \emph{completely} decentralized despite their dynamic and recommendation-based features. Individuals in search of a romantic partner can, at any time, decide to browse a dating platform and reach out to other individuals who appeal to them.   
 In contrast, the approaches we analyze in this paper follow a \emph{centralized protocol} despite the dynamic decentralized search component: Caseworkers perform repeat searches for a family on behalf of children, and they do so in approximately regular time intervals. Caseworkers, therefore, play an essential role throughout the process since they act as an intermediary to protect vulnerable children. Crucially, this introduces an asymmetry, both in the number of agents simultaneously active and the market power of agents on the two sides, that we have not observed in any other previously studied matching market. As a consequence, we develop a new model that allows us to capture the features of the two different search technologies in one model.

Purely random decentralized matching models have been widely studied under search frictions and homogeneous preferences, with transferable utility \citep{shimer_assortative_2000, shimer_matching_2001, atakan_assortative_2006}, and more relevant to our paper, with non-transferable utility \citep{eeckhout_bilateral_1999, chade_NTU_2001, smith_marriage_2006}. In all of these studies, unlike in our paper, preferences are aligned on each side of the market following a strict order of quality common for each individual on the same side. More recent studies have combined directional search rather than random search as an important feature \citep{lauermann_balance_2020, cheremukhin_targeted_2020}. 

Besides the directed versus random search distinction, simultaneous \citep{Stigler1961} versus sequential \citep{weitzman_optimal_1979} search have also been studied by classic search theory, with \citet{chade_simultaneous_2006} bridging the gap by characterizing optimal hybrid strategies. Other work extends this analysis to competitive environments, such as labor markets. \citet{AlbrechtGautierVroman2006} and \citet{Kircher2009} study workers applying to multiple firms, yielding approximately efficient outcomes through endogenous wage dispersion that coordinates search. More recent work on online platforms also focuses on the trade-off between the two approaches. \citet{HonkaChintagunta2017} documents that simultaneous consumer search intensifies price competition on insurance aggregators, while \citet{AusterGottardiWolthoff2025} cautions that frictionless simultaneous contacting can exacerbate adverse selection.  \citet{chade_sorting_2017} and \citet{wright2021directed} provide comprehensive surveys of various search models.

The closest to our model are the search theory papers by \citet{adachi_search_2003} and \citet{lauermann_stable_2014} on marriage markets, which consider non-symmetric preferences on both sides of the market. However, they only consider a single randomly chosen potential match in each period, while our problem requires that multiple matches with uncertain suitability may be investigated using either caseworker-driven or family-driven search in each period. Similarly, while \cite{immorlica_designing_2024} considers the problem of a centralized platform guiding search through a sequence of match proposals to agents in a two-sided market, they also only consider a single potential match per period. Additionally, they assume that the value for a match is symmetric between both sides. This is unrealistic for an adoption setting, where families' and children's desires can be at odds with each other. Consequently, we allow for non-symmetric values. 

Our work is also related to the literature on dynamic matching regarding the effects of congestion \citep{arnosti_managing_2021,Leshno_overload22} and different practical policies  \citep{unver_dynamic_2010, akbarpour2020thickness, sonmez_incentivized_2020, akbarpour_unpaired_2020, kerimov_optimality_OR_2023} in various market-design environments, including settings intended to help vulnerable populations \citep{ baccara2020optimal,kasy_adaptive_2020}. 
 Different studies investigate how matching platforms should be designed so that desirable outcomes can be achieved \citep[see, e.g.,][]{lee_propose_2015, fradkin_search_2017, akbarpour_unpaired_2020, altinok2023designing,  dierks2024child}. 
 The research in this area most closely related to our work is   \citet{shi_optimal_2020}, in which the author explores which side of the market should drive the search process depending on which side's preferences can easily be expressed or satisfied. The setting, however, is quite different from our work since we allow agents on \emph{both} sides of the market to arrive over time and let them face an optimal stopping problem as they can decide to remain unmatched until better future match opportunities arise.
 
\section{Descriptions of Approaches to Adoption Matching}\label{app:description_of_approaches}
Based on conversations with caseworkers and managers of a Florida agency that transitioned from a family-driven search to a caseworker-driven search approach, we provide a more detailed introduction to how both approaches function in practice. Although very little has been published about search approaches and their prevalence, subject-matter experts indicate that similar approaches are commonly used nationally, with some variation among states and agencies. In any system, a prospective adoptive family first undergoes an extensive vetting and training process, which usually entails a written \emph{home study}.
Through this report, a caseworker evaluates the family on various dimensions, gathering information from home visits, interviews with family members, third-party sources, and the caseworker's own judgment. 
 State regulations determine minimum requirements for home studies to assess the suitability of the intended adoptive parents for different types of children. Typically, home studies also include additional information about the family's environment and preferences.  More details on home studies can be found in online Appendix~~\ref{app:home_study}.

Regardless of the search approach, agencies may then employ scoring rules to assess family suitability for individual children, sometimes generating and using these scores as part of a recommender system \citep{slaugh_pennsylvania_2016}. These tools and others, such as those described by  \citet{hanna_innovative_2011}, provide suggestions but are not determinative; all investigation and matching decisions are in the purview of a committee comprising the caseworker, supervisors, and other agency staff.  

Once a family has gone through this process, they can apply for adoptive placements for any child for whom the parental rights of the birth family have been terminated. 
If an interested family is denied a requested placement for non-formal reasons  --- including the intent to place the child with a different family --- Florida Administrative Code 65C-16.005(9) requires that the case be additionally reviewed by a five-person Adoption Applicant Review Committee to ensure a fair evaluation. 

The core challenge is making the most suitable families aware of children who need adoptive placements. This is where the agency's search discipline becomes relevant: identifying and informing suitable families.  
It should be noted that although there is typically a single agency responsible for finding placements for a child, their search is non-exclusive: any qualified prospective family can, in theory, apply for an adoptive placement and be considered by the agency in the same way as families identified through the agency's own search. Consequently, a small percentage of children get adopted through other channels, e.g., by friends of their foster (or birth) family. 

\subsection{Family-driven Search}
Traditionally, most initiative to act rests with families. In a typical family-driven search protocol, the agency emails all approved families when a child becomes eligible for adoption after a judge's termination of parental rights (TPR) order. The email provides brief details about the child and invites families to express interest. After families have a chance to respond, the caseworker responsible for the child compiles a list of interested families. In most jurisdictions, caseworkers must give serious consideration to every responding family. The caseworker then begins reviewing those families to determine which family is the best suitable match. Obvious mismatches (e.g., a wheelchair-dependent child and a family living in a house not designed to accommodate wheelchairs) can be quickly identified and screened out, but most candidates require a careful review of the home study and follow-up interviews. This process is time-consuming and emotionally taxing for families. If multiple families are suitable, or the decision is complex, the case is referred to a committee, such as the Adoption Applicant Review Committee in Florida, for a final determination. When no family is deemed an adequate fit, the child remains in foster care, and the child's availability is re-advertised after some time.

\subsection{Caseworker-driven Search}

In caseworker-driven search, agencies do not issue broad announcements. Instead, once a given child becomes eligible, the child's caseworker sequentially informs specific families from the approved pool and invites them to apply to adopt the child.  Family selection strategies vary across caseworkers. Some rely more on scoring rules and recommender systems, while others may rely on informal networks to prioritize families they know or apply their own heuristics to identify families to consider. 
If an invited family expresses interest, the investigation mirrors the family-driven search process, including potential committee review. If the caseworker (or the review committee) feels that a family is reasonable but not ideal, they may also invite additional families to apply before coming to a final decision. If the caseworker --- who might be managing dozens of cases at different stages of the child welfare system ---  has exhausted the pool of eligible families without a suitable match emerging, the search is suspended until the family pool has sufficiently refreshed. This means that the child remains in foster care until the search is restarted after some time. 

\section{Preliminaries}\label{sec:model}

In this section, we develop an analytical model to contrast the two search disciplines and derive characterizations of agents' utilities.

\subsection{Model}

In our model, agents (children and families) have observable characteristics. Agents with the same characteristics are said to be of the same \emph{type}. 
We treat sibling sets of children who should be placed together as a single child. Furthermore, we view the caseworker as a direct representative of the child; thus, we consider a child-caseworker pair as a single child agent.
We let $C= \{ c_1,\ldots,c_n \}$ and $F=\{f_1,\ldots,f_m\}$ denote the set of all $n$ child types and the set of all $m$ family types, respectively. Individual agents are indifferent between agents of the same type, i.e., their preferences are over agent types: 
A child of type $c$ has a value $\val{c}{f} \in \RR$ for family type $f$, and 
a family of type $f$ has a value $\val{f}{c} \in \RR$ for child type $c$.
Preferences are assumed to be strict, i.e., $\val{c}{f} \neq \val{c}{f'}$ if $f \neq f'$ and $\val{f}{c} \neq \val{f}{c'}$ if $c \neq c'$.
Agents' valuations are summarized by a list of vectors $v=(v_{c_1},\ldots,v_{c_n},v_{f_1},\ldots,v_{f_m})$,
 and each agent has a value of 0 for remaining unmatched.
Given valuations $v$, we let $\bar v$ denote the maximum value of all $\val{c}{f}$ and $\val{f}{c}$.

There are infinitely many discrete time steps.
 At any time step, there is at most one agent of each type present in the system. 
 Thus, we can use $c \in C$ and $f \in F$ to refer to either individual agents and agent types without ambiguity.
 We refer to an agent (or agent type) from either set as $i \in A := C \cup F$. 
From now on, we will simply say that agent $i \in A$ is \emph{active} if they are present at the current time step. To account for multiple agents from the same type being present, we can introduce multiple agent types that are arbitrarily close in value, effectively creating a tie-breaker between agents of ``almost'' the same type.

At the beginning of each time step, all active agents are determined as follows:
 For each family type, one family of that type is active with probability $\lambda \in (0,1]$. This is determined independently for each family type.
 We call parameter $\lambda$ the \emph{market thickness indicator}, as it determines the expected number of active families at each time step: For small values of $\lambda$, there will be few active families in expectation. For large values of $\lambda$, it is quite likely that a family of each type will be active. 
 Further, exactly one child (type) is selected uniformly at random to be active. This feature is motivated by search processes used by adoption agencies: A caseworker works on the case of one child at a time, which we assume she selects randomly.\footnote{Our model does not endogenize the number of agents present in the system. 
This kind of instant replacement is a standard large-market assumption in the search-and-matching literature, and it is necessary to keep our model tractable.}

There is uncertainty regarding whether a specific child $c$ and a specific family $f$ are compatible: 
With probability $p\in (0,1)$, $f$ is a \emph{suitable} match for $c$ and \emph{unsuitable} with probability $1-p$.
Whether a match is suitable or not is determined independently at random for child-family pairs.
 We refer to $p$ as the \emph{match success probability}.
 Parameter $p$ captures the following aspect of adoption markets: When a family shows interest in a child, the family's decision is based on limited reported information, such as the sex, ethnicity, age of the child, and known disabilities. However, there are many other important characteristics of a child that determine whether the child is actually a good match for the family and whether there is mutual attraction. The same holds for a child (or his caseworker) showing interest in a family.
 Only if a family $f$ is a suitable match for child $c$ can a match between $c$ and $f$ be formed. If $c$ and $f$ form a match, they both obtain a value of $\val{c}{f}$ and $\val{f}{c}$, respectively.\footnote{Note that a uniform, binary suitability probability $p$ is an abstraction made to keep the model tractable. See online Appendix~\ref{app:theory_uncertainty_heterogeneity}.}
 Determining the suitability of a match is costly for both sides, as it is a time-consuming process. Children and families incur search costs $\kappa_C \in \RR_+$ and $\kappa_F \in \RR_+$, respectively, each time the suitability of a match including them is determined.
 All agents discount the future; however, they only discount time steps in which they are active.
Children's and families' discount factors are $\delta_C \in [0,1)$ and $\delta_F \in [0,1)$, respectively.\footnote{An alternative interpretation is to think of children and families leaving the process before the next time step when they are active with probability $1-\delta_C$ and $1-\delta_F$, respectively.} Although we keep homogeneous costs and discount factors to keep the analysis and notation concise, they can be made type-specific without loss of generality, and all our results go through. 

We model adoption matching as a dynamic process that we assume to be stationary. An instance $\instance$ together with a search technology (which we will introduce) induce a game.
 To reduce notation, we assume that instance $\instance$ is fixed unless stated otherwise.
Agents' strategies in this game are captured as follows: 
Child $c$ is either \emph{interested} in a family of type $f$ or not. Similarly, each family $f$ is either \emph{interested} in a child of type $c$ or not. We assume that all agents of the same type play the same strategy, and agents don't change their strategies in different time steps. Therefore, we can represent a strategy for a child $c$ as a vector $\strat{c}{} \in \{0,1\}^m$, where $\strat{c}{f}=1$ if $c$ is interested in matching with a family of type $f$. Similarly, a strategy for a family $f$ is given by a vector $\strat{f}{}\in \{0,1\}^n$, where $\strat{f}{c}$ indicates whether $f$ is interested in children of type $c$. A strategy profile is a tuple of vectors $s=(\strat{c_1}{},\ldots,\strat{c_n}{},\strat{f_1}{},\ldots,\strat{f_m}{})$, while we let $S$ denote the finite set of all possible strategy profiles. For $i \in A$ we let $\strat{-i}{}$ denote the tuple of all agents' strategies in $s$, except that agent $i$ is excluded. 
We say that $c$ and $f$ are \emph{mutually interested} in each other under strategy profile $\strat{}{}$ if $\strat{c}{f} = 1$ and $\strat{f}{c} = 1$. 
The set $M(\strat{}{}) = \{(c,f) \in C \times F \mid \strat{c}{f}= \strat{f}{c} = 1\}$ is called the \emph{matching correspondence of $s$}.
We use $M_i(\strat{}{})$ to denote the set of agents that agent $i$ is mutually interested in under $s$.

In the remainder of this section, we describe the two search technologies. We analyze the dynamic stochastic games induced by an instance and a search technology in a full information environment.

\textbf{Family-driven Search (FS):} 
At the beginning of each time step, after a child $c$ is randomly chosen to be active, each family that is active and interested in $c$ lets the child's caseworker know of their interest. This corresponds to families responding to an email announcement made by a caseworker. The caseworker immediately discards any families without mutual interest in $c$, i.e., where $s_c(f)=0$ or $s_f(c)=0$.
The caseworker then investigates all remaining families to determine whether they would actually be a suitable match. Recall that each investigated family is a suitable match for $c$ with probability $p$---which is determined independently for each family---and that each agent incurs cost for each investigation in which they are involved.
After all families have been processed by the caseworker, $c$ either matches with the most preferred choice from among those families identified as suitable matches or remains unmatched if no such family exists. We move on to the next time step.

\textbf{Caseworker-driven Search (CS):}
After a child $c$ is randomly chosen to be active at the beginning of a time step,
the child's caseworker can sequentially inform any of the active families. We assume families are informed in decreasing order of $\val{c}{f}$.\footnote{While this is trivially optimal in our model, higher degrees of uncertainty may warrant contacting families in a different order. See online Appendix~\ref{app:theory_uncertainty_heterogeneity} for a discussion.}
If there is mutual interest between $c$ and $f$, the suitability of a match between $c$ and $f$ is investigated. 
If the match turns out to be suitable, $c$'s search is over, $c$ and $f$ are matched and leave.\footnote{In practice, both sides still need to agree to the match. However, given that it would not be rational to investigate a match you are not willing to accept and that the remaining families all have lower values, refusing a suitable match at this point is never rational.}  We then move on to the next time step. 
Otherwise, the caseworker continues the search by selecting the next family in the list.
If all families have been processed, the child remains unmatched and we move on to the next time step. \Cref{proposition:decreasing_order} in online Appendix~\ref{app:theory_decreasing_order} shows that $c$'s utility is maximized if the caseworker processes families in decreasing order of $v_c(f)$.

\subsection{Utilities}\label{sec:model_utilities}
We define agents' utilities at a time step and characterize their flow utilities in both FS- and CS-induced stochastic games.
Assume child $c$ is active at the current time step and that active families are yet to be determined. Let $f$ be an arbitrary family.
For any $s \in S$, let $b_{cf}(\strat{}{}) = |\{f' \in M_c(\strat{}{}) \mid \val{c}{f'} > \val{c}{f} \}|$ denote the number of families in $M_c(\strat{}{})$ that $c$ likes better than $f$.
Further, let $\bp{c}{f}{s}$ denote the probability that $c$ will \emph{not} match with any other family that $c$ prefers over $f$ at the current time step. 
Noting that for any child $c$ the probability that a mutually interested family $f'$ is active at the current time step and a suitable match is $\lambda p$, it follows immediately that $\bp{c}{f}{s} = (1 - \lambda p)^{b_{cf}(\strat{}{})}$ for both FS and CS.

In FS, since investigations are conducted simultaneously, search costs are incurred for all of them, and therefore, the \emph{expected immediate utility} for the active child $c$  \emph{at an arbitrary time step} follows as
\begin{equation}
    \utt{FS}{c}{s} = \lambda \sum_{f' \in M_c(\strat{}{})} \Big( \bp{c}{f'}{s} p \val{c}{f'} - \kappa_C \Big).
\end{equation}
In CS, as investigations are conducted sequentially, search costs $\kappa_C$ are only incurred if the child has not successfully matched with a higher-valued family. Consequently, the \emph{expected immediate utility} of the active child $c$ follows as
\begin{equation}
    \utt{CS}{c}{s} = \lambda \sum_{f' \in M_c(\strat{}{})}\bp{c}{f'}{s} \left( p \val{c}{f'} - \kappa_C \right).
\end{equation}
In both cases above, $\lambda$ is the probability that a family is present to be investigated when the child is active.

Similarly, for a family $f$, we can express the expected immediate utilities for FS and CS at an arbitrary time step (conditional on $f$ being active) by 
\begin{equation}
    \utt{FS}{f}{s} = \frac{1}{n} \sum_{c' \in M_f(\strat{}{})} \Big( \bp{c'}{f}{s} p \val{f}{c'} - \kappa_F \Big)
\end{equation}
\begin{equation}
    \utt{CS}{f}{s} = \frac{1}{n} \sum_{c' \in M_f(\strat{}{})} \bp{c'}{f}{s} \left( p \val{f}{c'} - \kappa_F \right),
\end{equation}
 where $1/n$ is the probability that a given child is active in a timestep.

 From this, the crucial difference between FS and CS in our model becomes apparent: In FS, search costs are always incurred if there is mutual interest between agents. In CS, however, the sequential nature of the search means that search costs are only incurred if there is mutual interest and all previous match attempts at the time step have been unsuccessful.

We assume that each agent is risk-neutral and maximizes their expected (overall) utility, which is the expected discounted value of their eventual match minus the total discounted search costs they incur.\footnote{In practice, while caseworkers have multiple cases assigned to them, their goal with each case is to maximize the utility of the child belonging to that case. That is, based on our interviews with domain experts, the caseworker considers all assigned cases separately, and each child-caseworker pair is considered a separate self-interested agent.}
We use $(z)^+$ as shorthand notation for $\max\{z,0\}$.
$\Ind\big[ \cdot \big]$ is the indicator function, which has value $1$ if its argument is true and value $0$ otherwise.
We denote the expected (overall) utility of agent $i$ under strategy profile $s$ by $\ut[FS]{i}{\strat{}{}}$ in FS and $\ut[CS]{i}{\strat{}{}}$ in CS. Whenever it is clear from context whether we are referring to FS or CS we will simply write $\ut{i}{\strat{}{}}$.
\Cref{proposition:utility_characterization} characterizes children's and families' utilities in FS and CS via balance equations.

\begin{proposition}\label{proposition:utility_characterization}
	Given strategy profile $s$, child $c$'s utility in FS is the unique value $\ut[FS]{c}{s}$ that satisfies
	\begin{equation}\label{equation:utility_fs_c}
		\ut[FS]{c}{s} = \delta_C \ut[FS]{c}{s} + \lambda \sum_{f \in M_c(\strat{}{})} \Big( \bp{c}{f}{s} p \big( \val{c}{f} - \delta_C \ut[FS]{c}{s} \big) - \kappa_C \Big).
	\end{equation}
	Similarly, family $f$'s utility in FS is the unique value $\ut[FS]{f}{s}$ that satisfies		
	\begin{equation}\label{equation:utility_fs_f}
		\ut[FS]{f}{s} = \delta_F \ut[FS]{f}{s} + \frac{1}{n} \sum_{c \in M_f(\strat{}{})} \Big( \bp{c}{f}{s} p \big( \val{f}{c} - \delta_F \ut[FS]{f}{s} \big) - \kappa_F \Big).
	\end{equation}
	In CS, child $c$'s utility in CS is the unique value $\ut[CS]{c}{s}$ that satisfies
	\begin{equation}\label{equation:utility_cs_c}
		\ut[CS]{c}{s} = \delta_C \ut[CS]{c}{s} + \lambda \sum_{f \in M_c(\strat{}{})} \bp{c}{f}{s} \Big( p \big( \val{c}{f} - \delta_C \ut[CS]{c}{s} \big) -\kappa_C \Big).
	\end{equation}
	Similarly, family $f$'s utility in CS is the unique value $\ut[CS]{f}{s}$ that satisfies		
	\begin{equation}\label{equation:utility_cs_f}
		\ut[CS]{f}{s} = \delta_F \ut[CS]{f}{s} + \frac{1}{n} \sum_{c \in M_f(\strat{}{})} \bp{c}{f}{s} \Big( p \big( \val{f}{c} - \delta_F \ut[CS]{f}{s} \big) - \kappa_F \Big).
	\end{equation}
\end{proposition}

A formal proof can be found in online Appendix~\ref{app:utility_characterization}.
One difference between FS and CS is immediately apparent from the above balance equations. In CS, search costs only incur if previous match attempts have been unsuccessful. In FS, however, costs are always incurred if a family is present and mutually interested.

\section{Equilibria}\label{sec:equilibria}

We use a tie-breaking assumption, which allows us to exclude degenerate equilibria later on. After stating this assumption, we introduce two classes of strategies --- one for CS and one for FS --- and show that these classes capture agents' best responses. This will be helpful for obtaining results later on. We show that for both search technologies equilibria always exist and that equilibria form a lattice.

\subsection{Threshold Strategies}\label{sec:threshold_strategies}

For both search technologies, we make the following tie-breaking assumption: If agent $i$'s utility would (weakly) increase from mutual interest with agent $j$, then $i$ will be interested in $j$---even if $j$ is not interested in $i$. Similarly, if agent $i$'s utility would decrease from mutual interest with agent $j$, then $i$ will not be interested in $j$. 
This assumption allows us to exclude degenerate equilibria (e.g., no agent being interested in any other agent) later on without restricting agents in their endeavor to maximize their utility.

We now introduce \emph{threshold strategies} for FS and CS. As we will see, our tie-breaking assumption implies that agents' best responses belong to the class of threshold strategies.
Note that a best response always exists, because $S$ is finite.

\begin{definition} \label{definition:threshold-strategy}
	Child $c$ plays a \emph{CS threshold strategy} (CS-TS) \emph{with threshold} $y_c \in \RR$ \emph{in} $s$, if
	\begin{equation}
    	\strat{c}{f} = \Ind\big[p(\val{c}{f} - \delta_C y_c) \ge \kappa_C \big] \mbox{ for all } f \in F.
	\end{equation}
	Family $f$ plays a \emph{CS-TS with threshold} $y_f$ \emph{in} $s$, if 
	\begin{equation}
	    \strat{f}{c} = \Ind\big[p(\val{f}{c} - \delta_F y_f) \ge \kappa_F\big] \mbox{ for all } c \in C.
	\end{equation}
	Child $c$ plays an \emph{FS threshold strategy} (FS-TS) \emph{with threshold} $y_c$ \emph{in} $s$, if
	\begin{equation}
	    \strat{c}{f} = \Ind\big[\bp{c}{f}{s} p ( \val{c}{f} - \delta_C y_c) \ge \kappa_C\big] \mbox{ for all } f \in F.
	\end{equation} 
	Family $f$ plays an \emph{FS-TS with threshold} $y_f$ \emph{in} $s$, if 
	\begin{equation}
	    \strat{f}{c} = \Ind\big[ \bp{c}{f}{s} p (\val{f}{c} - \delta_F y_f ) \ge \kappa_F\big] \mbox{ for all } c \in C.
	\end{equation} 
\end{definition}

In a CS-TS or an FS-TS, the threshold $y_i$ can be interpreted as $i$'s reservation utility. However, note that these threshold strategies are more involved than standard \emph{simple threshold strategies} \citep{adachi_search_2003,immorlica_designing_2024}. While in a simple threshold strategy, an agent would be interested if the expected value is above their reserve utility, i.e., $p\val{i}{j}\geq y_i$, this does not suffice in this case. Instead, agents also have to account for their costs and, in the case of FS, for the likelihood that the child may find another match with a higher value during the same period.\footnote{See \Cref{proposition:simple_thresholds} in online Appendix~\ref{app:theory_simple_thresholds} for an illustration of the non-existence of simple threshold best responses.}

Let $u^{FS*}_i(\strat{-i}{})$ and $u^{CS*}_i(\strat{-i}{})$ denote $i$'s utility from a best response to $\strat{-i}{}$ in FS and CS, respectively.
As before, we simply write $\ut[*]{i}{\strat{-i}{}}$ if there is no ambiguity.
\Cref{proposition:ts_best_response} shows that agents' best responses always have the form of a threshold strategy.

\begin{proposition}\label{proposition:ts_best_response}
	Let $i \in A$ and $\strat{-i}{}$ be an arbitrary strategy profile of all agents excluding $i$. In both FS and CS, a best response of $i$ to $\strat{-i}{}$ corresponds to a threshold strategy with threshold $\ut[*]{i}{\strat{-i}{}}$.
\end{proposition}

A formal proof can be found in online Appendix~\ref{app:ts_best_response}.
To derive results later on, it will prove useful to switch between thresholds and strategies. We therefore provide the following definition.

\begin{definition} \label{definition:threshold-profile}
	In CS, a \emph{strategy profile induced by threshold profile} $y \in \RR^{n+m}$ is denoted by $s^{CS}(y)$ and satisfies for each $i \in A$, $s^{CS}_i(y)$ is a CS-TS with threshold $y_i$ in $s^{CS}(y)$.
	In FS, a \emph{strategy profile induced by threshold profile} $y \in \RR^{n+m}$ is denoted by $s^{FS}(y)$ and satisfies for each $i \in A$, $s^{FS}_i(y)$ is a FS-TS with threshold $y_i$ in $s^{FS}(y)$.
\end{definition}

We again omit the superscript if this does not lead to ambiguity.
It is trivial to obtain $s^{CS}(y)$ by inserting $y$ in the corresponding equations in \Cref{definition:threshold-strategy}.
Algorithm~\ref{alg:fs_thresholds} from online Appendix~\ref{app:alg_fs_ts} can be used to compute $s^{FS}(y)$. From now on, we use $\bp[FS]{c}{f}{y}$ and $\bp[CS]{c}{f}{y}$ as shorthand-notation for $\bp{c}{f}{s^{FS}(y)}$ and $\bp{c}{f}{s^{CS}(y)}$, respectively. Whenever it is clear from the context, we simply write $\bp{c}{f}{y}$.

\subsection{Equilibrium Existence and Lattice Structure}\label{sec:equilibrium_existence}

In this section, we show that Nash equilibria always exist under both search technologies.\footnote{Technically, we have a stochastic game model, and therefore, these are Nash equilibria of stochastic games. They correspond to the Markov-perfect Nash-equilibrium selection among subgame-perfect Nash equilibria when considered as a repeated game.}
We say that strategy profile $\strat{}{}$ is an \emph{equilibrium in FS (FSE)} if $\strat{}{}$ is a Nash equilibrium in the game induced by FS. Analogously, strategy profile $\strat{}{}$ is an \emph{equilibrium in CS (CSE)} if $\strat{}{}$ is a Nash equilibrium in the game induced by CS.
We use $S^{FS}$ to denote the set of FSE, and let $Y^{FS} = \{(\ut{i}{s})_{i \in A}\}_{s \in S^{FS}}$ be the corresponding set of \emph{equilibrium threshold profiles in FS}. For CS, those sets are defined analogously.
Before we can prove that these sets are never empty, we need to define a partial order $\le_C$ on $Y = [0,\bar v]^{n+m}$. Note that if agents only play individually rational strategies, their utility is always lower bounded by $0$ and upper bounded by $\bar v$.
\begin{definition}
	Let $\le_C$ be the partial order on $Y$, where for all $y,y' \in Y$ it holds that $y \le_C y'$ if and only if $y_c \le y'_c$ for all $c \in C$ and $y_f \ge y'_f$ for all $f \in F$.
\end{definition}

Having defined partial order $\le_C$, we can now prove that equilibria always exist in both settings. We state this result in \Cref{proposition:equilibria_existence}.

\begin{proposition}\label{proposition:equilibria_existence}
	The set of FS and CS equilibrium threshold profiles $Y^{FS}$ and $Y^{CS}$ is non-empty and both $(Y^{FS},\le_C)$ and $(Y^{CS},\le_C)$ form complete lattices.
\end{proposition} 

A formal proof can be found in online Appendix~\ref{app:equilibria_existence}. The proof proceeds by defining a best-response mapping and showing fixed-point existence using  Tarski's fixed-point theorem \citep{tarski_lattice-theoretical_1955}. Such fixed points coincide with pure-strategy equilibria. 

In general, there can be more than one FSE or CSE for a fixed instance $\instance$. \Cref{proposition:equilibria_existence} not only guarantees that equilibria always exist in both settings, but also highlights that there is a special ordering over equilibria: there exists a child-optimal equilibrium that children unanimously prefer over all other equilibria; i.e., their utility is weakly higher compared to any other equilibrium. Similarly, there exists a family-optimal equilibrium that families prefer. From now on, we let $\se{co-CS}$ denote the \emph{child-optimal CSE (co-CSE)}, $\se{co-FS}$ the \emph{child-optimal FSE (co-FSE)}, $\se{fo-CS}$ the \emph{family-optimal CSE (fo-CSE)}, and $\se{fo-FS}$ the \emph{family-optimal FSE (fo-FSE)}. This is reminiscent of the structure of the set of stable matchings in standard two-sided matching markets \citep{knuth_stable_1997}.

\section{Comparison of Family-driven Search and Caseworker-driven Search}\label{sec:theory_comparison}

In this section, we investigate the impact of the two search technologies on equilibrium outcomes. Here, we present our main theoretical result: An FSE can never Pareto dominate a CSE, as any increase in utility for one agent can only arise if another agent lowers their interest threshold, corresponding to a decrease in that agent's utility. There exist instances, however, where each CSE is a Pareto improvement over all FSEs.
We further find that no approach is always preferable for either children or families.

\subsection{Pareto Comparison}\label{sec:theory_comparison_pareto}

A natural way to determine which equilibrium outcomes are preferable is to check whether one equilibrium is a Pareto improvement over the other. We first formalize the Pareto dominance relationship for strategy profiles in our model.

\begin{definition}
	Strategy profile $s \in S$ is a Pareto improvement over strategy profile $s' \in S$ if $\ut{i}{s'} \le \ut{i}{s}$ for all $i \in A$ and there exists $j \in A$, such that $\ut{j}{s'} < \ut{j}{s}$.
\end{definition}

Note that $\ut{i}{s}$ either denotes $\ut[CS]{i}{s}$ or $\ut[FS]{i}{s}$, depending on whether we refer to $s$ as a CSE or an FSE.
We find that FSEs can never Pareto dominate CSEs, but there are instances where each CSE Pareto dominates all FSEs. Before we can formally show this, we need to state two lemmas. The first lemma is useful for understanding why FSEs cannot Pareto dominate CSEs. \Cref{lemma:additional_match_fs} shows that if there is a pair with mutual interest in FS that is not present in CS, then at least one of the two agents in the pair must be strictly worse off in FS compared to CS.

\begin{lemma}\label{lemma:additional_match_fs}
	Let $\se{FS}\in S^{FS}$ and $\se{CS} \in S^{CS}$. If there exists $c\in C$ and $f \in F$, such that $(c,f) \in M(\se{FS})$ and $(c,f) \notin M(\se{CS})$, then either $\ut{c}{\se{FS}} < \ut{c}{\se{CS}}$ or $\ut{f}{\se{FS}} < \ut{f}{\se{CS}}$.
\end{lemma}
A formal proof can be found in online Appendix~\ref{app:additional_match_fs}. 
Intuitively, if two agents are not mutually interested in each other in CS but are in FS, then at least one of them had to lower their interest threshold in FS---which means that their optimal reservation utility is lower in FS. However, since the optimal reservation utility corresponds to the once-discounted utility, this agent must be strictly worse off.

The next lemma almost immediately follows from \Cref{lemma:additional_match_fs} and is used for the proof of \Cref{theorem:pareto_dominance} as well as for later results.

\begin{lemma}\label{lemma:same_matches_cs_weakly_better}
	Let $\se{FS}\in S^{FS}$, $\se{CS} \in S^{CS}$, and $c \in C$. If $M_c(\se{FS}) \subseteq M_c(\se{CS})$, then $\ut{c}{\se{FS}} \le \ut{c}{\se{CS}}$.
\end{lemma}
A formal proof can be found in online Appendix~\ref{app:same_matches_cs_weakly_better}.

It is quite intuitive that a child $c$ cannot be worse off under CS if all families that are mutually interested in $c$ under FS are also interested in $c$ under CS.
 This allows us to show the following theorem as our main result.  

\begin{theorem}\label{theorem:pareto_dominance}
	An FSE can never be a Pareto improvement over a CSE. On the other hand, there exists an instance where all CSEs are Pareto improvements over all FSEs.
\end{theorem}
A formal proof can be found in online Appendix~\ref{PROOF:pareto_dominance}.
The main intuition for why an FSE can never be a Pareto improvement over a CSE is that the only way an agent can be better off in FS compared to CS is to have a higher chance of matching with someone they like. But with \Cref{lemma:additional_match_fs}, this implies that some other agent had to lower their interest threshold, which means that their utility decreased.
There are two reasons why a CSE can Pareto dominate an FSE. First, CS can save agents' search costs. Second, because search costs are only incurred in CS if previous match attempts at the current time step have been unsuccessful, agents do not have to worry about accumulating search costs as much in CS as in FS. Therefore, agents are incentivized to express interest in more potential match candidates in CS compared to FS.

\subsection{No Approach Dominates the Other}\label{sec:theory_comparison_no_domination}

Even though CSEs can be Pareto improvements over FSEs, we find that CS is not always better for everyone compared to FS.
In fact, a CSE might yield arbitrarily higher (or lower) utility for all children or families compared to an FSE.

\begin{proposition}\label{proposition:multiplicity_utility_gaps}
	For any $L > 0$ and $0 < \epsilon < L$, there exists an instance where
	\begin{enumerate}
		\item the child-optimal equilibrium, which we denote as $\se{co}$, is the same in both CS and FS, and similarly, the family-optimal equilibrium, which we denote as $\se{fo}$, is the same in both CS and FS,
		\item $\ut{c}{\se{co}} = L$ for all $c \in C$ and $\ut{f}{\se{co}} \le \epsilon$ for all $f \in F$, and
		\item $\ut{c}{\se{fo}} \le \epsilon$ for all $c \in C$ and $\ut{f}{\se{fo}} = L$ for all $f \in F$.
	\end{enumerate}
\end{proposition}
A formal proof can be found in online Appendix~\ref{PROOF:multiplicity_utility_gaps}.
It proceeds by constructing examples where child and family utilities are mismatched, causing the utility gap between child-optimal and family-optimal equilibria to be arbitrarily large in both FS and CS. Thus, depending on which equilibria are realized, both approaches can be arbitrarily better for either side of the market.	

Both a single child and a single family can be arbitrarily worse off in equilibrium under CS, even if FS and CS admit only one equilibrium each. This is highlighted by the following two propositions.

\begin{proposition}\label{proposition:child_worse_off_in_cs}
	There exists an instance where a child is strictly worse off under the unique CSE compared to the unique FSE.
\end{proposition}
A formal proof can be found in online Appendix~\ref{PROOF:child_worse_off_in_cs}. It proceeds by constructing an example with two child and family types where child $c_1$ is significantly preferred over $c_2$ by all families, while all children slightly prefer $f_1$ over $f_2$.  This implies that family $f_2$ is only matched with child $c_1$ if no $f_1$ family is currently present. The higher search costs in FS can then make $f_2$ lose interest in $c_1$, regardless of patience levels, causing $f_2$ to settle for $c_2$. This allows $c_2$ to be matched. Conversely, in CS, family $f_2$ does not incur high search costs for waiting until they are matched with a $c_1$. If they are patient enough, $f_2$ therefore prefers waiting for their preferred choice $c_1$. This leaves $c_2$ without any interested family and, therefore, unmatched.  
 
Similarly, a family can be worse off in CS when FS and CS each only admit one equilibrium.

\begin{proposition}\label{proposition:family_worse_off_in_cs}
	There exists an instance where a family is strictly worse off under the unique CSE compared to the unique FSE.
\end{proposition}

A formal proof can be found in online Appendix~\ref{PROOF:family_worse_off_in_cs}.
Just as children can benefit from families that decide to settle for a less preferred child, so can other families. It can be the case that a family $f$ is interested in a child $c$ in a CSE but $f$ is not interested in $c$ in an FSE, because the associated expected costs would be too high. Not having $f$ as competition might be enough incentive for another family $f'$ to be interested in $c$ under the FSE. As a result, $f'$ can be strictly better off in FS.

\section{Effects of Model Parameters}\label{sec:theory_parameters}

We showed that FSEs cannot be Pareto improvements over CSEs, but CSEs can be Pareto improvements over FSEs.
Additionally, we found that some agents can be better off under an FSE compared to a CSE. In order to better understand the conditions under which one of the two approaches might be preferable, we explore the effects that different parameters have on equilibrium outcomes in FS and CS. 
We provide two more results in favor of CS: 
First, we show that as families' patience decreases, at some point all children will be weakly better off in any CSE compared to any FSE.
Second, increasing supply on the family side, i.e., increasing the market thickness indicator $\lambda$, can negatively affect children's utilities in FS but not in CS.
Finally, as a sanity check, we investigate the effect of certain parameters or parameter combinations in the limit; all of these latter results can be found in \Cref{app:theory_limit_results}.

\subsection{Discount Factors}\label{sec:parameter_effects_discounting}

For this subsection, let $S^{CS}(\delta_F')$ and $S^{FS}(\delta_F')$ denote the set of CSEs and FSEs when $\delta_F = \delta_F'$, respectively.
We now demonstrate that as families' patience decreases below a certain threshold, all children will always be better off in CS than in FS.

\begin{proposition}\label{proposition:impatient_families}
	For each instance there exists $\bar \delta_F \in [0,1)$, such that for all $\delta_F' \in [0, \bar \delta_F]$ it holds that $\ut{c}{\strat[FS]{}{}} \le \ut{c}{\strat[CS]{}{}}$ for all $c \in C$, $\strat[CS]{}{} \in S^{CS}(\delta_F')$, $\strat[FS]{}{} \in S^{FS}(\delta_F')$. 
\end{proposition}
 A formal proof can be found in online Appendix~\ref{PROOF:impatient_families}. Intuitively, the statement follows because in CS, families' interest in a very unlikely match incurs lower search costs than in FS. While patient families may still not be interested in some children in CS that they are interested in under FS (which drives Proposition~\ref{proposition:child_worse_off_in_cs}), any sufficiently impatient family will be unwilling to wait.

However, as can be seen in the proof of \Cref{proposition:family_worse_off_in_cs}, an analogous statement for families' utilities and children's patience level does not hold.
Intuitively, a family $f$ might be worse off in CS, because another family $f'$ is not shying away from $c$, as $f'$ does not have to worry about wasted search efforts in CS.

\subsection{Market Thickness}\label{sec:parameter_effects_market_thickness}

Adoption agencies might intuitively prefer to have a larger pool of available families to choose from. Here, we present another result that suggests this might be generally beneficial in CS but not always in FS when it comes to children's utilities. Increasing supply on the family side, i.e., increasing the market thickness indicator $\lambda$, can negatively affect children's utilities in FS but not in CS.
For the remainder of \Cref{sec:parameter_effects_market_thickness}, assume that all instance parameters are fixed except for $\lambda$. Let $s^{co-CS,\lambda}$ denote the child-optimal CSE given market thickness indicator $\lambda$. Definitions for $s^{fo-CS,\lambda}$, $s^{co-FS,\lambda}$, and $s^{fo-FS,\lambda}$ are analogous. 
\Cref{proposition:increased_lambda_fs} shows that increasing $\lambda$ can lead to some children being worse off in FS.

\begin{proposition}\label{proposition:increased_lambda_fs}
	There exists an instance with a child $c \in C$ and $\lambda, \lambda' \in (0,1]$ with $\lambda < \lambda'$, such that $u_c(s^{co-FS,\lambda}) > u_c(s^{co-FS,\lambda'})$.
\end{proposition}

A formal proof can be found in online Appendix~\ref{PROOF:increased_lambda_fs}. Effectively, what is happening is that if multiple families are interested in a child, then increased market thickness $\lambda$ increases competition and, therefore, the search costs for less preferred families. If the child is close to being indifferent between families and some families lose interest due to the higher cost, then the resulting decrease in the child's utility can be larger than the increase caused by a higher chance to match with a slightly more preferred family.

In CS, on the other hand, increasing $\lambda$ can only have a positive effect on children's utilities in equilibrium. 

\begin{proposition}\label{proposition:increased_lambda_cs}
	Let $\lambda, \lambda' \in (0,1]$, such that $\lambda \le \lambda'$. Then $u_c(s^{co-CS,\lambda}) \le u_c(s^{co-CS,\lambda'})$ and $u_c(s^{fo-CS,\lambda}) \le u_c(s^{fo-CS,\lambda'})$ for all $c \in C$.
\end{proposition}

A formal proof can be found in online Appendix~\ref{app:increased_lambda_cs}.
The reason why this holds in CS is that, unlike in FS, families will not shy away from children in whom they are interested just because the probability of matching with them decreases. This result is reminiscent of a similar result in standard two-sided matching markets, as the number of agents in one side increases, the other side agents become all unambiguously better off under side-optimal stable matchings \citep{gale/sotomayor:85}. However, it only holds for CS and only for the children's welfare.

\section{Numerical Evaluation}\label{sec:num_eval}

We previously established that CSEs can be Pareto improvements over FSEs while FSEs cannot be Pareto improvements over CSEs, and that agents can be better off in either approach (see \Cref{theorem:pareto_dominance} and \Cref{proposition:multiplicity_utility_gaps}).
Additionally, we have shown that all children will be better off in CS compared to FS if families are sufficiently impatient.
Here, we present numerical results to further investigate the conditions under which children and families will be better off in CS or FS.
Our results suggest that CS is almost always preferable for both sides of the market. Only when agents' preferences are perfectly correlated and families are very patient do we find that, on average, there are more child types better off in FS than in CS in equilibrium.

\Cref{sec:num_eval_setup} describes how our numerical experiments are set up. We then explain how equilibria are computed in \Cref{sec:equilibrium_computation}. In \Cref{sec:num_eval_results}, we compare FS and CS in terms of their Pareto dominance relationship. We further quantify how many agents are typically better off in either approach.

\subsection{Setup}\label{sec:num_eval_setup}
We now describe the setup of our numerical evaluation. 
We set the number of agent types on each side to be $n=m=50$. 

\textbf{Valuations:}
For the generation of agents' valuations, we follow other approaches from the matching literature \citep{abdulkadiroglu_expanding_2015, mennle_power_2015}.
Each agent type $i \in A$ is uniformly assigned a ``quality'' $q_i$ at random from $[0,1]$.
Then, for each child-family pair $(c,f) \in C \times F$, idiosyncratic values $\eta_c(f)$ and $\eta_f(c)$ are randomly drawn from $[0,1]$. For a given value $\alpha \in [0,1]$, we obtain the preliminary valuations $v'_c(f) = \alpha q_f + (1-\alpha) \eta_c(f)$ and $v'_f(c) = \alpha q_c + (1-\alpha) \eta_f(c)$. Note that as $\alpha$ increases, agents' preferences become more similar and end up being identical (vertical) for $\alpha=1$. 
Final valuations $v$ are obtained by normalizing $v'$, such that the minimal and maximal value that each agent has for a match is $0$ and $1$, respectively.
Although we consider various values for $\alpha$, note that the practical level of verticality in child welfare tends to be relatively low. For example, data provided by the platform that we examine in Section~\ref{sec:empirical_casestudy} shows notable variation in families' stated preferences on several dimensions, which, taken together, suggests that preferences are not very aligned. See online
Appendix~\ref{app:fam_pref} for details. 

\textbf{Data:}
We generated $200$ quality-value pairs $(q^{(1)},\eta^{(1)}),\ldots,(q^{(200)},\eta^{(200)})$ as described above. 
Parameters $p$ and $\lambda$ are chosen to be $p = \lambda = 0.5$, and we let $\delta := \delta_C = \delta_F$ and $\kappa := \kappa_C = \kappa_F$
For each pair $(q^{(k)},\eta^{(k)})$, we computed the child-optimal CSE/FSE and the family-optimal CSE/FSE for each combination of $\alpha$, $\delta$, and $\kappa$, where $\alpha \in \{0, 1/3, 2/3, 1.0\}$, $\delta \in \{0.8,0.9,0.975,0.99\}$, and $\kappa \in \{0.01,0.02,0.05,0.1\}$.
Thus, we consider $200 \cdot 4 \cdot 4 \cdot 4 = 12800$ different instances and compute a total of $51200$ equilibria. Unless specified otherwise, results are averaged over all instances.

\subsection{Equilibrium Computation}\label{sec:equilibrium_computation}

The mapping $T$ defined in the proof of \Cref{proposition:equilibria_existence} can be used to find the child-optimal and family-optimal equilibria. The following procedure converges to an equilibrium threshold profile: Start from the $\le_C$-minimal element in $Y$, i.e., the minimum point of the lattice spanned by the partial order $\le_C$ over the threshold vectors in the game, and recursively apply $T$ to it. The $\le_C$-minimal element in $Y$ is the threshold vector $\underline{y}$ where $\underline{y}_c=0$ for each child $c$ and $\underline{y}_f=\bar v $ for each family $f$. This produces a sequence $y^0,y^1,y^2,\ldots$ of threshold profiles, which converges to the fo-FSE. Starting from the $\le_C$-maximal element in $Y$, i.e., the maximum point of the lattice spanned by the partial order $\le_C$ over the threshold vectors in the game, yields the co-FSE. The $\le_C$-maximal element is the threshold vector $\overline{y}$ where $\overline{y}_c=\bar v$ for each child $c$ and $\overline{y}_f=0$ for each family $f$. In order to terminate after a finite number of steps, we force the procedure to stop once $|y^k_i - y^{k+1}_i| \le \epsilon$ for all $i \in A$ for some previously chosen small parameter $\epsilon > 0$. The threshold profiles obtained by this procedure can then be mapped to the corresponding strategy profiles. We have performed additional checks to validate that the computed strategy profiles are indeed equilibria.

\subsection{Results}\label{sec:num_eval_results}

Before comparing FS and CS, we first note that family- and child-optimal equilibria in FS coincide roughly $97\%$ of the time. The same holds for CS. For simplicity, we only consider family-optimal equilibria in our analysis. As equilibria are almost always unique within each search technology, results for child-optimal equilibria do not differ markedly.  We do, however, observe substantial differences between FS and CS. Of all cases considered, the CSE and the FSE only coincide once in the sense that the same agents are mutually interested in each other. 

\begin{figure}[ht]
	\centering
	\includegraphics[scale=0.55]{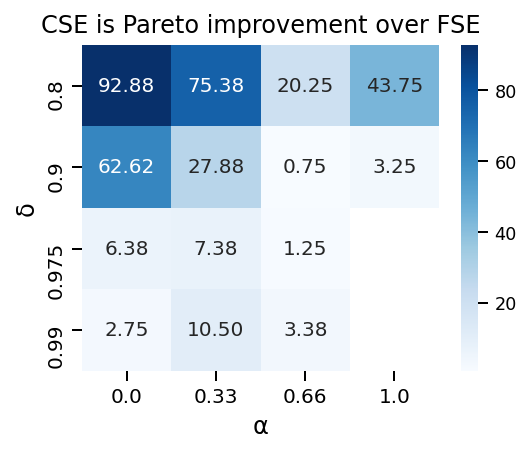}
	\caption{Percentage of instances in which the CSE is a Pareto improvement over the corresponding FSE for different combinations of agents' patience levels and degree of preference correlation.}
	\label{fig:pareto_dominance}
\end{figure}

Consistent with \Cref{theorem:pareto_dominance}, the family-optimal FSE never represents a Pareto improvement upon the corresponding family-optimal CSE. However, for approximately 22\% of all instances, the CSE Pareto dominates the corresponding FSE. \Cref{fig:pareto_dominance} shows the distribution of cases in which the CSE dominates an FSE for different discount factors and levels of correlation among preferences. Two insights emerge from this analysis: First, as agents become more impatient, CSEs more frequently constitute Pareto improvements over FSEs. As indicated by \Cref{proposition:impatient_families}, once families become sufficiently impatient, any CSE will Pareto dominate all FSEs.
Second, when preferences exhibit high correlation, CSEs rarely Pareto dominate FSEs. The case of vertical preferences --- i.e., $\alpha = 1$ --- helps to explain this effect. If agents are patient enough, a family $f$ in the CS regime might wait for an opportunity to match with a high-type child $c$, even if $f$ is not $c$'s first choice and $f$ must wait a long time until getting matched. In FS, however, if there are enough other families that $c$ prefers over $f$, $f$ or $c$ might shy away from being interested in order to avoid accumulating search costs for such an ``unlikely'' match. In that case, $f$ might settle for another low-type child or multiple low-type children instead (see the example from the proof of \Cref{proposition:child_worse_off_in_cs}). These low-type children now benefit from FS, while $f$ will be worse off in FS compared to CS.

\begin{figure}[ht]
	\centering
	\includegraphics[scale=0.50]{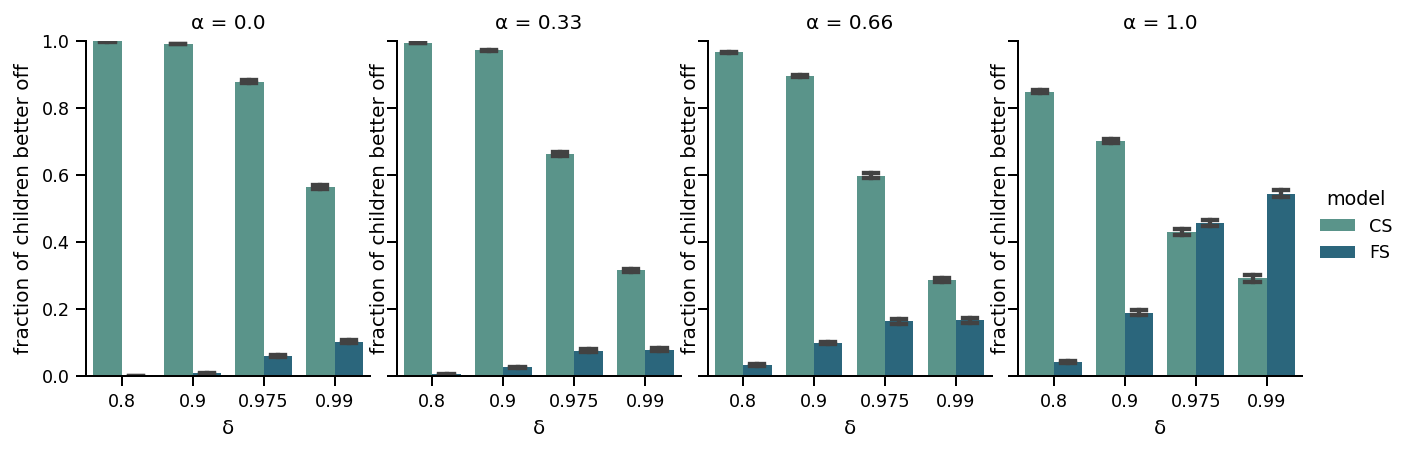}
	\caption{The ratio of children on average (strictly) better off in terms of utilities in either approach in the family-optimal equilibrium for different combinations of agents' patience and preference correlation level.}
	\label{fig:children_better_off_utilities}
\end{figure}

The previous Pareto comparison only allows for a very high-level comparison of FS and CS.
To better understand the conditions under which certain agents benefit from FS or CS, we compare the number of agents who are better off in either search discipline. Our numerical experiments show that all families are almost always better off in CS. We refer the reader to online Appendix~\ref{app:num_eval_families} for more details on families' statistics. For children, the combination of model parameters affects which approach appears more appealing.
\Cref{fig:children_better_off_utilities} shows how many children are (strictly) better off (in terms of utilities) in CS and FS for different parameter combinations.

CS provides higher utility than FS for almost all children when agents are sufficiently impatient (e.g., $\delta = 0.8$) because CS allows agents to express interest in more potential match partners without risking wasted search efforts. Being interested in more agents increases the probability of getting matched at each time step, which is especially valuable to children when patience is low. On the other hand, FS incentivizes agents to focus on a smaller set of match candidates due to higher expected total search costs. When $\delta = 0.8$, families will, on average, be interested in $35.9$ and $43.5$ child types in FS and CS, respectively.\footnote{The probability of matches occurring is another metric of interest to stakeholders. Our findings on match probabilities are presented in online Appendix~\ref {app:num_eval_match_probabilities}.}
Interestingly, more children benefit from FS than CS when agents are extremely patient and agents' preferences are almost completely aligned. Although unlikely to occur in practical child welfare settings, this explains why CSEs are less frequently Pareto improvements over FSEs under these conditions, as we previously saw in \Cref{fig:pareto_dominance}. 

\begin{figure}[ht]
	\centering
	\includegraphics[scale=0.50]{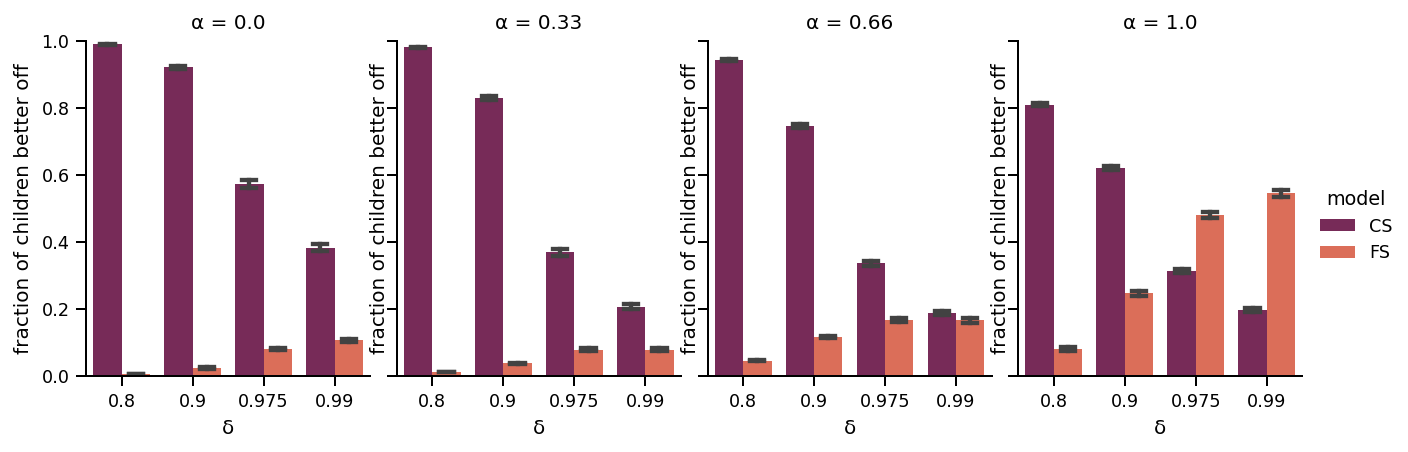}
	\caption{The ratio of children on average (strictly) better off in terms of (expected discounted) match value in either approach in the family-optimal equilibrium for different combinations of agents' patience and the level of preference correlation.}
	\label{fig:children_better_off_match_values}
\end{figure}

\Cref{fig:children_better_off_match_values} shows that CS not only reduces wasted search efforts in many cases but also enables children to match with more preferred families. We calculate a child's match value as the child's utility, ignoring the expected search costs, which might be less relevant to a policymaker trying to improve child outcomes.

\section{Empirical Evidence from a Field Implementation}\label{sec:empirical_casestudy}

To complement our model and understand the real-world implications of switching from an FS to a CS approach, we analyze children's outcomes for a multi-county \textit{circuit} in Florida that started implementing a CS approach on July 1, 2018, by adopting a technology platform, which we simply refer to as the \textit{platform}. Florida's child welfare system is organized into 20 circuits, each of which is independently administered on the state's behalf by a nonprofit community-based care (CBC) organization contracted by the state; our dataset contains no instances of a child's case transferring between circuits. The focal circuit's CBC --- which we refer to as the \textit{agency} --- was among the first circuits to adopt the platform, which became available to other circuits on a rolling basis throughout the remainder of 2018 and beyond as each agency completed the required licensing agreements and user training. While the platform was made available across circuits, the level of leadership buy-in varied widely: most circuits made it available as a supplementary tool for harder-to-place children, but none, except the focal agency, mandated its use for all cases.

The agency's previous FS approach relied on regular email announcements to the full pool of registered families to advertise children in need of adoption. Out of frustration with difficulties in finding adoptive placements, the agency's leadership decided to implement the platform and follow a CS discipline. In the new approach, caseworkers contact individual families listed on the platform whom they consider a good match for a child. It is important to distinguish the search discipline --- the agency's managerial decision to transition from FS to CS --- from the technology platform that supports it. The platform provides a searchable, ranked database of pre-registered families; it did not recruit families, manage cases, or automate matching decisions throughout the study period. The agency retained full control over family recruitment throughout the study period; the platform's only recruitment-adjacent role was sending postcards to families already approved to adopt, inviting them to create a profile and complete a registration questionnaire.

\begin{figure}[t!]
	\begin{center}
	\includegraphics[scale=0.55]{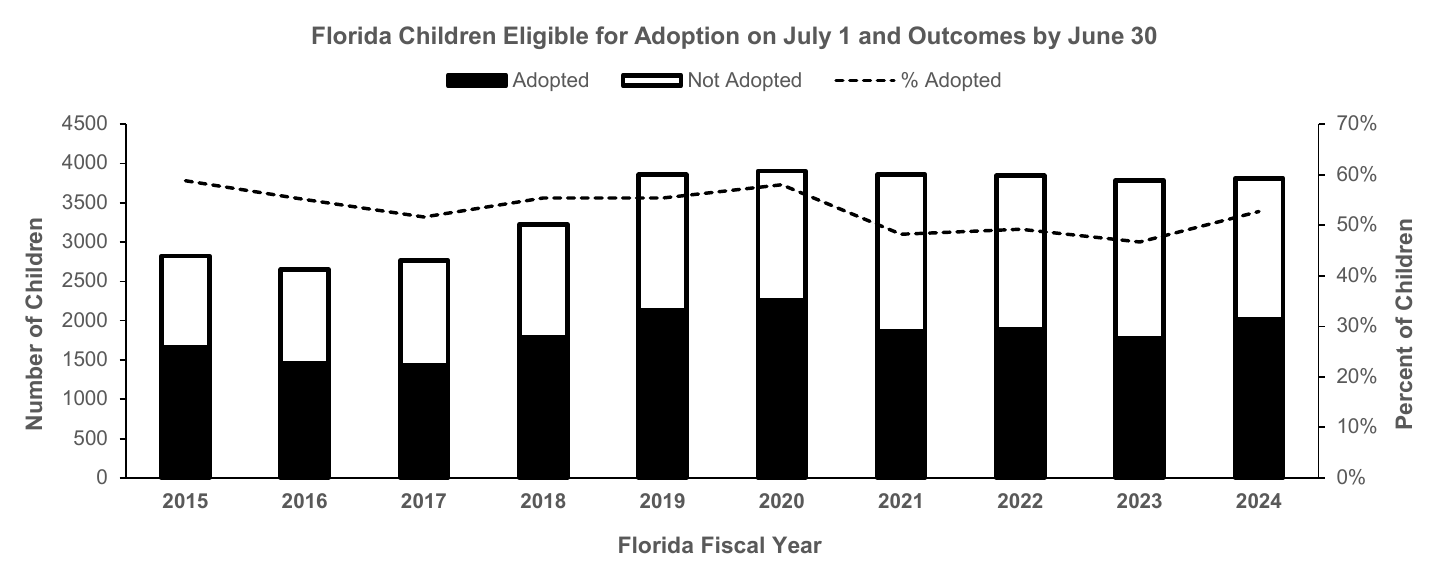}
 \end{center}
	\caption{Children eligible for adoption in Florida on the first day (July 1) of each state fiscal year and whether they are adopted by the last day (June 30), where each fiscal year runs from July 1 of the previous calendar year through June 30 of the labeled year \citep{fl2019,fl2024}.  
 }
	\label{fig:fl}
\end{figure}

\subsection{Data}\label{sec:data_description}

Our analysis compares the focal agency's performance to outcomes for children in Florida reported in the Adoption and Foster Care Analysis and Reporting System (AFCARS) from the US \citet{afcarsAdopt,afcarsFoster} for FY2015 to FY2021, which is the most recently available data on all foster care cases and finalized adoptions. From 766,527 AFCARS foster care 6-month update records for Florida children during this period, we identified 9,544 children under a termination of parental rights order who needed adoption placement search services after October 1, 2014. While only 2,596 ($27.2\%$) of these children were adopted before the end of the time horizon, we note that this low rate of adoption finalizations is partially driven by data censoring (i.e., the end of the observed horizon) and children leaving the system for other reasons than adoption (e.g., emancipation). Conditional on remaining in the system, 16.1\% of children finalized an adoption within one year of TPR, 38.0\% within two years, and 49.0\% within three years.
While the platform's usage was free to all agencies in Florida during the study period, the overwhelming majority of placements during the observed period were still conducted via traditional methods, with only the focal agency utilizing CS as its primary method. While some other agencies occasionally employed the platform's tools to help identify families for harder-to-place children, these placements constitute relatively few adoptions in the AFCARS dataset.\footnote{About $3\%$ of children in other agencies with an active adoption search after July 1, 2018, found adoptions via CS through the platform. A closer look at other agencies' use of the platform can be found in online Appendix~\ref{app:further_otherAg_usage}.}

For the agency, we obtained case data from the platform about 335 children in need of adoptive search with TPR before the cut-off date of October 1, 2021. The platform provided its first matches around July 1, 2018. While the agency performed CS for all children in need of an adoptive placement, a small number of these children found placement through other channels, such as word of mouth within the agency, Florida's online photolisting websites, and other contracted recruitment efforts for specific children.  We include these placements in our analysis because serendipitous matches occur outside the primary (FS or CS) search channel in any system and cannot be identified in the AFCARS dataset. Details of how the data is assembled and pre-processed can be found in online Appendix~\ref{app:further_data_description}.

\begin{table}[t!]
\footnotesize
\centering
\caption{Summary Statistics for AFCARS and Focal Agency (Platform) Data}
\label{tab:summ}
\begin{tabular}{llrrrr} 
  &   & \multicolumn{1}{r}{\textbf{AFCARS   }}   &
& \multicolumn{1}{r}{\textbf{Focal Agency  }} &        \\
  &   & \multicolumn{1}{r}{\textbf{ (N=9,544)}}   &
& \multicolumn{1}{r}{\textbf{ (N=335)}} &  \\
\hline
{\textbf{\textit{Attribute}}}                                 & {\textbf{\textit{Covariate}}} & {\textbf{\textit{Mean (SD) or  \%}}} &
& {\textbf{\textit{Mean (SD) or \%}}} & 
\\ \hline
{\textbf{Case Duration} (years)}                     & {}                       & {{$1.41$}}  {$(1.26)$}       &           & {$1.57$}   {$(0.91)$}      &           \\ 
{\textbf{Adopted before End of Data Horizon}}             & {}                       & {{$27.2\%$}}         & {}              & {{$49.3\%^\dagger$}}         & {}              \\ 
{\textbf{Age at TPR} (years)}             & {\textit{Age}}                    & {{$8.08$}}   {$(5.15)$}      &           & {{$8.42$}}  {$(4.71)$}        &          \\ 
\multicolumn{1}{l}{\textbf{Sex}}                                 & {}                       & {{}}             & {}              & {{}}             & {}              \\ 
{\quad \textbf{Female}}                                    & {\textit{Female}}                 & {$48.3\%$}       & {}              & {$43.0\%$}       & {}              \\ 
{\quad \textbf{Male}}                                      & {}                       & {$51.7\%$}       & {}              & {$57.0\%$}       & {}              \\ 
\multicolumn{1}{l}{\textbf{Race} (may be multiple)}              & {}                       & {{}}             & {}              & {{}}             & {}              \\ 
{\quad \textbf{American Indian or Alaskan Native}}         & {}                       & {$0.3\%$}        & {}              & {$0.0\%$}        & {}              \\ 
{\quad \textbf{Asian}}                                     & {}                       & {$0.6\%$}        & {}              & {$0.0\%$}        & {}              \\ 
{\quad \textbf{Black or African American}}                 & {\textit{Black}}                  & {$36.0\%$}       & {}              & {$23.3\%$}       & {}              \\ 
{\quad \textbf{Native Hawaiian/Other Pacific Islander}} & {}                       & {$0.2\%$}        & {}              & {$0.0\%$}             & {}              \\ 
{\quad \textbf{White}}                                     & {}                       & {$69.8\%$}       & {}              & {$60.9\%$}       & {}              \\ 
{\quad \textbf{Other}}                                     & {}                       & {{N/A}}          & {}              & {$9.0\%$}       & {}              \\ 
{\textbf{Hispanic or Latino Ethnicity}}              & {\textit{Hispanic}}               & {$15.0\%$}       & {}              & {$3.3\%$}        & {}              \\ 
{\textbf{Clinical Disability Diagnosis}}             & {\textit{Disability}}             & {$32.0\%$}       & {}              & {$35.8\%$}       & {}              \\ \hline
\multicolumn{6}{l}{$^\dagger$ \textit{Includes 146 adoptions via the platform and 19 children registered on the platform but adopted by}}\\ \multicolumn{6}{l}{\textit{non-relative/non-foster families found off-platform.}}\\ \hline
\end{tabular}
\end{table}

Table~\ref{tab:summ} presents child statistics for the AFCARS and the agency's platform data, highlighting some key differences: The focal agency's population of children tends to be slightly older, more predominantly male, and slightly more likely to have a clinically diagnosed disability --- all factors associated with greater difficulty in placing children. However, the difference in ages is not statistically significant, as indicated by both a Welch’s t-test and a Mann–Whitney U test ($p>0.05$). A chi-squared test suggests that the proportion of boys among children listed on the platform is higher than in the statewide AFCARS data, although this difference is not statistically significant at the 0.05 level. Similarly, the difference in the prevalence of the disability designation between the two datasets is not statistically significant ($p>0.05$).
Fewer children served by the focal agency are Black, which reflects regional variations in Florida's population. Due to differences in how the datasets treat multi-racial children, we provide analysis in Section~\ref{app:further_excl_black} of the Appendix that shows how estimates of the platform's performance improve if the Black variable refers only to children with the Black or African American variable exclusively selected as a race variable in the AFCARS dataset. 

It should be noted that Florida experienced at least two dramatic shocks to its child welfare system in the years over which the platform was implemented. As shown in Figure~\ref{fig:fl}, the statewide population of children legally free for adoption --- which also includes children on a path to adoption by relatives and foster parents --- increased by nearly 50\% from 2015 to 2018 \citep{fl2019,fl2024}. \citet{quast2018opioid} document a relationship between opioid prescriptions and child welfare system entries in Florida in the early 2010s that could partially explain this increase in children in need of adoption as the opioid crisis worsened. Despite an increase in children needing adoptive placements, the state's reported metric of the percentage of children eligible for adoption on July 1 of a given year and adopted by June 30 of the following year remained above 55\% from 2017 until 2019. However, that metric dropped below 50\% in the years after the coronavirus pandemic and during the height of the opioid crisis \citep{hackworth2025florida}, as caseworker turnover, staffing shortages, an increased inflow of children, and slower judicial processing times may have hindered adoptive searches. After reaching the lowest reported value of 46.7\% in 2022-2023, the metric did not return to above 50\% until the most recent report covering 2023-2024, that is, after the end of our study period.

\subsection{Empirical Strategy and Embedded Bias Against a Strong Focal Agency Effect}\label{sec:empirical_approaches}

Using the AFCARS and the agency's platform datasets, we assess the focal agency's relative performance using two statistical approaches that leverage the datasets' similar structures. First, we construct a benchmark for the set of agency children using the Florida AFCARS case data.
Second, we statistically estimate the focal agency's effect by appending the agency's case data to the AFCARS case data and including a focal agency indicator variable. This approach dampens the estimated focal agency's effect and its statistical significance, as the 335 children of the focal agency are also separately included in the AFCARS dataset without the focal agency indicator.
 Thus, we expect the true focal agency effect to be larger than the estimates suggest.

We also note a second factor that would work against the focal agency effect being significant. In the AFCARS dataset, we explicitly filter out most of the children for whom no adoptive search was required; i.e., those adopted by foster parents, relatives, or non-relatives with whom placement was already in effect at the date of the termination of parental rights (TPR) judicial process (see Appendix \ref{app:further_data_description}). However, for a small number of children, a non-relative adoptive placement may have been pre-identified but was not yet in effect at the time of TPR, and they remain in our analysis. As these children are adopted quickly without a search, their inclusion favors the Florida-wide benchmark, leading to a downward bias in the assessment of the focal agency's performance relative to it.

Both of our statistical approaches model children's time to adoptions using a Cox proportional hazards model \citep{cox1972regression}. The Cox proportional hazards model includes a baseline hazard function that describes how the likelihood of adoption changes over time, along with a parameter for each covariate that affects the baseline hazard. The following characteristics were available in both AFCARS and the agency's platform data, allowing us to control for them in the hazards model (for discrete variables, one of the categories is omitted for statistical identification):
\begin{enumerate}
    \item female (versus male, the omitted category);
    \item Black or African-American (versus no such designation, the omitted category);
    \item Hispanic or Latino ethnicity designation (versus no such designation, the omitted category);
    \item clinical disability diagnosis (versus  no diagnosis, the omitted category);
    \item age in years upon termination of parental rights and its square; and
    \item federal fiscal year (e.g., October 1, 2014, to September 30, 2015, for FY2015) of the TPR order, with FY2015 as the omitted category.
\end{enumerate}

Focusing on Model 1 --- the Cox proportional hazards model without a focal agency effect that is used for benchmarking ---  let \(t=0\) denote the TPR date for some child \(i\). The hazard of adoption finalization at time \(t\) is given
by the semi-parametric Cox model
\begin{align}{ \small 
h\!\left(t \mid X_i\right)
      = h_0(t)\;
        \exp\!\Bigl(}&{ \small
            \beta_1 \text{Female}_i
          + \beta_2 \text{Black}_i
          + \beta_3 \text{Hispanic}_i
          + \beta_4 \text{Disability}_i
          + \beta_5 \text{Age}_i
          + \beta_6 \text{Age}_i^{2}} \nonumber\\
    & \textstyle{\small           
 + \!\!\sum_{y=2016}^{2021}\!\beta_y \,\mathbbm{1}\{\text{FY}_i=y\}
        \Bigr),
        }
\label{eq:cox}
\end{align}
where \(h_0(t)\) is a baseline hazard rate common to all children\footnote{The baseline hazard rate is estimated using the methodology of  \citet{breslow1972discussion}.} and 
\(X_i\) collects the covariates listed above. The indicator $\mathbbm{1}\{\text{FY}_i=y\}$ flags the fiscal year in which TPR occurred. Estimation relies on Cox's partial likelihood, computed from the hazard rates of all children.
Maximizing the log-partial likelihood yields the coefficient estimates.
An exponentiated coefficient $\exp(\beta_k)>1$ indicates that covariate~$k$ is associated with a faster‐than‐baseline adoption rate, whereas $\exp(\beta_k)<1$ signals a slower rate. We explain in Section~\ref{sec:empirical_add_cox} how the model in Equation \eqref{eq:cox} is extended to include the focal agency effect.

\subsection{Benchmark Against Statewide Outcomes}\label{sec:empirical_benchmark}

Using only the statewide AFCARS data, we construct a benchmark for the outcomes of children served by the agency. Model 1 in Table~\ref{tab:cox2} shows how demographic factors affect the time to adoption using the AFCARS dataset for Florida children, i.e., the maximum-likelihood estimation of the Cox model in Equation \eqref{eq:cox}. Note that this dataset does not identify the agency's children; therefore, this model does not account for focal-agency effects. Controlling for the fiscal year in which the search started, the child's gender is not statistically significant; factors associated with a slower time to or diminished likelihood of adoption are being older, having a disability, and being Hispanic or Black. Using this fitted model, we collected the necessary covariate data on each child served by the agency and predicted the likelihood of a finalized adoption at monthly intervals, as explained below. If a child is adopted, the benchmark adoption probability continues to accumulate as if the child's search continued until the earlier of February 1, 2023, or the child's 18th birthday.

\begin{table}[t]
\centering
\footnotesize
\caption{Cox Proportional Hazards Models of Time Until Adoption}
\label{tab:cox2}
\begin{tabular}{l  d{4.6}  d{4.6}}
                       & \multicolumn{1}{c}{Model 1} & \multicolumn{1}{c}{Model 2}  \\ \hline
Female                 & 1.041   & 1.051 \\
                       & (1.008)   & (1.288) \\
Black                  & 0.671^{\ast\ast\ast} & 0.665^{\ast\ast\ast} \\
                       & (-9.175)   & (-9.567) \\
Hispanic               & 0.751^{\ast\ast\ast}   & 0.764^{\ast\ast\ast}\\
                      & (-4.903) & (-4.656)\\
Age at TPR (years)     & 0.900^{\ast\ast\ast}    & 0.904^{\ast\ast\ast}\\
                       & (-6.229)  & (-6.144)\\
(Age at TPR)$^2$       & 0.998   & 0.998\\
                       & (-1.539)  & (-1.882)\\
Disability             & 0.892^{\ast\ast}   & 0.869^{\ast\ast}\\
                       & (-2.648)  & (-3.336) \\
TPR in FY2016 
& 0.976     & 0.952 \\
                      & (-0.358)       & (-0.727) \\
TPR in FY2017 
& 1.010     & 0.973\\
                      & (0.144)       & (-0.401) \\
TPR in FY2018 
& 0.795^{\ast\ast}      & 0.748^{\ast\ast\ast}\\
     & (-3.426)      & (-4.379) \\
TPR in FY2019             & 0.585^{\ast\ast\ast}     & 0.593^{\ast\ast\ast}\\
                      & (-7.521)    & (-7.511)\\
TPR in FY2020          & 0.312^{\ast\ast\ast}     & 0.328^{\ast\ast\ast}\\
                       & (-13.264)    & (-13.261)\\
TPR in FY2021 
 & 0.174^{\ast\ast\ast}     & 0.206^{\ast\ast\ast}\\
                       & (-11.144)       & (-11.712)\\
Focal Agency               & \multicolumn{1}{c}{-}     & 1.540^{\ast\ast\ast}\\
                     &     & (5.205)\\          
                       \hline
Focal agency children & \multicolumn{1}{c}{-}  & \multicolumn{1}{c}{335}  \\
N & \multicolumn{1}{c}{9,544}  & \multicolumn{1}{c}{9,879} \\
Concordance & \multicolumn{1}{c}{0.706}  &  \multicolumn{1}{c}{-}  \\
Log-likelihood ratio test & \multicolumn{1}{c}{1383.319}   & \multicolumn{1}{c}{1446.877}\\ 
 & \multicolumn{1}{c}{on 12 d.f.}   & \multicolumn{1}{c}{on 13 d.f.} \\ \hline
\multicolumn{3}{l}{\textit{Note: all coefficients are exponentiated with z-statistics in parentheses.}}\\
\multicolumn{3}{l}{$^\ast p < 0.05$, $^{\ast\ast} p < 0.01$, $^{\ast\ast\ast} p < 0.001$}\\
\end{tabular}
\end{table}

Each predicted adoption probability equals the expected value of a Bernoulli random variable that takes the value 1 if the child is adopted and 0 if not. Therefore, we can establish a benchmark for the expected number of adoptions by summing the individual probabilities. We then compare the agency's actual number of adoptions to this benchmark. Let $\mathcal{C}$ represent the set of children of the agency. 
For any child $i \in \mathcal{C}$ with attributes $X_i$, we can derive the adoption probability from the estimated survival function using the estimated hazard function $\hat{h}(\cdot|X_i)$ in Model 1
\begin{align}
\hat{\pi}(X_i,t):= 1-\exp\!\left(- \int_{0}^{t} \hat{h}\!(u\mid X_i)\, \mathrm{d}u\right).
\end{align}
This gives the probability that child 
$i$'s adoption is finalized within $t$
years.

A child's maximum possible search horizon from TPR until the child turns 18 is denoted by $\tau_i^e$. We use $\tau_i^d$ to represent the time between the TPR date and the earliest possible adoption finalization date after registration on the platform (that is, after the agency began their search). We calculate $\tau_i^d$ as the time between the case creation date and the TPR date, plus an additional three months to account for the legally required period in Florida during which a child must reside with the adoptive family before the adoption is finalized.\footnote{We provide an alternate analysis in which we assume $\tau_i^d=0$, as well as an analysis excluding agency children whose placement was identified through other channels than the platform in online Appendix~\ref{app:further_benchmark}.}

To account for the time that a child might have already been eligible for adoption before being registered on the platform, we calculate a conditional survival probability for child $i$ at any time $t>\tau_i^d$ since TPR  as 
\begin{align}\label{eq:predictprob}
\tilde{\pi}(X_i,t,\tau_i^e,\tau_i^d):=\frac{\hat{\pi}(X_i,\min\{t,\tau^e_i\})-\hat{\pi}(X_i,\tau_i^d)}{1-\hat{\pi}(X_i,\tau_i^d)}, 
\end{align}
which provides the AFCARS benchmark $\hat{\mu}(t)$ for the number of expected adoptions by time $t$:
\begin{align}\label{eq:benchmark}
\hat{\mu}(t):=\textstyle \sum_{i\in\mathcal{C}} \tilde{\pi}(X_i,t,\tau_i^e,\tau_i^d).
\end{align}

\begin{figure}[t!]
	\begin{center}
	\includegraphics[scale=0.55]{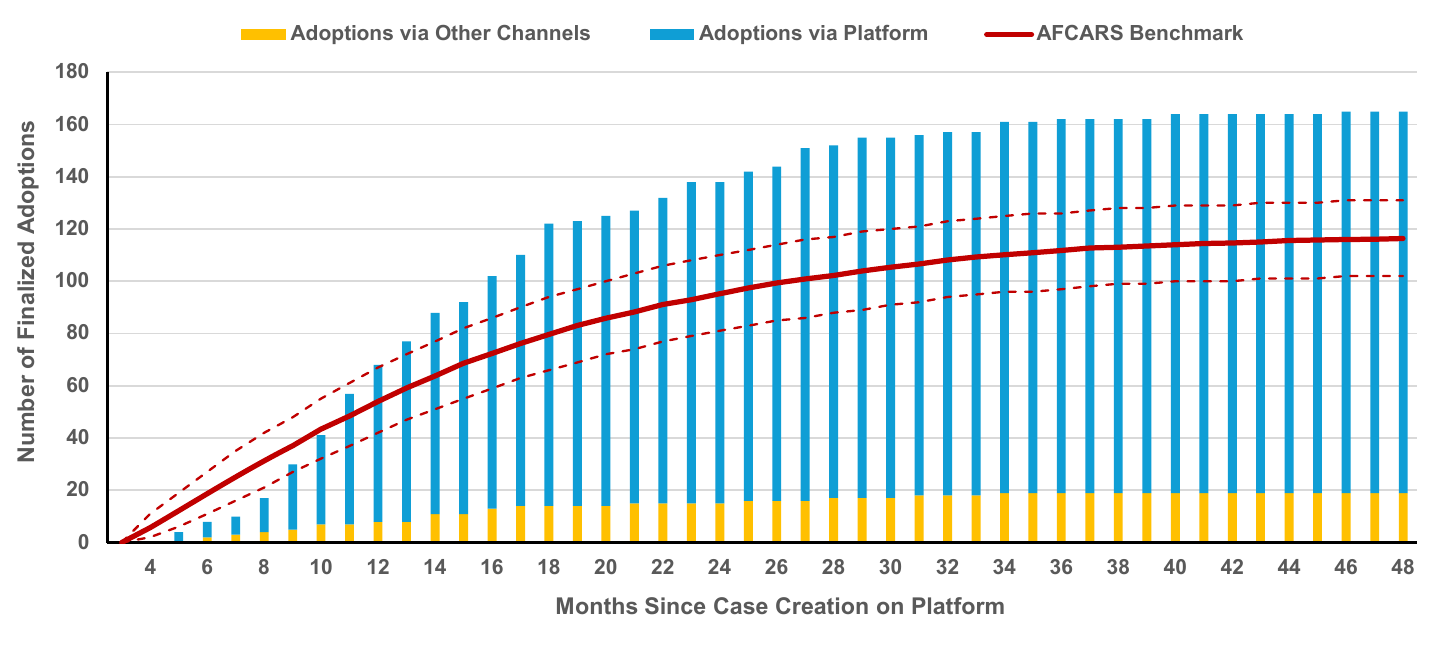}
 \end{center}
	\caption{Actual adoptions by the agency using the platform and other channels compared to Florida AFCARS benchmark model. Dashed lines represent the 95\% confidence interval defined by the Poisson binomial distribution.
 }
	\label{fig:fmyears}
\end{figure}

Figure~\ref{fig:fmyears} shows the adoptions achieved by the agency by using the platform compared to this benchmark from the AFCARS proportional hazards model. The results suggest that the agency has significantly outperformed the commonly used two-year and three-year search window benchmarks.  For children registered on the platform during the study period, 138 adoptions were finalized within two years --- 123 enabled by the platform and 15 through other channels --- while the benchmark model predicted only 95.3 for the agency. These additional adoptions constitute a 44.8\% improvement over the benchmark. At the three-year mark, this difference between adoptions achieved and the predicted number is 50.2 adopted children, a 44.9\% increase over the benchmark. While the focal agency already outperformed the benchmark at the 12-month mark, it underperformed during the first 10 months. This is likely due to many children not being registered on the platform immediately after TPR. While we account for this via the conditional survival probability \eqref{eq:predictprob}, the time until the earliest possible adoption date via the platform, $\tau_i^d$, only accounts for the legal minimum of three months after placement. This is exceedingly conservative, as even if a family was identified day one, it realistically takes time to contact them and place the child. A less conservative approach without conditional survival probability (i.e., effectively assuming timelines start at registration rather than at TPR), presented in Appendix \ref{app:further_no_varying_cox}, finds that the focal agency never significantly underperforms the benchmark.

\subsection{Focal Agency Effect in Cox Proportional Hazards Model}
\label{sec:empirical_add_cox}

To provide additional evidence of the focal agency's performance using CS relative to statewide outcomes, we re-estimated the Cox proportional hazards model on a combined dataset to assess whether the focal agency is associated with faster finalization of adoption.

Since there can be a substantial difference between the TPR date for some children and the date on which they were added to the platform, we treat the focal agency variable as time-varying: the child's timeline starts at TPR, with the focal agency variable's value switching from zero to one at a point three months after the platform case is created. The three-month delay accounts for the minimum period Florida law requires before an adoption can be finalized after placement. Thus, any adoption within the three months after listing cannot be attributed to the focal agency's search. 

More precisely, when compared with the Cox proportional hazards model introduced in Equation \eqref{eq:cox}, we add a binary covariate indicating whether the child has been served by the focal agency using the platform. As mentioned in the previous paragraph, the term $\textit{FocalAgency}_i(t)$ is time-dependent and becomes 1 three months after a child is registered on the platform. Each case's timeline still begins on the TPR date.\footnote{In online Appendix~\ref{app:further_benchmark}, we provide two alternative analyses with a non-time-varying focal agency variable; in these specifications, the timeline for agency children begins once they are registered on the platform and at TPR, respectively.}

We denote this Model 2, with resulting coefficients found in Table~\ref{tab:cox2}.
The control variables behave as expected.  Girls experience a nominally higher but statistically insignificant adoption hazard compared to boys, while Black and Hispanic children tend to be adopted more slowly than children of other racial groups. A clinical disability diagnosis, a child's increasing age, and TPR in pandemic years or immediately before are also linked to decreased adoption hazards.

This analysis finds that the focal agency is consistently associated with faster adoptions.  The estimated focal agency hazard ratio is 1.540, indicating a 54.0\% increase in hazard rates relative to other circuits that do not primarily employ CS (i.e., faster adoption on average).  This ratio is significant at the 0.1\% level, despite the biases against the focal agency built into the dataset construction mentioned above. Taken together, this suggests that the agency has seen accelerated adoption finalization rates compared to statewide outcomes after implementing the platform.

\subsection{Discussion of Empirical Analysis} \label{sec:connecting-theory-empirics}

While our empirical analysis cannot formally test the theoretical model's predictions, several patterns in the data are consistent with the mechanisms identified by the theory.
First, our model predicts that CS leads to more and faster placements when family preferences are heterogeneous, and families are moderately impatient (Figures~\ref{fig:children_better_off_utilities}~and~\ref{fig:children_better_off_match_values} in
Section~\ref{sec:num_eval_results} and Theorem~\ref{theorem:pareto_dominance} in Section~\ref{sec:theory_comparison_pareto}). Both conditions appear to hold in practice: Appendix~\ref{app:fam_pref} documents substantial variation in families' stated preferences across multiple dimensions, and practitioners report that families typically desire timely placements, particularly since home studies expire after one year.
This is consistent with the benchmark analysis finding that the only agency that primarily employs CS in Florida has a $44.9\%$ higher likelihood of adoption finalization within three years, and the hazard model similarly estimates a $54\%$ increase in the agency's instantaneous adoption rate.

However, several specific limitations constrain the causal interpretation of these empirical findings. First, we cannot fully separate platform effects from search-method effects. The platform is designed for CS and also provides a simple scoring-based recommendation tool to rank families for a child.\footnote{Recommendation systems are not unique to the CS implementation; for example, \citet{slaugh_pennsylvania_2016} documents the use of a recommendation system in the statewide adoption agency of Pennsylvania, which employs only an FS method in adoptions.} We do not know which, if any, recommendation systems were used by other circuits that primarily utilized FS. Second, since the AFCARS dataset lacks circuit-level identifiers, we cannot account for unobserved institutional differences across circuits (i.e., circuit fixed effects). Consequently, we cannot determine which children were registered and placed by the focal agency in the AFCARS data prior to the platform's adoption. This precludes a difference-in-differences design, which would have yielded more robust estimates of the effect of switching search methods. Third, the AFCARS data covers a limited post-implementation horizon, and the study period coincides with significant external shocks --- most notably the COVID-19 pandemic --- that may have differentially affected circuits. While we provide an additional robustness check in Appendix~\ref{sec:platform_year} that separately examines the platform's effect in each year, fixed effects cannot capture circuit-specific differential responses to such shocks.
Finally, while our Cox proportional hazards model controls for observable child characteristics such as race, age, gender, and disability, we cannot rule out that unobserved differences between the focal circuit and the rest of the state contribute to the observed performance gap. That said, all caseworkers in the focal agency were required to use the platform, mitigating concerns about within-circuit selection on caseworker characteristics.

However, the agency's suggested over-performance over longer time horizons may also indicate that the platform technology helps caseworkers to be more persistent in their search efforts for hard-to-place children. When \citet{avery2000perceptions} investigated the longest-waiting children in New York, caseworkers were found to be pessimistic about the children's chances at adoption. They also neglected to use the search tools available to them. However, in addition to evidence of usage from the platform's actual placements, a survey of the platform's 73 active users in Florida suggests a positive assessment of caseworker satisfaction with the new CS-driven system. A total of 51 respondents --- mostly caseworkers but also some supervisors and recruitment specialists --- responded to the survey in 2023. About 30\% of the respondents had been in their current role for at least 5 years, so many were familiar with other search methods besides the platform. When asked how satisfied they were with the platform, approximately 33\% were very satisfied, and 31\% were somewhat satisfied. Only 5\% were very dissatisfied, and only 9\% were somewhat dissatisfied.  

Similarly, we note that network effects among families, such as positive word of mouth about the CS experience that encourages other families to participate, may also have contributed to the focal agency's improved performance. This is consistent with the benefits of caseworker-driven search shown in our analytical model, where lower search costs and higher utility for families are central drivers of improvements. 

Establishing direct empirical links between the theoretical parameters --- such as search costs, family patience, and preference heterogeneity --- and placement outcomes remains an important direction for future research.

\section{Conclusion} \label{sec:conclusion}

Treating the search for adoptive families as an operations challenge, we develop the first formal game-theoretic model for the child welfare system adoption process and introduce a novel search-and-matching framework to compare caseworker-driven and family-driven search paradigms. We characterize the Nash equilibria of these models, showing that agents adopt threshold strategies under a mild tie-breaking condition and that equilibria form a non-empty complete lattice. 

Our model shows that caseworker-driven search more effectively avoids wasted search efforts and, in most settings, improves outcomes. While caseworker-driven equilibria can Pareto dominate all family-driven equilibria, the converse does not hold. Although caseworker-driven equilibria do not always Pareto-dominate those of family-driven search, they typically perform better for most agents, as confirmed by extensive numerical analysis. Within our model, when families are sufficiently impatient, caseworker-driven search unambiguously benefits all children. However, we also identify conditions --- notably, highly patient agents with strongly correlated preferences --- under which caseworker-driven search may harm children. Yet such conditions are rare in practice: families usually desire a timely placement after completing the adoption approval process, and preferences tend to be idiosyncratic. 
This is consistent with our empirical study, which found that an agency using CS significantly outperformed statewide benchmarks; however, data limitations prevent a causal attribution of that difference to the search discipline alone.

However, there are some practical circumstances under which 
caseworker-driven search may not be the preferred approach. First, CS concentrates 
matching decisions in the hands of individual caseworkers, which could propagate 
personal biases --- whether conscious or unconscious --- regarding family suitability. 
Under FS, a broader pool of families self-selects into consideration, which may 
partially mitigate the effects of any single caseworker's biases, though FS introduces 
its own fairness concerns if families with greater resources or information are better 
positioned to identify and pursue children. Second, the effectiveness of CS depends 
on the availability of a structured, searchable database of prospective families. In 
settings where such data infrastructure is lacking or where family information is 
incomplete, caseworkers may not have sufficient information to identify strong matches.

Our work provides a foundation for future empirical research into managing the operation of adoptive placement searches, especially how to allocate caseworkers' efforts across different search channels. While existing research in child welfare literature emphasizes the effectiveness of highly specialized recruiters, our work suggests that technology may help frontline caseworkers achieve better outcomes for children. Establishing a causal relationship between search technology and improved placement outcomes remains an important goal for future research, though limitations stemming from data privacy concerns regarding child welfare, decentralized agencies collecting different or no data, and adoption journeys spanning multiple years present significant obstacles.
Additional research on caseworker behavior, family engagement, and children's outcomes across search practices is essential to improving services for this vulnerable population.
\newpage

 \bibliographystyle{informs2014}

 \bibliography{references}

@article{Stigler1961,
  author  = {Stigler, George J.},
  title   = {The Economics of Information},
  journal = {Journal of Political Economy},
  year    = {1961},
  volume  = {69},
  number  = {3},
  pages   = {213--225}
}

@article{AlbrechtGautierVroman2006,
  author  = {Albrecht, James W. and Gautier, Pieter A. and Vroman, Steven B.},
  title   = {Equilibrium Directed Search with Multiple Applications},
  journal = {Review of Economic Studies},
  year    = {2006},
  volume  = {73},
  number  = {4},
  pages   = {869--891}
}

@article{wright2021directed,
  title={Directed search and competitive search equilibrium: A guided tour},
  author={Wright, Randall and Kircher, Philipp and Julien, Beno{\^\i}t and Guerrieri, Veronica},
  journal={Journal of Economic Literature},
  volume={59},
  number={1},
  pages={90--148},
  year={2021},
  publisher={American Economic Association 2014 Broadway, Suite 305, Nashville, TN 37203-2425}
}

@article{Kircher2009,
  author  = {Kircher, Philipp},
  title   = {Efficiency of Simultaneous Search},
  journal = {Journal of Political Economy},
  year    = {2009},
  volume  = {117},
  number  = {5},
  pages   = {861--913}
}

@report{hackworth2025florida,
  author       = {Hackworth, Benjamin T.},
  title        = {Patterns of and Trends in Substance Use in Florida: 2025 Annual Report},
  year         = {2025},
  institution  = {Florida Department of Children and Families},
  url          = {https://cdn.ymaws.com/www.fadaa.org/resource/resmgr/files/resource_center/reports/PatternsofAndTrendsinSUinFlo.pdf}
}

@article{HonkaChintagunta2017,
  author  = {Honka, Elisabeth and Chintagunta, Pradeep},
  title   = {Simultaneous or Sequential? Search Strategies in the {U.S.} Auto Insurance Industry},
  journal = {Marketing Science},
  year    = {2017},
  volume  = {36},
  number  = {1},
  pages   = {21--42}
}

@article{HalaburdaPiskorskiYildirim2018,
  author  = {Halaburda, Hanna and Piskorski, Mikolaj Jan and Yildirim, Pinar},
  title   = {Competing by Restricting Choice: The Case of Matching Platforms},
  journal = {Management Science},
  year    = {2018},
  volume  = {64},
  number  = {8},
  pages   = {3574--3594}
}

@article{AusterGottardiWolthoff2025,
  title={Simultaneous search and adverse selection},
  author={Auster, Sarah and Gottardi, Piero and Wolthoff, Ronald},
  journal={Review of Economic Studies},
  volume={92},
  number={6},
  pages={3541--3573},
  year={2025},
  publisher={Oxford University Press UK}
}

@article{gale/sotomayor:85,
author = {David Gale and Marilda Sotomayor},
 journal = {The American Mathematical Monthly},
 number = {4},
 pages = {261--268},
 publisher = {Mathematical Association of America},
 title = {Ms. {M}achiavelli and the Stable Matching Problem},
 urldate = {2022-09-23},
 volume = {92},
 year = {1985}
}

@article{dierks2024child,
  title={When Family Incentives Meet Caseworker Scarcity: Improving Adoption Outcomes for Children with Disabilities},
  author={Dierks, Ludwig and Slaugh, Vincent and {\"U}nver, M Utku},
  journal={Available at SSRN 4778791},
  year={2026}
}

@article{baccara2020optimal,
  title={Optimal dynamic matching},
  author={Baccara, Mariagiovanna and Lee, SangMok and Yariv, Leeat},
  journal={Theoretical Economics},
  volume={15},
  number={3},
  pages={1221--1278},
  year={2020},
  publisher={Wiley Online Library}
}

@article{altinok2023designing,
  title={Designing the Menu of Licenses for Foster Care},
  author={Altinok, Ahmet and MacDonald, Diana},
  journal={Available at SSRN 4466506},
  year={2023}
}

@article{Leshno_overload22,
Author = {Leshno, Jacob D.},
Title = {Dynamic Matching in Overloaded Waiting Lists},
Journal = {American Economic Review},
Volume = {112},
Number = {12},
Year = {2022},
Month = {December},
Pages = {3876-3910}}

@article{akbarpour2020thickness,
  title={Thickness and information in dynamic matching markets},
  author={Akbarpour, Mohammad and Li, Shengwu and Gharan, Shayan Oveis},
  journal={Journal of Political Economy},
  volume={128},
  number={3},
  pages={783--815},
  year={2020},
  publisher={The University of Chicago Press Chicago, IL}
}

@article{slaugh_pennsylvania_2016,
	title = {The {Pennsylvania} {Adoption} {Exchange} {Improves} {Its} {Matching} {Process}},
	volume = {46},
	language = {en},
	number = {2},
	urldate = {2020-03-03},
	journal = {Interfaces},
	author = {Slaugh, Vincent W. and Akan, Mustafa and Kesten, Onur and Ünver, M. Utku},
	month = apr,
	year = {2016},
	pages = {133--153},
}

@techreport{robinson-cortes_who_2019,
	address = {315 Baxter Hall, Pasadena, CA 91125, USA},
	title = {Who {Gets} {Placed} {Where} and {Why}? {An} {Empirical} {Framework} for {Foster} {Care} {Placement}},
	language = {en},
	institution = {California Institute of Technology},
	author = {Robinson-Cortés, Alejandro},
	month = nov,
	year = {2019},
}

@article{hanna_innovative_2011,
	title = {Innovative {Practice} {Approaches} to {Matching} in {Adoption}},
	volume = {5},
	language = {en},
	number = {1},
	urldate = {2020-03-03},
	journal = {Journal of Public Child Welfare},
	author = {Hanna, Michele and McRoy, Ruth},
	month = jan,
	year = {2011},
	pages = {45--66},
}

@article{sonmez_incentivized_2020,
  title={Incentivized kidney exchange},
  author={S{\"o}nmez, Tayfun and {\"U}nver, M Utku and Yenmez, M Bumin},
  journal={American Economic Review},
  volume={110},
  number={7},
  pages={2198--2224},
  year={2020},
  publisher={American Economic Association 2014 Broadway, Suite 305, Nashville, TN 37203}
}

@article{kerimov_optimality_OR_2023,
  title={On the optimality of greedy policies in dynamic matching},
  author={Kerimov, S{\"u}leyman and Ashlagi, Itai and Gurvich, Itai},
  journal={Operations Research},
  year={2023},
  publisher={INFORMS}
}

@inproceedings{akbarpour_unpaired_2020,
  title={Unpaired kidney exchange: Overcoming double coincidence of wants without money},
  author={Akbarpour, Mohammad and Combe, Julien and He, Yinghua and Hiller, Victor and Shimer, Robert and Tercieux, Olivier},
  booktitle={Proceedings of the 21st ACM Conference on Economics and Computation},
  pages={465--466},
  year={2020}
}

@article{unver_dynamic_2010,
	title = {Dynamic {Kidney} {Exchange}},
	volume = {77},
	language = {en},
	number = {1},
	urldate = {2020-03-03},
	journal = {Review of Economic Studies},
	author = {Ünver, M. Utku},
	month = jun,
	year = {2010},
	pages = {372--414},
}

@techreport{andersson_dynamic_2018,
	address = {Lund, Sweden},
	title = {Dynamic {Refugee} {Matching}},
	language = {en},
	institution = {Lund University},
	author = {Andersson, Tommy and Ehlers, Lars and Martinello, Alessandro},
	year = {2018},
}

@article{bansak_improving_2018,
	title = {Improving refugee integration through data-driven algorithmic assignment},
	volume = {359},
	language = {en},
	number = {6373},
	urldate = {2020-03-03},
	journal = {Science},
	author = {Bansak, Kirk and Ferwerda, Jeremy and Hainmueller, Jens and Dillon, Andrea and Hangartner, Dominik and Lawrence, Duncan and Weinstein, Jeremy},
	month = jan,
	year = {2018},
	pages = {325--329},
}

@article{hitsch_matching_2010,
	title = {Matching and {Sorting} in {Online} {Dating}},
	volume = {100},
	language = {en},
	number = {1},
	urldate = {2020-03-03},
	journal = {American Economic Review},
	author = {Hitsch, Günter J. and Hortaçsu, Alı and Ariely, Dan},
	month = mar,
	year = {2010},
	pages = {130--163},
}

@article{weitzman_optimal_1979,
	title = {Optimal {Search} for the {Best} {Alternative}},
	volume = {47},
	language = {en},
	number = {3},
	urldate = {2020-03-26},
	journal = {Econometrica},
	author = {Weitzman, Martin L.},
	month = may,
	year = {1979},
	pages = {641},
}

@article{chade_simultaneous_2006,
	title = {Simultaneous {Search}},
	volume = {74},
	issn = {0012-9682, 1468-0262},
	url = {http://doi.wiley.com/10.1111/j.1468-0262.2006.00705.x},
	doi = {10.1111/j.1468-0262.2006.00705.x},
	language = {en},
	number = {5},
	urldate = {2020-04-13},
	journal = {Econometrica},
	author = {Chade, Hector and Smith, Lones},
	month = sep,
	year = {2006},
	pages = {1293--1307},
}

@article{shimer_assortative_2000,
	title = {Assortative {Matching} and {Search}},
	volume = {68},
	language = {en},
	number = {2},
	urldate = {2020-05-15},
	journal = {Econometrica},
	author = {Shimer, Robert and Smith, Lones},
	month = mar,
	year = {2000},
	pages = {343--369},
}

@article{adachi_search_2003,
	title = {A search model of two-sided matching under nontransferable utility},
	volume = {113},
	language = {en},
	number = {2},
	urldate = {2020-05-17},
	journal = {Journal of Economic Theory},
	author = {Adachi, Hiroyuki},
	month = dec,
	year = {2003},
	pages = {182--198},
}

@article{hitsch_what_2010,
	title = {What makes you click?—{Mate} preferences in online dating},
	volume = {8},
	language = {en},
	number = {4},
	urldate = {2020-06-11},
	journal = {Quantitative Marketing and Economics},
	author = {Hitsch, Günter J. and Hortaçsu, Ali and Ariely, Dan},
	month = dec,
	year = {2010},
	pages = {393--427},
}

@article{chade_sorting_2017,
	title = {Sorting through {Search} and {Matching} {Models} in {Economics}},
	volume = {55},
	doi = {10.1257/jel.20150777},
	language = {en},
	number = {2},
	urldate = {2020-06-16},
	journal = {Journal of Economic Literature},
	author = {Chade, Hector and Eeckhout, Jan and Smith, Lones},
	month = jun,
	year = {2017},
	pages = {493--544},
}

@article{arnosti_design_2020,
  title={Design of lotteries and wait-lists for affordable housing allocation},
  author={Arnosti, Nick and Shi, Peng},
  journal={Management Science},
  volume={66},
  number={6},
  pages={2291--2307},
  year={2020},
  publisher={INFORMS}
}

@article{chade_NTU_2001,
  title={Two-sided search and perfect segregation with fixed search costs},
  author={Chade, Hector},
  journal={Mathematical Social Sciences},
  volume={42},
  number={1},
  pages={31--51},
  year={2001},
  publisher={Elsevier}
}

@inproceedings{you_strategy_2022,
  title={Strategy-proof house allocation with existing tenants over social networks},
  author={You, Bo and Dierks, Ludwig and Todo, Taiki and Li, Minming and Yokoo, Makoto},
  booktitle={Proceedings of the 21st International Conference on Autonomous Agents and Multiagent Systems},
  pages={1446--1454},
  year={2022}
}

@inproceedings{kawasaki_mechanism_2021,
  title={Mechanism design for housing markets over social networks},
  author={Kawasaki, Takehiro and Wada, Ryoji and Todo, Taiki and Yokoo, Makoto},
  booktitle={Proceedings of the 20th International Conference on Autonomous Agents and Multiagent Systems},
  pages={692--700},
  year={2021}
}

@unpublished{combe_market_2025,
  title={Market design for distributional objectives in (re) assignment: An application to improve the distribution of teachers in schools},
    year={2025},
  author={Combe, Julien and Dur, Umut and Tercieux, Olivier and Terrier, Camille and {\"U}nver, M. Utku},
  note={Working Paper, SSRN: 5112041}
}

@article{lauermann_stable_2014,
	title = {Stable marriages and search frictions},
	volume = {151},
	language = {en},
	urldate = {2020-06-16},
	journal = {Journal of Economic Theory},
	author = {Lauermann, Stephan and Nöldeke, Georg},
	month = may,
	year = {2014},
	pages = {163--195},
}

@article{lee_propose_2015,
	title = {Propose with a rose? {Signaling} in internet dating markets},
	volume = {18},
	language = {en},
	number = {4},
	urldate = {2020-07-08},
	journal = {Experimental Economics},
	author = {Lee, Soohyung and Niederle, Muriel},
	month = dec,
	year = {2015},
	pages = {731--755},
}

@article{cheremukhin_targeted_2020,
	title = {Targeted search in matching markets},
	volume = {185},
	language = {en},
	urldate = {2020-09-11},
	journal = {Journal of Economic Theory},
	author = {Cheremukhin, Anton and Restrepo-Echavarria, Paulina and Tutino, Antonella},
	month = jan,
	year = {2020},
	pages = {104956},
}

@article{shimer_matching_2001,
	title = {Matching, {Search}, and {Heterogeneity}},
	volume = {1},
	language = {en},
	number = {1},
	urldate = {2020-09-14},
	journal = {Advances in Macroeconomics},
	author = {Shimer, Robert and Smith, Lones},
	month = jan,
	year = {2001},
}

@article{smith_marriage_2006,
	title = {The {Marriage} {Model} with {Search} {Frictions}},
	volume = {114},
	issn = {0022-3808, 1537-534X},
	language = {en},
	number = {6},
	urldate = {2020-09-14},
	journal = {Journal of Political Economy},
	author = {Smith, Lones},
	month = dec,
	year = {2006},
	pages = {1124--1144},
}

@techreport{macdonald_foster_2019,
	address = {Phoenix, AZ, USA},
	title = {Foster {Care}: {A} {Dynamic} {Matching} {Approach}},
	language = {en},
	institution = {Arizona State University},
	author = {MacDonald, Diana E.},
	month = nov,
	year = {2019},
}

@article{eeckhout_bilateral_1999,
	title = {Bilateral {Search} and {Vertical} {Heterogeneity}},
	volume = {40}, 
        language = {en},
	number = {4},
	urldate = {2020-10-12},
	journal = {International Economic Review},
	author = {Eeckhout, Jan},
	month = nov,
	year = {1999},
	pages = {869--887},
}

@article{kasy_adaptive_2020,
	title = {Adaptive {Combinatorial} {Allocation}},
	language = {en},
	urldate = {2020-12-15},
	journal = {arXiv:2011.02330 [econ, stat]},
	author = {Kasy, Maximilian and Teytelboym, Alexander},
	month = nov,
	year = {2020},
	note = {arXiv: 2011.02330},
}

@article{arnosti_managing_2021,
	title={Managing congestion in matching markets},
  author={Arnosti, N and Johari, R and Kanoria, Y},
  journal={Manufacturing \& Service Operations Management},
  volume={23},
  number={3},
  pages={620--636},
  year={2021},
  publisher={INFORMS}
}

@inproceedings{mennle_power_2015,
	series = {{IJCAI}'15},
	title = {The {Power} of {Local} {Manipulation} {Strategies} in {Assignment} {Mechanisms}},
	isbn = {978-1-57735-738-4},
	booktitle = {Proceedings of the 24th {International} {Conference} on {Artificial} {Intelligence}},
	author = {Mennle, Timo and Weiss, Michael and Philipp, Basil and Seuken, Sven},
	year = {2015},
	pages = {82--89},
}

@article{tarski_lattice-theoretical_1955,
	title = {A lattice-theoretical fixpoint theorem and its applications},
	volume = {5},
	doi = {10.2140/pjm.1955.5.285},
	language = {en},
	number = {2},
	urldate = {2021-02-12},
	journal = {Pacific Journal of Mathematics},
	author = {Tarski, Alfred},
	month = jun,
	year = {1955},
	pages = {285--309},
}

@article{ma_spatio-temporal_2020,
  title={Spatio-temporal pricing for ridesharing platforms},
  author={Ma, Hongyao and Fang, Fei and Parkes, David C},
  journal={Operations Research},
  volume={70},
  number={2},
  pages={1025--1041},
  year={2022},
  publisher={INFORMS}
}

@article{abdulkadiroglu_expanding_2015,
	title = {Expanding “{Choice}” in {School} {Choice}},
	volume = {7},
	language = {en},
	number = {1},
	urldate = {2021-02-15},
	journal = {American Economic Journal: Microeconomics},
	author = {Abdulkadiroğlu, Atila and Che, Yeon-Koo and Yasuda, Yosuke},
	month = feb,
	year = {2015},
	pages = {1--42},
}

@article{avery_adoptuskids_2009,
	title = {{AdoptUSKids} national photolisting service: {Characteristics} of listed children and length of time to placement},
	volume = {31},
	language = {en},
	number = {1},
	urldate = {2021-03-10},
	journal = {Children and Youth Services Review},
	author = {Avery, Rosemary J. and Butler, J.S. and Schmidt, Ellie Bradsher and Holtan, Barbara A.},
	month = jan,
	year = {2009},
	pages = {140--154},
}

@article{immorlica_designing_2024,
author = {Immorlica, Nicole and Lucier, Brendan and Manshadi, Vahideh and Wei, Alexander},
title = {Designing Approximately Optimal Search on Matching Platforms},
journal = {Management Science},
volume = {69},
number = {8},
pages = {4609-4626},
year = {2023},
doi = {10.1287/mnsc.2022.4601},
}

@article{lauermann_balance_2020,
	title = {The {Balance} {Condition} in {Search}‐and‐{Matching} {Models}},
	volume = {88},
	language = {en},
	number = {2},
	urldate = {2021-06-09},
	journal = {Econometrica},
	author = {Lauermann, Stephan and Nöldeke, Georg and Tröger, Thomas},
	year = {2020},
	pages = {595--618},
}

@article{baccara_child-adoption_2014,
	title = {Child-{Adoption} {Matching}: {Preferences} for {Gender} and {Race}},
	volume = {6},
	language = {en},
	number = {3},
	urldate = {2021-06-09},
	journal = {American Economic Journal: Applied Economics},
	author = {Baccara, Mariagiovanna and Collard-Wexler, Allan and Felli, Leonardo and Yariv, Leeat},
	month = jul,
	year = {2014},
	pages = {133--158},
}

@article{roth_evolution_1984,
	title = {The {Evolution} of the {Labor} {Market} for {Medical} {Interns} and {Residents}: {A} {Case} {Study} in {Game} {Theory}},
	volume = {92}, 
        language = {en},
	number = {6},
	urldate = {2021-06-10},
	journal = {Journal of Political Economy},
	author = {Roth, Alvin E.},
	month = dec,
	year = {1984},
	pages = {991--1016},
}

@article{roth_natural_1991,
	title = {A {Natural} {Experiment} in the {Organization} of {Entry}-{Level} {Labor} {Markets}: {Regional} {Markets} for {New} {Physicians} and {Surgeons} in the {United} {Kingdom}},
	volume = {81},
	number = {3},
	journal = {The American Economic Review},
	author = {Roth, Alvin E.},
	year = {1991},
	pages = {415--440},
}

@misc{childrens_bureau_afcars_2020,
	title = {The {AFCARS} Report: Preliminary {FY2021} Estimates},
	publisher = {U.S. Department of Health and Human Services},
	author = {{Children's Bureau}},
	month = nov,
	year = {2022},
        url ={https://www.acf.hhs.gov/sites/default/files/documents/cb/afcars-report-29.pdf},
note = {Accessed: 2024-02-28}
}

@article{triseliotis_long-term_2002,
	title = {Long-term foster care or adoption? {The} evidence examined},
	volume = {7},
	number = {1},
	journal = {Child \& Family Social Work},
	author = {Triseliotis, John},
	year = {2002},
	pages = {23--33},
}

@article{courtney_midwest_2010,
title={Homelessness and health care access after emancipation: Results from the Midwest Evaluation of Adult Functioning of Former Foster Youth},
  author={Kushel, Margot B and Yen, Irene H and Gee, Lauren and Courtney, Mark E},
  journal={Archives of Pediatrics \& Adolescent Medicine},
  volume={161},
  number={10},
  pages={986--993},
  year={2007},
  publisher={American Medical Association}
}

@article{atakan_assortative_2006,
	title = {Assortative {Matching} with {Explicit} {Search} {Costs}},
	volume = {74},
	number = {3},
	journal = {Econometrica},
	author = {Atakan, Alp E.},
	year = {2006},
	pages = {667--680},
}

@article{kanoria_facilitating_2021,  
    author  = {Kanoria, Yash and Saban, Daniela},
    title   = {Facilitating the Search for Partners on Matching Platforms},
    journal = {Management Science},
    year    = {2021},
    volume  = {67},
    number  = {10},
    pages   = {5990--6029}
}

@book{knuth_stable_1997,
	title = {Stable marriage and its relation to other combinatorial problems: {An} introduction to the mathematical analysis of algorithms},
	volume = {10},
	publisher = {American Mathematical Soc.},
	author = {Knuth, Donald Ervin},
	year = {1997},
}

@techreport{fradkin_search_2017,
	title = {Search, {Matching}, and the {Role} of {Digital} {Marketplace} {Design} in {Enabling} {Trade}: {Evidence} from {Airbnb}},
	shorttitle = {Search, {Matching}, and the {Role} of {Digital} {Marketplace} {Design} in {Enabling} {Trade}},
	language = {en},
	urldate = {2021-09-21},
	institution = {Boston University Questrom School of Business},
	author = {Fradkin, Andrey},
	year = {2017},
}

@article{combe_mechanism_2021,
  title={The design of teacher assignment: Theory and evidence},
  author={Combe, Julien and Tercieux, Olivier and Terrier, Camille},
  journal={The Review of Economic Studies},
  volume={89},
  number={6},
  pages={3154--3222},
  year={2022},
  publisher={Oxford University Press}
}

@article{delacretaz_matching_2020,
  title={Matching mechanisms for refugee resettlement},
  author={Delacr{\'e}taz, David and Kominers, Scott Duke and Teytelboym, Alexander},
  journal={American Economic Review},
  volume={113},
  number={10},
  pages={2689--2717},
  year={2023},
  publisher={American Economic Association 2014 Broadway, Suite 305, Nashville, TN 37203}
}

@article{shi_optimal_2020,
	title={Optimal matchmaking strategy in two-sided marketplaces},
  author={Shi, P},
  journal={Management Science},
  volume={69},
  number={3},
  pages={1323--1340},
  year={2023},
  publisher={INFORMS}
}

@misc{afcarsFoster,
	title = {{AFCARS Foster Care File, 6-month periods (FY2016A - 2022B)}},
	url = {https://doi.org/10.34681/yjxr-zz92},
	author = {{Children’s Bureau, Administration on Children, Youth and Families}},
shorthand = {Children’s Bureau},
	year = {2023},
note = {Accessed: 2024-02-28},
}

@misc{afcarsAdopt,
	title = {{AFCARS Adoption File 2021 {[Data set]}}},
	url = {https://doi.org/10.34681/psb7-a026},
	author = {{Children’s Bureau, Administration on Children, Youth and Families}},
shorthand = {Children’s Bureau},
	year = {2023},
note = {Accessed: 2024-02-28},
}

@article{roby2010adoption,
  title={Adoption activities on the Internet: A call for regulation},
  author={Roby, Jini L and White, Holly},
  journal={Social Work},
  volume={55},
  number={3},
  pages={203--212},
  year={2010},
  publisher={Oxford University Press}
}

@article{avery2000perceptions,
  title={Perceptions and practice: Agency efforts for the hardest-to-place children},
  author={Avery, Rosemary J},
  journal={Children and Youth Services Review},
  volume={22},
  number={6},
  pages={399--420},
  year={2000},
  publisher={Elsevier}
}

@article{feldman2016not,
  title={Not too late: Effects of a diligent recruitment program for hard to place youth},
  author={Feldman, Sara Wolf and Price, Kerry Monahan and Ruppel, Joanne},
  journal={Children and Youth Services Review},
  volume={65},
  pages={26--31},
  year={2016},
  publisher={Elsevier}
}

@article{vandivere2015experimental,
  title={Experimental evaluation of a child-focused adoption recruitment program for children and youth in foster care},
  author={Vandivere, Sharon and Malm, Karin E and Zinn, Andrew and Allen, Tiffany J and Mcklindon, Amy},
  journal={Journal of Public Child Welfare},
  volume={9},
  number={2},
  pages={174--194},
  year={2015},
  publisher={Taylor \& Francis}
}

@misc{fl2019,
  author = "{Florida Department of Children and Families}",
  title = "Adoption Incentive Annual Report",
  howpublished = "Office of Child Welfare",
  year = "2019",
  url = "https://www.myflfamilies.com/sites/default/files/2023-02/2019%2520Adoption%2520Incentive%2520Report.pdf",
note = {Accessed: 2024-02-28}
}

@misc{fl2024,
  author = "{Florida Department of Children and Families}",
  title = "Adoption Incentive Annual Report",
  howpublished = "Office of Child and Family Well-Being",
  year = "2024",
  url = "https://www.myflfamilies.com/sites/default/files/2024-12/2024%20Annual%20Adoption%20Incentive%20Report.pdf",
note = {Accessed: 2025-03-11}
}

@article{cox1972regression,
  title={Regression models and life-tables},
  author={Cox, David R},
  journal={Journal of the Royal Statistical Society: Series B (Methodological)},
  volume={34},
  number={2},
  pages={187--202},
  year={1972},
  publisher={Wiley Online Library}
}

@article{riley2019, title={Adoptions Powered by Algorithms}, url={https://www.wsj.com/articles/adoptions-powered-by-algorithms-11546620390}, journal={Wall Street Journal}, author={Riley, N S}, year={2019}, month={Jan}, day={4},note = {Accessed: 2024-02-28}}

@article{yamatani2009child,
  title={Child welfare worker caseload: {W}hat's just right?},
  author={Yamatani, Hide and Engel, Rafael and Spjeldnes, Solveig},
  journal={Social Work},
  volume={54},
  number={4},
  pages={361--368},
  year={2009},
  publisher={Oxford University Press}
}

@article{lushin2023burdened,
  title={A burdened workforce: Exploring burnout, job satisfaction and turnover among child welfare caseworkers in the era of {COVID}-19},
  author={Lushin, Victor and Katz, Colleen C and Julien-Chinn, Francie J and Lalayants, Marina},
  journal={Children and Youth Services Review},
  volume={148},
  pages={106910},
  year={2023},
  publisher={Elsevier}
}

@article{berenguer2023managing,
  title={Managing volunteers and paid workers in a nonprofit operation},
  author={Berenguer, Gemma and Haskell, William B and Li, Lei},
  journal={Management Science},
  volume={70},
  number={8},
  pages={5298--5316},
  year={2024},
  publisher={INFORMS}
}

@article{spindler1970social,
  title={Social and Rehabilitation Services: A Challenge to Operations Research},
  author={Spindler, Arthur},
  journal={Operations Research},
  volume={18},
  number={6},
  pages={1112--1124},
  year={1970},
  publisher={INFORMS}
}

@article{quast2018opioid,
  title={Opioid prescription rates and child removals: Evidence from Florida},
  author={Quast, Troy and Storch, Eric A and Yampolskaya, Svetlana},
  journal={Health Affairs},
  volume={37},
  number={1},
  pages={134--139},
  year={2018}
}

@article{breslow1972discussion,
  title={Contribution to discussion of paper by DR Cox},
  author={Breslow, Norman E},
  journal={Journal of the Royal Statistical Society, Series B},
  volume={34},
  pages={216--217},
  year={1972}
}

@techreport{fl_dcf_2022,
  title        = {A Comprehensive, Multi-Year Review of the Revenues, Expenditures, and Financial Position of All Community-Based Care Lead Agencies with System of Care Analysis},
  author       = {{Florida Department of Children and Families}},
  year         = {2022},
  month        = {dec},
  institution  = {Florida Department of Children and Families},
  address      = {Tallahassee, FL},
  note         = {State Fiscal Years 2020--2021 and 2021--2022},
  url          = {https://www.myflfamilies.com/programs/childwelfare/dashboard/},
}

@article{gypen2017outcomes,
  title={Outcomes of children who grew up in foster care: Systematic-review},
  author={Gypen, Laura and Vanderfaeillie, Johan and De Maeyer, Skrallan and Belenger, Laurence and Van Holen, Frank},
  journal={Children and Youth Services Review},
  volume={76},
  pages={74--83},
  year={2017},
  publisher={Elsevier}
}

@article{slaugh2025child,
  title={Child welfare services in the united states: An operations research perspective},
  author={Slaugh, Vincent W and Dierks, Ludwig and Scheller-Wolf, Alan and Trapp, Andrew C and {\"U}nver, M Utku},
  journal={Nonprofit Operations and Supply Chain Management: Theory and Practice},
  pages={321--347},
  year={2025},
  publisher={Springer}
}

@misc{acf_afcars_dashboard,
  author       = {{U.S. Administration for Children and Families}},
  title        = {{AFCARS Adoption and Foster Care Analysis and Reporting System Dashboard}},
  year         = {2024},
  url          = {https://tableau-public.acf.gov/views/afcars_dashboard_main_page/mainpage},
  note         = {Accessed: 2026-03-05}
}
 \clearpage

\clearpage

\renewcommand\thesection{\Alph{section}}
\pagenumbering{arabic}
\setcounter{section}{0}
\section*{e-Companion}
 \section{Home Studies}\label{app:home_study}
Before a family can be considered for adoption, they have to complete a home study. 
 State regulations determine minimum requirements for home studies. For example, 2024 Florida Statutes Chapter 63 Section 092 states:
\begin{quote}
\emph{
The preliminary home study must be made to determine the suitability of the intended adoptive parents and may be completed prior to the identification of a prospective adoptive child. The study must include, at a minimum, the following:
\begin{itemize}
   \item An interview with the intended adoptive parents.
 \item  Records checks of the department's central abuse registry, which the department shall provide to the entity conducting the preliminary home study, and criminal records correspondence checks under s. 39.0138 through the Department of Law Enforcement on the intended adoptive parents.
\item  An assessment of the physical environment of the home.
\item  A determination of the financial security of the intended adoptive parents.
\item Documentation of counseling and education of the intended adoptive parents on adoptive parenting, as determined by the entity conducting the preliminary home study. The training specified in s. 409.175(14) shall only be required for persons who adopt children from the department.
\item Documentation that information on adoption and the adoption process has been provided to the intended adoptive parents.
\item Documentation that information on support services available in the community has been provided to the intended adoptive parents.
\item A copy of each signed acknowledgment of receipt of disclosure required by s. 63.085.
    \end{itemize}
    }
    \end{quote}

However, most agencies perform more comprehensive studies than required by law to ensure a smoother process. In practice, home studies typically also include additional information on 
\begin{itemize}
    \item Family background, including education and health 
    \item Relationships and social environment 
    \item Parenting experiences
    \item A description of daily life routines
    \item Details about home and neighborhood 
    \item Reasons for seeking an adoptive placement
    \item Recommendations by the caseworker about what types of children the family is suited for (e.g., especially regarding special needs, but also other factors such as gender or age)
\end{itemize}

\section{Increased Uncertainty or Heterogeneity}\label{app:theory_uncertainty_heterogeneity}
Our analytical model assumes a relatively limited amount of variability and uncertainty: the value of every match is known a priori. The only uncertainty is whether the child is compatible with the considered family, with the suitability probability $p$ being the same for all matches. This is not without loss of generality. In practice, compatibility is a spectrum, and values may be uncertain. This means that while caseworkers and families may have prior beliefs based on their own experiences and potentially supplemented by recommender systems, the actual value of a match is not known until the investigation is conducted. Unfortunately, explicitly modeling such higher degrees of uncertainty is intractable in a model where both market sides are strategic and more than one match may be investigated per time period. 

This is because if values are uncertain, family utility is no longer monotonic in terms of how selective a child is. While children being less selective (i.e., considering families it has lower expected value for) still increases a family's utility if it causes the child to be interested in them, it often reduces a family's utility if the child was already interested in them (as the child has some probability of instead matching with a different family they didn't even consider before). This may lead to non-convergence of best responses and non-existence of equilibria in pure strategy, even if only either the suitability or market thickness is heterogeneous.

\begin{proposition}
    If suitability probabilities are heterogeneous, i.e., if there exist different $p_{c,f}$  for different child/family pairs, or if families have heterogeneous probabilities $\lambda_f$ to be present, then pure strategy equilibria do not always exist. 
\end{proposition}
\begin{proof}{Proof.}\label{PROOF:UNCERTAINTY}
	Consider the following example in FS where each child/family pair has its own probability $p_{c,f}$ for being suitable. 
	Let $C = \{c_1,c_2\}$, $F = \{f_1, f_2, f_3\}$, and let valuations and suitability probabilities be according to the following tables
   
	\begin{table}[H]
            \begin{center}
		{}
		{\begin{tabular}{ |c|c|c|c| } \hline
				$\val{c}{f}$ & $f_1$ & $f_2$ & $f_3$ \\ \hline
				$c_1$ & $2$ & $1$ & $0.5$ \\ \hline
                $c_2$ & $0.5$ & $0$ & $1$ \\ \hline
			\end{tabular}\quad
			\begin{tabular}{ |c|c|c|c| } \hline
				$\val{f}{c}$ & $f_1$ & $f_2$ & $f_3$ \\ \hline
				$c_1$ & $20$ & $0.11$ & $2$ \\ \hline
                $c_2$ & $5$ & $0$ & $2$ \\ \hline
		\end{tabular}
            \quad
		\begin{tabular}{ |c|c|c|c| } \hline
				$p_{c,f}$ & $f_1$ & $f_2$ & $f_3$ \\ \hline
				$c_1$ & $0.1$ & $0.99$ & $1$ \\ \hline
                $c_2$ & $0.5$ & $0.5$ & $0.5$ \\ \hline
		\end{tabular}}
		{}
            \end{center}
	\end{table}
    
	Let $\lambda=1$ (i.e., all family types are always present). Further, assume children are very impatient and have negligible search costs, i.e. $\delta_C = \kappa_C= 0$, while families are very patient  $\delta_F = 1$ and have search cost $\kappa_F=0.1$. Note that this implies that a) both children are so impatient, that they will be interested in all families and b) families will be so patient, that they are at most interested in a single-child type (unless both children types give almost the same expected utility, which cannot happen in pure strategy with these parameters). Note that setting these extreme values is not required for equilibrium non-existence and is only done for brevity.

    In this setting, it is not rational for $f_2$ to be interested in $c_2$, independent of the other agent's strategies. Further, if $f_1$ is interested in $c_1$ in equilibrium, $f_2$ cannot be interested in $c_1$ as doing so would result in negative utility, as $c_1$ prefers $f_1$ to $f_2$: the probability that $c_1$ doesn't successfully match with $f_1 $ times the probability that $f_2$ and $c_1$ are compatible times $ v_{f_2}(c_1)$ equals  $(1-0.1) \times0.99\times 0.11=0.09801\leq 0.1 =\kappa_F$. However, if $f_1$ is not interested in $c_1$, $f_2$ would obtain positive utility (i.e., $0.99\times 0.11=0.1089\geq 0.1$) and therefore would be interested. 
    
    Thus, we only have 4 possible equilibrium candidates, depending on in which child families $f_1$ and $f_3$ are interested. 

    \begin{enumerate}
        \item If $f_1$ and $f_3$ are interested in $c_1$ in equilibrium, then $f_1$ will obtain utility $0.1\times 20-0.1=1.9$. However, $f_1$ deviating to be (solely) interested in $c_2$ would yield utility $0.5\times 5-0.1=2.4>1.9$, a contradiction. 
        \item If $f_1$ is interested in $c_1$ and $f_3$ is interested in $c_2$ in equilibrium, then 
         $f_3$ will obtain utility $0.5\times 2-0.1=0.9$. However, $f_3$ deviating to be (solely) interested in $c_1$ would yield utility $(1-0.1)\times 1\times  2-0.1=1.7>0.9$, a contradiction. 
         \item If $f_1$ and $f_3$ are interested in $c_2$ in equilibrium, then $f_1$ will only be chosen if $f_3$ is incompatible, thus obtaining expected utility $0.5\times0.5\times 5-0.1=1.15$. However, $f_1$ deviating to be (solely) interested in $c_1$ would yield utility $0.1\times 20-0.1=1.9>1.15$, a contradiction. 
        \item If $f_1$ is interested in $c_2$ and $f_3$ is interested in $c_1$ in equilibrium, then $f_2$ is interested in $c_1$. Since $c_1$ prefers $f_2$ and has a $p_{c_1,f_2}=0.99$ likelihood of being compatible, this implies that $f_3$ only obtains utility $(1-0.99)\times1 \times 2-0.1=-0.08$, an immediate contradiction. 
    \end{enumerate}

    In conclusion, there can be no equilibrium in pure strategies, because $f_1$ always prefers the child type $f_3$ is not interested in, while $f_3$ always prefers the child type $f_1$ is interested in.

To see the nonexistence of equilibria for heterogeneous $\lambda_f$, consider the following example. 
	Let $C = \{c_1,c_2\}$, $F = \{f_1, f_2, f_3\}$ and let valuations and $\lambda_f$ be according to the following tables
	\begin{table}[H]
             \begin{center}		
		{}
		{\begin{tabular}{ |c|c|c|c| } \hline
				$\val{c}{f}$ & $f_1$ & $f_2$ & $f_3$ \\ \hline
				$c_1$ & $2$ & $1$ & $0.5$ \\ \hline
                $c_2$ & $0.5$ & $0$ & $1$ \\ \hline
			\end{tabular}\quad
			\begin{tabular}{ |c|c|c|c| } \hline
				$\val{f}{c}$ & $f_1$ & $f_2$ & $f_3$ \\ \hline
				$c_1$ & $4$ & $0.11$ & $2$ \\ \hline
                $c_2$ & $5$ & $0$ & $1$ \\ \hline
		\end{tabular}
        \quad
			\begin{tabular}{ |c|c|c|c| } \hline
				 & $f_1$ & $f_2$ & $f_3$ \\ \hline
				$\lambda_{f}$ & $0.1$ & $0.99$ & $1$ \\ \hline
		\end{tabular}}
		{}
            \end{center}
	\end{table}
	Let $p=1$ (i.e., all families are suitable). Further, assume children are impatient and have negligible search costs, i.e. $\delta_C = \kappa_C= 0$, while families are very patient  $\delta_F = 1$ and have search cost $\kappa_F=0.1$. Note that this implies that a) both children are so impatient, that they will be interested in all families and b) families will be so patient, that they are at most interested in a single-child type (unless both children types give almost the same expected utility, which cannot happen in pure strategy with these parameters). Note that setting these extreme values are not required for equilibrium non-existence and is only done for brevity.

    In this setting, it is not rational for $f_2$ to be interested in $c_2$, independent of the other agent's strategies. Further, if $f_1$ is mutually interested in $c_1$ in equilibrium, $f_2$ cannot be interested in $c_1$ as doing so would result in negative utility: due to $c_1$ preferring $f_1$ whenever present ($10\%$ of the time), the immediate expected value of showing interest is only $(1-0.1)\times 0.11=0.099$, less than the search cost of $0.1$. However, if $f_1$ is not mutually interested in $c_1$, $f_2$ is the child's first choice and would obtain positive utility (i.e., $ 0.11 \geq 0.1$). Therefore, $f_2$ would be interested. 
    
    Thus, we only have 4 possible equilibrium candidates, depending on which child families $f_1$ and $f_3$ are interested in. 

    \begin{enumerate}
        \item If $f_1$ and $f_3$ are interested in $c_1$ in equilibrium, $f_1$ will obtain utility $4-0.1=3.9$. However, $f_1$ deviating to be (solely) interested in $c_2$ would yield utility $5-0.1=4.9>3.9$, a contradiction. 
        \item If $f_1$ is interested in $c_1$ and $f_3$ is interested in $c_2$ in equilibrium, 
         $f_3$ will obtain utility $1-0.1=0.9$. However, $f_3$ deviating to be (solely) interested in $c_1$, it would successfully match whenever $f_1$ is not present, yielding utility $(1-0.1)\times  2-0.1=1.7>0.9$, a contradiction. 
         \item If $f_1$ and $f_3$ are interested in $c_2$ in equilibrium, $f_1$ will obtain no utility, as $c_2$ will always match successfully with $f_3$, a contradiction since $f_1$ can guarantee positive utility by being interested in $c_1$. 
        \item If $f_1$ is interested in $c_2$ and $f_3$ is mutually interested in $c_1$ in equilibrium, $f_2$ is interested in $c_1$. Since $c_1$ prefers $f_2$ over $f_3$, $f_3$ only obtains utility $(1-0.99)\times1 \times 2-0.1=-0.08$, an immediate contradiction. 
    \end{enumerate}

    In conclusion, there can be no equilibrium in pure strategies, because $f_1$ always prefers the child type $f_3$ is not interested in, while $f_3$ always prefers the child type $f_1$ is interested in.

    \Halmos
\end{proof}

As the same effect can be recreated by uncertainty about match values, we immediately obtain the following.  
\begin{corollary}
       If values are uncertain, pure strategy equilibria do not always exist.
\end{corollary}

While equilibria still exist in mixed strategy (i.e., where agents play randomized strategies), it is intractable to fully characterize or calculate mixed strategy equilibria in such a complex system.  

While we, therefore, restrict our formal analysis to a low level of uncertainty, it is important to note that uncertainty typically favors CS: In CS, the caseworker can decide whether to investigate another family based on the expected utility gain while already knowing the true value of previously investigated matches. Additionally, investigating a family does not prevent the caseworker from going back and matching with previously investigated families. This allows caseworkers more flexibility than FS, where all investigation decisions must be made simultaneously, increasing the chance of ``wasted'' investigations. 

 However, our insight that neither approach always dominates the other still holds. Similar to before, the lower cost of being mutually interested in CS changes strategic considerations on both sides of the market and can lead to more matches in FS (e.g., if families are so patient that they are only interested in a small set of very attractive children in CS).

\section{Additional Lemmas and Propositions}

\subsection{\Cref{proposition:decreasing_order}}\label{app:theory_decreasing_order}

Here, we show that processing families in decreasing order of $v_c(f)$ is optimal for children in CS.

\begin{proposition}\label{proposition:decreasing_order}
	In CS, $c$'s utility is maximized if the caseworker processes families in decreasing order of $\val{c}{f}$.
\end{proposition}

\begin{proof}{Proof.}
Assume child $c$ is active at the current time step. Note that families without interest in $c$ do not affect $c$'s utility in any way. For the remaining families, $c$ faces a Pandora's box problem \citep{weitzman_optimal_1979} where $c$ receives a payoff of $\val{c}{f}$ with probability $p$ and a payoff of 0 with probability $1-p$ when the box corresponding to family $f$ is opened. Notice that the reservation value of the box corresponding to family $f$ is higher than for family $f'$ if and only if $\val{c}{f} > \val{c}{f'}$.
\end{proof}

\subsection{\Cref{proposition:simple_thresholds}}\label{app:theory_simple_thresholds}

Strategy $s_f$ is a simple threshold strategy with threshold $z \in \RR$ for family $f$ if $\strat{f}{} = \Ind[\val{f}{c} \ge z]$, $\forall c \in C$.
Here, we show that there exist instances where families cannot best respond with a simple threshold strategy.

\begin{proposition}\label{proposition:simple_thresholds}
	In FS, there exists an instance with a family $f$ and other agents' strategies $\strat{-f}{}$, such that no simple threshold strategy $\strat{f}{}$ is a best response for $f$.
\end{proposition}

\begin{proof}{Proof.}
Consider the following example:
Let $C=\{c_1,c_2\}$, $F=\{f_1,f_2\}$ and let valuations be according to the following tables for some $\epsilon > 0$.
\begin{table}[H]
	{}
	{\begin{tabular}{ |c|c|c|c| } \hline
			$\val{c}{f}$ & $f_1$ & $f_2$ \\ \hline
			$c_1$ & 1 & $1-\epsilon$ \\ \hline
			$c_2$ & 1 & $1-\epsilon$ \\ \hline
		\end{tabular}\quad
		\begin{tabular}{ |c|c|c|c| } \hline
			$\val{f}{c}$ & $f_1$ & $f_2$ \\ \hline
			$c_1$ & 1 & 1 \\ \hline
			$c_2$ & $1-\epsilon$ & $1-\epsilon$ \\ \hline
	\end{tabular}}
	{}
\end{table}
Suppose strategy profile $s$ is such that $\strat{c}{f} = \strat{f}{c} = 1$ for all $c \in C$, $f \in F$. If $\epsilon$ is small enough and $(1-\lambda p)p < \kappa_F \le p$, then it is optimal for $f_2$ to only be interested in $c_2$, even though $f_2$ strictly prefers $c_1$.
\end{proof}

\section{Remaining Proofs}\label{app:proofs}

\subsection{Proof of \Cref{proposition:utility_characterization}}\label{app:utility_characterization}

In FS, one way to express $c$'s utility is as follows:
\begin{equation}
	\ut[FS]{c}{s} = \Bigg[ 1 - \lambda p \sum_{f \in M_c(\strat{}{})} \bp{c}{f}{s} \Bigg] \delta_C \ut[FS]{c}{s} + \lambda \sum_{f \in M_c(\strat{}{})} \Big[ \bp{c}{f}{s} p \val{c}{f} - \kappa_C \Big].
\end{equation}
The probability of a match forming between $c$ and $f$ at the current time step is $\lambda p \bp{c}{f}{s}$ if there is mutual interest, in which case $c$ obtains a value of $v_c(f)$ and leaves the process. For any active family that showed interest, $c$ incurs search costs $\kappa_C$. If $c$ remains unmatched, then $c$ receives $\delta_C \ut[FS]{c}{s}$. 
By pulling $\delta_C \ut[FS]{c}{s}$ out of the sums, we get
\begin{equation}
	\ut[FS]{c}{s} = \delta_C \ut[FS]{c}{s} + \lambda \sum_{f \in M_c(\strat{}{})} \Big( \bp{c}{f}{s} p \big( \val{c}{f} - \delta_C \ut[FS]{c}{s} \big) - \kappa_C \Big).
\end{equation}
The proof for families' utilities in FS and agents' utilities in CS is analogous and omitted.

\subsection{Proof of \Cref{proposition:ts_best_response}}\label{app:ts_best_response}

First of all, notice that whether agent $i$ is interested in some agent $j$ or not does not affect agent $i$'s utility if $j$ is not interested in $i$.
Further, for an arbitrary family $f$ in FS, $\bp{c}{f}{s}$ does not depend on $\strat{f}{}$. By slightly modifying \Cref{equation:utility_fs_f}, we can see that when $f$ plays a best response in $\strat{}{}$ it must hold that
\begin{equation}\label{equation:best_response_fs_f}
	\ut[FS*]{f}{\strat{-f}{}} = \delta_F \ut[FS*]{f}{\strat{-f}{}} + \frac{1}{n} \sum_{c \in M_f(\strat{}{})} \Big( \bp{c}{f}{s} p \big( \val{f}{c} - \delta_F \ut[FS*]{f}{\strat{-f}{}} \big) - \kappa_F \Big).
\end{equation}
By \Cref{equation:best_response_fs_f} it must hold for all $c\in C$ that
\begin{equation}
	\strat{f}{c} = \Ind[\bp{c}{f}{s} p \big( \val{f}{c} - \delta_F \ut[FS*]{f}{\strat{-f}{}} \big) \ge \kappa_F]
\end{equation} 
when there is mutual interest between $c$ and $f$, as $\strat{f}{}$ would otherwise not be a best response. 
That is, because all children $c$ contribute non-negatively to $f$'s utility if and only if $\bp{c}{f}{s} p \big( \val{f}{c} - \delta_F \ut[FS*]{f}{\strat{-f}{}} \big) \ge \kappa_F$.
By our tie-breaking assumption, the claim of the proposition follows for families in FS. The proof for families in CS is analogous and thus omitted.

In the remainder of the proof, we show that the statement holds for children in FS. The proof for CS is again analogous and therefore omitted. Let $c$ be an arbitrary child. For a best response $\strat{c}{}$ to $\strat{-c}{}$ we have that
\begin{equation}\label{equation:best_response_fs_c}
	\ut[FS*]{c}{\strat{-c}{}} = \delta_C \ut[FS*]{c}{\strat{-c}{}} + \lambda \sum_{f \in F} \strat{c}{f} \strat{f}{c} \Big( \bp{c}{f}{s} p \big( \val{c}{f} - \delta_C \ut[FS*]{c}{\strat{-c}{}} \big) - \kappa_C \Big).
\end{equation}
As for families, it must be the case that
\begin{equation}
	\strat{c}{f} = \Ind[\bp{c}{f}{s} p \big( \val{c}{f} - \delta_C \ut[FS*]{c}{\strat{-c}{}} \big) \ge \kappa_C]
\end{equation} 
because for all $f,f'\in F$
\begin{equation}
	\bp{c}{f}{s} \ge \bp{c}{f'}{s} \iff \val{c}{f} \ge \val{c}{f'},
\end{equation}
and otherwise $\strat{c}{}$ would not be a best response to $\strat{-c}{}$.
Again, by our tie-breaking assumption, the claim of the proposition follows for children in FS.

\subsection{Proof of \Cref{proposition:equilibria_existence}}\label{app:equilibria_existence}
For FS, define a mapping $T^{FS}: Y \rightarrow Y$ as follows: $T^{FS}=(T^{FS}_i)_{i \in A}$, where 
	\begin{equation}
		T^{FS}_c(y) = \delta_C y_c + \lambda \sum_{f \in F} \Ind[\bp{c}{f}{y} p (\val{f}{c} - \delta_F y_f) \ge \kappa_F] \Big( \bp{c}{f}{y} p ( \val{c}{f} - \delta_C y_c) - \kappa_C \Big)^+
	\end{equation}
	for all $c \in C$ and
	\begin{equation}
		T^{FS}_f(y) = \delta_F y_f + \frac{1}{n} \sum_{c \in C} \Ind[\bp{c}{f}{y} p (\val{c}{f} - \delta_C y_c) \ge \kappa_C] \Big( \bp{c}{f}{y} p (\val{f}{c} - \delta_F y_f) - \kappa_F \Big)^+.
	\end{equation}
	for all $f \in F$.
	Note that any fixed point of $T^{FS}$ (i.e., any $y$ with $T^{FS}(y)=y$) is an equilibrium threshold profile in FS.
	We now show that $T^{FS}$ is monotonically increasing according to $\le_C$. Let $c \in C$, $y,y' \in Y$, and $y \le_C y'$. We have
	\begin{align}
		T^{FS}_c(y) &= \delta_C y_c + \lambda \sum_{f \in F} \Ind[\bp{c}{f}{y} p (\val{f}{c} - \delta_F y_f) \ge \kappa_F] \Big( \bp{c}{f}{y} p ( \val{c}{f} - \delta_C y_c) - \kappa_C \Big)^+\\
		&\le \delta_C y'_c + \lambda \sum_{f \in F} \Ind[\bp{c}{f}{y'} p (\val{f}{c} - \delta_F y'_f) \ge \kappa_F] \Big( \bp{c}{f}{y'} p ( \val{c}{f} - \delta_C y'_c) - \kappa_C \Big)^+\\ &= T^{FS}_c(y').
	\end{align}
	The inequality holds for the following reason: Suppose a family $f$ is interested in $c$ under $s(y)$ but not under $s(y')$. Since $f$ is weakly less selective in $s(y')$, it must be the case that there exists another family $f'$ with $\val{c}{f'} > \val{c}{f}$ that is not mutually interested in $c$ under $s(y)$ but under $s(y')$. Note that for any family that loses interest in $c$ under $s(y')$, there must exist such a unique family that replaces it and is preferred by $c$.
	
	Since each child $c$ is weakly more selective under $s(y')$ than $s(y)$, we have for all $f \in F$
	\begin{align}
		T^{FS}_f(y) &= \delta_F y_f + \frac{1}{n} \sum_{c \in C} \Ind[\bp{c}{f}{y} p (\val{c}{f} - \delta_C y_c) \ge \kappa_C] \Big( \bp{c}{f}{y} p (\val{f}{c} - \delta_F y_f) - \kappa_F \Big)^+\\
		&\ge \delta_F y'_f + \frac{1}{n} \sum_{c \in C} \Ind[\bp{c}{f}{y'} p (\val{c}{f} - \delta_C y'_c) \ge \kappa_C] \Big( \bp{c}{f}{y'} p (\val{f}{c} - \delta_F y'_f) - \kappa_F \Big)^+\\ 
		&= T^{FS}_f(y').
	\end{align}
	Note that $T^{FS}$ maps elements from $Y$ to $Y$ and $(Y,\le_C)$ is a complete lattice. By Tarski's fixed point theorem, the claim follows for FS.
 
For CS, we define a mapping $T^{CS}: Y \rightarrow Y$ as follows: $T^{CS}=(T^{CS}_i)_{i \in A}$, where 
\begin{equation}
	T^{CS}_c(y) = \delta_C y_c + \lambda \sum_{f \in F} \Ind[p (\val{f}{c} - \delta_F y_f) \ge \kappa_F] \bp{c}{f}{y} \Big( p ( \val{c}{f} - \delta_C y_c) - \kappa_C \Big)^+
\end{equation}
for all $c \in C$ and
\begin{equation}
	T^{CS}_f(y) = \delta_F y_f + \frac{1}{n} \sum_{c \in C} \Ind[p (\val{c}{f} - \delta_C y_c) \ge \kappa_C] \bp{c}{f}{y} \Big( p (\val{f}{c} - \delta_F y_f) - \kappa_F \Big)^+.
\end{equation}
for all $f \in F$. Note that any fixed point of $T^{CS}$ is an equilibrium threshold profile in CS.
We now show that $T^{CS}$ is monotonically increasing according to $\le_C$. Let $c \in C$, $y,y' \in Y$, and $y \le_C y'$. It holds that
\begin{align}
	T^{CS}_c(y) &= \delta_C y_c + \lambda \sum_{f \in F} \Ind[p (\val{f}{c} - \delta_F y_f) \ge \kappa_F] \bp{c}{f}{y} \Big( p ( \val{c}{f} - \delta_C y_c) - \kappa_C \Big)^+\\
	&\le \delta_C y'_c + \lambda \sum_{f \in F} \Ind[p (\val{f}{c} - \delta_F y'_f) \ge \kappa_F] \bp{c}{f}{y'} \Big( p ( \val{c}{f} - \delta_C y'_c) - \kappa_C \Big)^+\\ &= T^{CS}_c(y'),
\end{align}
because each family is weakly less selective under $s(y')$ than $s(y)$.

Similarly, since each child is weakly more selective under $s(y')$ than $s(y)$, we have for all $f \in F$
\begin{align}
	T^{CS}_f(y) &= \delta_F y_f + \frac{1}{n} \sum_{c \in C} \Ind[p (\val{c}{f} - \delta_C y_c) \ge \kappa_C] \bp{c}{f}{y} \Big(p (\val{f}{c} - \delta_F y_f) - \kappa_F \Big)^+\\
	&\ge \delta_F y'_f + \frac{1}{n} \sum_{c \in C} \Ind[p (\val{c}{f} - \delta_C y_c) \ge \kappa_C] \bp{c}{f}{y'} \Big( p (\val{f}{c} - \delta_F y'_f) - \kappa_F \Big)^+\\ 
	&= T^{CS}_f(y').
\end{align}
Note that $T^{CS}$ maps elements from $Y$ to $Y$ and $(Y,\le_C)$ is a complete lattice. By Tarski's fixed point theorem, the claim for CS follows.

\subsection{Proof of Lemma~\ref{lemma:additional_match_fs}}
\begin{proof}{Proof.}\label{app:additional_match_fs}
If $(c,f) \in M(\se{FS})$ and $(c,f) \notin M(\se{CS})$, then it holds that $\bp{c}{f}{\se{FS}} p(\val{c}{f} - \delta_C \ut{c}{\se{FS})} \ge \kappa_C$ and $\bp{c}{f}{\se{FS}} p(\val{f}{c} - \delta_F \ut{f}{\se{FS})} \ge \kappa_F$.
Further, either $p(\val{c}{f} - \delta_C \ut{c}{\se{CS}}) < \kappa_C$ or $p(\val{f}{c} - \delta_F \ut{f}{\se{CS}}) < \kappa_F$. If the former is true we have 
\begin{equation}
	\delta_C \ut{c}{\se{FS}} \le \val{c}{f} - \frac{\kappa_C}{\bp{c}{f}{\se{FS}} p} \le \val{c}{f} - \frac{\kappa_C}{ p} < \delta_C \ut{c}{\se{CS}},
\end{equation} 
since $0 < \beta_c(f,\se{FS}) \le 1$. Otherwise, we get 
\begin{equation}
	\delta_F \ut{f}{\se{FS}} \le \val{f}{c} - \frac{\kappa_F}{\bp{c}{f}{\se{FS}} p} \le \val{f}{c} - \frac{\kappa_F}{ p} < \delta_F \ut{f}{\se{CS}}.
\end{equation} 
\end{proof}

\subsection{Proof of Lemma~\ref{lemma:same_matches_cs_weakly_better}.}
\begin{proof}{Proof.}\label{app:same_matches_cs_weakly_better}
If $c$ responds to $\se{CS}_{-c}$ with interest in all families they were mutually interested in under $\se{FS}$, $c$ is mutually interested in the same families as under $\se{FS}$ since $M_c(\se{FS}) \subseteq M_c(\se{CS})$. Further, $c$'s expected costs are weakly lower in CS compared to FS. Hence, there exists a strategy for $c$ in CS where $c$'s utility is weakly higher than $\ut{c}{\se{FS}}$. The fact that $c$ plays a best response in $\se{CS}$ completes the proof.
\end{proof}

\subsection{Proof of Theorem~\ref{theorem:pareto_dominance}}
\begin{proof}{Proof.}\label{PROOF:pareto_dominance}
It is easy to see why all CSEs can be Pareto improvements over all FSEs:
Consider the following example:
Let $C = \{c\}$, $F = \{f_1, f_2\}$ and let valuations be according to the following tables for some $\epsilon > 0$.
\begin{table}[H]
	{}
	{\begin{tabular}{ |c|c|c|c| } \hline
			$\val{c}{f}$ & $f_1$ & $f_2$ \\ \hline
			$c_1$ & 1 & $1-\epsilon$ \\ \hline
		\end{tabular}\quad
		\begin{tabular}{ |c|c|c|c| } \hline
			$\val{f}{c}$ & $f_1$ & $f_2$ \\ \hline
			$c_1$ & 1 & 1 \\ \hline
	\end{tabular}}
	{}
\end{table} 
 If $p(1-\lambda p) \ge \max\{\kappa_C, \kappa_F\}$ and $\epsilon$ and $\delta_C$ are sufficiently small, the child is mutually interested with both families, that is, the matching correspondences of the unique FSE and the unique CSE are identical. However, $c_1$ and $f_2$ only incur search costs for their mutual interest in CS if the match between $f_1$ and $c_1$ is unsuitable or $f_1$ is absent. Both $c_1$ and $f_2$ thus strictly prefer CS. Thus, the CSE is a Pareto improvement over the FSE.

Let $\se{FS}\in S^{FS}$ and $\se{CS} \in S^{CS}$. We now show that if there exists an agent $i \in A$, such that $\ut{i}{\se{FS}} > \ut{i}{\se{CS}}$, then there exists another agent $j \in A$ with $\ut{j}{\se{FS}} < \ut{j}{\se{CS}}$.
Fix consider that $i=c\in C$. If $\ut{c}{\se{FS}} > \ut{c}{\se{CS}}$, we get by \Cref{lemma:same_matches_cs_weakly_better} that $M_c(\se{FS}) \not\subseteq M_c(\se{CS})$. Then, by \Cref{lemma:additional_match_fs}, the claim immediately follows.

Now suppose $i=f \in F$. Further, for the sake of contradiction, assume $M_c(\se{FS}) \subseteq M_c(\se{CS})$ for all $c \in C$, as otherwise by \Cref{lemma:additional_match_fs} the claim would immediately follow. Therefore, $f$'s increase in $\ut{f}{\se{FS}}$ can only come from the fact that there exists a child $c \in M_f(\se{FS})$ and another family $f' \in F$ with $\val{c}{f'} > \val{c}{f}$ such that $f'$ is not mutually interested in $c$ under $\se{FS}$ but under $\se{CS}$. 
Now suppose $c$ responds to $\strat[CS]{-c}{}$ with $\strat[FS]{c}{}$ in CS. By \Cref{lemma:same_matches_cs_weakly_better}, this would imply $\ut[FS]{c}{\se{FS}} \le \ut[CS]{c}{(\strat[FS]{c}{},\strat[CS]{-c}{})}{}$. However, note that $c$'s utility would strictly increase further if $c$ would show interest in $f'$ instead of $f$ since $\val{c}{f'} > \val{c}{f}$, that is, 
\begin{align}
    \ut[FS]{c}{\se{FS}} \le \ut[CS]{c}{(\strat[FS]{c}{},\strat[CS]{-c}{})}{} <\ut[CS]{c}{\se{CS}},
\end{align}
and therefore $c$ is worse off under FS, and the claim follows.\end{proof}

\subsection{Proof of Proposition~\ref{proposition:multiplicity_utility_gaps}}
\begin{proof}{Proof.}\label{PROOF:multiplicity_utility_gaps}
	Consider the following example:
	Let $C=\{c_1,c_2\}$, $F=\{f_1,f_2\}$ and let valuations be according to the following tables.
	\begin{table}[H]
		{\label{tab:val_multiplicity}}
		{\begin{tabular}{ |c|c|c|c| } \hline
				$\val{c}{f}$ & $f_1$ & $f_2$ \\ \hline
				$c_1$ & $L$ & $\epsilon$ \\ \hline
				$c_2$ & $\epsilon$ & $L$ \\ \hline
			\end{tabular}\quad
			\begin{tabular}{ |c|c|c|c| } \hline
				$\val{f}{c}$ & $f_1$ & $f_2$ \\ \hline
				$c_1$ & $\epsilon$ & $L$ \\ \hline
				$c_2$ & $L$ & $\epsilon$ \\ \hline
		\end{tabular}}
		{}
	\end{table}	
	If agents are patient enough (i.e., $\delta_C$ and $\delta_F$ are close enough to $1$) and search costs are small enough, there exist only two equilibria, independent of search technology. In the child-optimal equilibrium, $\se{co}$, children are only interested in their preferred choice (i.e., the agent type for which they have a match value of $L$) and in the family-optimal equilibrium $\se{fo}$ families are only interested in their preferred choice.
	Note that if each child is mutually interested in at most one family in strategy profile $s$, then $\ut[CS]{i}{s} = \ut[FS]{i}{s}$ for all $i \in A$.
	As $p, \delta_F, \delta_C \rightarrow 1$, children's utilities will be weakly less than $\epsilon$ under $\se{fo}$ while being positive and converging to $L$ under $\se{co}$. Similarly, families' utilities will be weakly less than $\epsilon$ under $\se{co}$ while being positive and converging to $L$ under $\se{fo}$.
\end{proof}

\subsection{Proof of Proposition~\ref{proposition:child_worse_off_in_cs}}
\begin{proof}{Proof.}\label{PROOF:child_worse_off_in_cs}
	Consider the following example:
	Let $C=\{c_1,c_2\}$, $F=\{f_1,f_2\}$ and let valuations be according to the following tables for some $\epsilon > 0$.
	\begin{table}[H]
		{\label{tab:val_children_worse}}
		{\begin{tabular}{ |c|c|c|c| } \hline
				$\val{c}{f}$ & $f_1$ & $f_2$ \\ \hline
				$c_1$ & 1 & $1-\epsilon$ \\ \hline
				$c_2$ & 1 & $1-\epsilon$ \\ \hline
			\end{tabular}\quad
			\begin{tabular}{ |c|c|c|c| } \hline
				$\val{f}{c}$ & $f_1$ & $f_2$ \\ \hline
				$c_1$ & 1 & 1 \\ \hline
				$c_2$ & 0 & $\kappa_F/p$ \\ \hline
		\end{tabular}}
		{}
	\end{table}		
	If $\epsilon > 0$ is small enough and $p(1 - \lambda p) < \kappa_F$, then in the unique FSE $\se{FS}$, only $c_1$ and $f_1$ will be mutually interested in each other and $c_2$ and $f_2$.  Family $f_2$ will not be interested in $c_1$ in FS, independent of how patient $f_2$ is. That is, because the probability of actually matching with $c_1$ while facing competition from $f_1$ is too small, yet $f_2$ would have to incur search costs every time $c_1$ is active. However, if $\delta_F$ is sufficiently large, then the unique CSE matching correspondence is $M(\se{CS}) = \{(c_1,f_1), (c_1,f_2)\}$. In CS, $c_2$, the ``low-type'' child, will remain unmatched.
\end{proof}

\subsection{Proof of Proposition~\ref{proposition:family_worse_off_in_cs}}
\begin{proof}{Proof.}\label{PROOF:family_worse_off_in_cs}
	Consider the following example:
	Let $C = \{c\}$, $F = \{f_1, f_2, f_3\}$ and let valuations be according to the following tables for some $\epsilon > 0$.
	\begin{table}[H]
		{}
		{\begin{tabular}{ |c|c|c|c| } \hline
				$\val{c}{f}$ & $f_1$ & $f_2$ & $f_3$ \\ \hline
				$c$ & $1$ & $1-\epsilon$ & $1-2\epsilon$ \\ \hline
			\end{tabular}\quad
			\begin{tabular}{ |c|c|c|c| } \hline
				$\val{f}{c}$ & $f_1$ & $f_2$ & $f_3$ \\ \hline
				$c$ & $1$ & $\kappa_F/p + \epsilon$ & $1$ \\ \hline
		\end{tabular}}
		{}
	\end{table}
	Choose $\delta_C = \delta_F = 0$, $\lambda =1$ and $p<1$.
	If $\epsilon$, $\kappa_C$ and $\kappa_F$ are sufficiently small, but positive and $\kappa_F <  p(1-p)/(2-p)$ we get the following: In the unique FSE $\se{FS}$, child $c$ will be mutually interested in both $f_1$ and $f_3$, but family $f_2$ is not interested in $c$ due immediate utility being lower than search costs, that is,  \begin{align}
	    (1-p)p(\kappa_F/p+\epsilon) \\
        =  (1-p) \kappa_F +\epsilon < \kappa_F.
	\end{align} However, in the unique CSE, all agents will have mutual interest in each other, and therefore $f_3$'s utility strictly decreases, that is, 
    \begin{align}
	\ut[CS]{f_3}{s^CS} = &  (1-p)^2 (p - \kappa_F)\\
        <&  (1-p) p - \kappa_F \ut[FS]{f_3}{s^FS}
	\end{align}
    since $\kappa_F< p(1-p)/(2-p)$.
\end{proof}
\subsection{Proof of Proposition~\ref{proposition:impatient_families}}
\begin{proof}{Proof.}\label{PROOF:impatient_families}
Suppose all instance parameters except for $\delta_F$ are fixed.
Let $\MM$ be the set of all child-family pairs $(c,f)$ for which there exists $\delta_F \in [0,1)$, such that $c$ and $f$ are mutually interested in each other under some FSE $\strat[FS]{}{}$. 
Further, let $v_{\min}$ denote the smallest value a family in $\MM$ has for a mutually interested child, i.e., $v_{\min} = \min_{(c,f) \in \MM} \val{f}{c}$. 
For $\delta_F$, such that
\begin{align}\label{equation:delta_f_small}
	\delta_F \le \frac{v_{\min} - \kappa_F/p}{\bar v},
\end{align}
it follows that $p(\val{f}{c} - \delta_F \bar v) \ge \kappa_F$ for all $(c,f) \in \MM$. Since $\bar v$ is an upper bound for agents' utilities, for any such $\delta_F$ and any $(c,f) \in \MM$, being interested in $c$ is (weakly) advantageous for $f$ under CS, independent of all other strategies. As in the proof of \Cref{lemma:same_matches_cs_weakly_better}, this implies that $c$ must be weakly better off in CS compared to FS.
\end{proof}

\subsection{Proof of Proposition~\ref{proposition:increased_lambda_fs}}
\begin{proof}{Proof.}\label{PROOF:increased_lambda_fs}
	Consider the following example:
	Let $C=\{c\}$, $F=\{f_1,f_2\}$ and let valuations be according to the following tables for some $\epsilon > 0$.
	\begin{table}[H]
		{\label{tab:children_worse_off_fs_lambda}}
		{\begin{tabular}{ |c|c|c|c| } \hline
				$\val{c}{f}$ & $f_1$ & $f_2$ \\ \hline
				$c$ & 1 & $1-\epsilon$ \\ \hline
			\end{tabular}\quad
			\begin{tabular}{ |c|c|c|c| } \hline
				$\val{f}{c}$ & $f_1$ & $f_2$ \\ \hline
				$c$ & 1 & 1 \\ \hline
		\end{tabular}}
		{}
	\end{table}
	Choose $p$ and $\kappa_F$, such that $p(1-\lambda' p) < \kappa_F \le p(1-\lambda p)$.
	If $\kappa_C$, $\delta_C$, and $\delta_F$ are sufficiently small, child $c$ will be mutually interested in $f_1$ and $f_2$ in the unique FSE for $\lambda$. But in the unique FSE for $\lambda'$, only $c$ and $f_1$ will be mutually interested in each other. However, if the difference between $\lambda$ and $\lambda'$ is very small, $c$'s utility can be smaller in the case where the market thickness indicator is $\lambda$, as the increased probability of $f_1$ being active might not compensate for the loss of $f_2$'s interest.
\end{proof}

\subsection{Proof of \Cref{proposition:increased_lambda_cs}}\label{app:increased_lambda_cs}

Let $u = (u_i(s^{co-CS,\lambda}))_{i\in A}$ denote the vector of agents' utilities under $s^{co-CS,\lambda}$ and let $Y'=[u_c, \bar v]^n \times [0, u_f]^m \subseteq [0, \bar v]^{n+m} = Y$.
We now define a mapping $T^\lambda: Y \rightarrow Y$ as follows: $T^\lambda=(T^\lambda_i)_{i \in A}$, where 
\begin{equation}
	T^\lambda_c(y) = \delta_C y_c + \lambda \sum_{f \in F} \Ind[p (\val{f}{c} - \delta_F y_f) \ge \kappa_F] \bp{c}{f}{y} \Big( p ( \val{c}{f} - \delta_C y_c) - \kappa_C \Big)^+
\end{equation}
for all $c \in C$ and
\begin{equation}
	T^\lambda_f(y) = \delta_F y_f + \frac{1}{n} \sum_{c \in C} \Ind[p (\val{c}{f} - \delta_C y_c) \ge \kappa_C] \bp{c}{f}{y} \Big( p (\val{f}{c} - \delta_F y_f) - \kappa_F \Big)^+.
\end{equation}

As we have seen in a previous proof, $T^\lambda$ is $\le_C$-monotone on $Y$. We now show that $T^{\lambda'}$ maps from $Y'$ to $Y'$, which by Tarski's fixed-point theorem yields the result. Since $T^{\lambda}$ is $\le_C$-monotone and $u$ is the $\le_C$-minimal element of $Y'$, it is sufficient to show that $T^{\lambda'}(u) \in Y'$.

Because $u$ is an equilibrium threshold profile, we have that $T^{\lambda}(u)=u \in Y'$. Further, it can easily be verified that $T^{\lambda}(y) \le_C T^{\lambda'}(y)$ for all $y \in Y$. Hence, $T^{\lambda'}(u) \in Y'$, which completes the proof.

\section{Algorithm to Compute FS-TSs from Threshold Profiles}\label{app:alg_fs_ts}

Algorithm~\ref{alg:fs_thresholds} can be used to compute $\strat[FS]{}{y}$ for a given threshold profile $y\in \RR^{n+m}$.
Since $\bp{c}{f}{s}$ only depends on families $f' \in F$ with $\val{c}{f'} > \val{c}{f}$ and for each child families are processed in decreasing order of $\val{c}{f}$, the final strategy profile satisfies the equations for FS from \Cref{proposition:utility_characterization}. 

\begin{algorithm}[ht]
   	\SetAlgoNoLine
   	\KwIn{$y \in \RR^{n+m}$}
   	\KwOut{Strategy profile $s \in S$}
   	$\strat{c}{f} :=0$ and $\strat{f}{c} :=0$ for all $c \in C$, $f \in F$
   	
   	\For{$c \in C$}{
   		$U := F$
   		
   		\While{$U \neq \emptyset$} {
   			$f := \arg\max_{f' \in U} \val{c}{f'}$
   			
   			\If{$\bp{c}{f}{s} p (\val{f}{c} - \delta_F y_f) \ge \kappa_F$} {
   				$\strat{f}{c} := 1$
   			}
   			\If{$\bp{c}{f}{s} p (\val{c}{f} - \delta_C y_c) \ge \kappa_C$} {
   				$\strat{c}{f} := 1$
   			}
   			$U := U \setminus \{f\}$
   		}
   	}
   	\caption{Thresholds to strategy profile}
   	\label{alg:fs_thresholds}
\end{algorithm}

\section{Limit Results}\label{app:theory_limit_results}
    
Here, we provide a collection of limit results that all illustrate how the differences between FS and CS disappear as certain parameters take on extreme values.

\subsection{Negligible Search Costs}

The next proposition shows that the games induced by CS and FS become identical as search costs become negligible.

\begin{proposition}\label{proposition:limit_kappa_small}
   	As $\kappa_C \rightarrow 0$ and $\kappa_F \rightarrow 0$, $|\ut[FS]{i}{s} - \ut[CS]{i}{s}| \rightarrow 0$ for all $s \in S$, $i \in A$.
\end{proposition}
\begin{proof}{Proof.}
Assume that $\kappa_C = \kappa_F =: \kappa$ and let $\strat{}{} \in S$.
For each family $f$, define $T_{\strat{}{},f}^{FS}: Y \rightarrow \RR$ and $T_{\strat{}{},f}^{CS}: Y \rightarrow \RR$ as follows:
\begin{equation}
   	T_{\strat{}{},f}^{FS}(y) = \delta_F y_f + \frac{1}{n} \sum_{c \in C} \strat{c}{f} \strat{f}{c} \Big( \bp{c}{f}{\strat{}{}} p ( \val{f}{c} - \delta_F y_f) - \kappa \Big)
\end{equation}
and
\begin{equation}
   	T_{\strat{}{},f}^{CS}(y) = \delta_F y_f + \frac{1}{n} \sum_{c \in C} \strat{c}{f} \strat{f}{c} \bp{c}{f}{\strat{}{}} \Big( p ( \val{f}{c} - \delta_F y_f) - \kappa \Big).
\end{equation}
Similarly, for all $c \in C$, let
\begin{equation}
   	T_{\strat{}{},c}^{FS}(y) = \delta_C y_c + \lambda \sum_{f \in F} \strat{c}{f} \strat{f}{c} \Big( \bp{c}{f}{\strat{}{}} p ( \val{c}{f} - \delta_C y_c) - \kappa \Big)
\end{equation}
and
\begin{equation}
   	T_{\strat{}{},c}^{CS}(y) = \delta_C y_c + \lambda \sum_{f \in F} \strat{c}{f} \strat{f}{c} \bp{c}{f}{\strat{}{}} \Big( p ( \val{c}{f} - \delta_C y_c) - \kappa \Big).
\end{equation}
Notice that the unique fixpoints $y^{FS}, y^{CS} \in Y$ of $T_{\strat{}{}}^{FS}(y)$ and $T_{\strat{}{}}^{CS}(y)$ define agents' utilities under $\strat{}{}$ in FS and CS, respectively.
For all $y \in Y$ and $f \in F$ it holds that
\begin{align} 
   	&\lim_{\kappa \rightarrow 0} \left( \delta_F y_f + \frac{1}{n} \sum_{c \in C} 	\strat{c}{f} \strat{f}{c} \Big( \bp{c}{f}{\strat{}{}} p ( \val{f}{c} - \delta_F y_f) - \kappa \Big) \right)\\
   	= &\delta_F y_f + \frac{1}{n} \sum_{c \in C} \strat{c}{f} \strat{f}{c} \Big( \bp{c}{f}{\strat{}{}} p ( \val{f}{c} - \delta_F y_f) \Big)\\
   	= &\lim_{\kappa \rightarrow 0} \left( \delta_F y_f + \frac{1}{n} \sum_{c \in C} 	\strat{c}{f} \strat{f}{c} \bp{c}{f}{\strat{}{}} \Big( p ( \val{f}{c} - \delta_F y_f) - \kappa \Big) \right).
\end{align}
Similarly, for all $y \in Y$ and $c \in C$ we have that
\begin{align} 
   	&\lim_{\kappa \rightarrow 0} \left( \delta_C y_c + \lambda \sum_{f \in F} \strat{c}{f} \strat{f}{c} \Big( \bp{c}{f}{\strat{}{}} p ( \val{c}{f} - \delta_C y_c) - \kappa \Big) \right)\\
   	= &\delta_C y_c + \lambda \sum_{f \in F} \strat{c}{f} \strat{f}{c} \Big( \bp{c}{f}{\strat{}{}} p ( \val{c}{f} - \delta_C y_c) \Big)\\
   	= &\lim_{\kappa \rightarrow 0} \left( \delta_C y_c + \lambda \sum_{f \in F} \strat{c}{f} \strat{f}{c} \bp{c}{f}{\strat{}{}} \Big( p ( \val{c}{f} - \delta_C y_c) - \kappa \Big) \right),
\end{align}
and hence $\lim_{\kappa \rightarrow 0} |\ut[FS]{i}{\strat{}{}} - \ut[CS]{i}{\strat{}{}}| = 0$ for all $i \in A$.
\end{proof}

FS and CS do not necessarily become identical if only one side has negligible search costs. In both cases---i.e., if only $\kappa_C \rightarrow 0$ or only $\kappa_F \rightarrow 0$---we can create instances where the sets of equilibria differ from each other in the two approaches.

\subsection{High Match Success Probability and Market Thickness}

Match success probability and market thickness indicator are strongly connected in our model, such that we cannot make any insightful statements about the limit behavior if only one of them approaches 1. However, if it is certain that a family of each type will be present at each time step and that each family would be a suitable match, we observe once more that FS and CS become equivalent in a slightly different way.

\begin{proposition}\label{proposition:limit_p_lambda}
   	As $\lambda p \rightarrow 1$, $|u^{FS}_i(\se{FS}(y)) - u^{CS}_i(\se{CS}(y))| \rightarrow 0$ for all $y \in Y$, $i \in A$.
\end{proposition}
\begin{proof}{Proof.} 
For each family $f$, define $T_{f}^{FS}: Y \rightarrow \RR$ and $T_{f}^{CS}: Y \rightarrow \RR$ as follows:
\begin{equation}
   	T_{f}^{FS}(y) = \delta_F y_f + \frac{1}{n} \sum_{c \in C} \Ind[\bp[FS]{c}{f}{y} p (\val{c}{f} - \delta_C y_c) \ge \kappa_C] \Big( \bp[FS]{c}{f}{y} p ( \val{f}{c} - \delta_F y_f) - \kappa_F \Big)^+
\end{equation}
and
\begin{equation}
   	T_{f}^{CS}(y) = \delta_F y_f + \frac{1}{n} \sum_{c \in C} \Ind[p (\val{c}{f} - \delta_C y_c) \ge \kappa_C] \bp[CS]{c}{f}{y} \Big( p ( \val{f}{c} - \delta_F y_f) - \kappa_F \Big)^+.
\end{equation}
Similarly, for all $c \in C$, let
\begin{equation}
   	T_{c}^{FS}(y) = \delta_C y_c + \lambda \sum_{f \in F} \Ind[\bp[FS]{c}{f}{y} p (\val{f}{c} - \delta_F y_f) \ge \kappa_F] \Big( \bp[FS]{c}{f}{y} p ( \val{c}{f} - \delta_C y_c) - \kappa_C \Big)^+
\end{equation}
and
\begin{equation}
   	T_{c}^{CS}(y) = \delta_C y_c + \lambda \sum_{f \in F} \Ind[p (\val{f}{c} - \delta_F y_f) \ge \kappa_F] \bp[CS]{c}{f}{y} \Big( p ( \val{c}{f} - \delta_C y_c) - \kappa_C \Big)^+.
\end{equation}
Notice that the unique fixpoints $y^{FS}, y^{CS} \in Y$ of $T^{FS}(y)$ and $T^{CS}(y)$ define agents' utilities under an FS-TS and CS-TS with threshold $y$ in FS and CS, respectively.
For all $y \in Y$ and $f \in F$, it holds that
\begin{align} 
   	&\lim_{\lambda p \rightarrow 1} \left( \delta_F y_f + \frac{1}{n} \sum_{c \in C} \Ind[\bp[FS]{c}{f}{y} p (\val{c}{f} - \delta_C y_c) \ge \kappa_C] \Big( \bp[FS]{c}{f}{y} p ( \val{f}{c} - \delta_F y_f) - \kappa_F \Big)^+ \right)\\
   	= &\delta_F y_f + \frac{1}{n} \sum_{c \in C} \Ind \left[f = \text{argmax}_{f' \in F: \val{f'}{c} - \delta_F y_{f'} > \kappa_F \land \val{c}{f'} - \delta_C y_c > \kappa_C} \val{c}{f'} \right] \Big( \val{f}{c} - \delta_F y_f - \kappa_F \Big)\\
   	= &\lim_{\lambda p \rightarrow 1} \left( \delta_F y_f + \frac{1}{n} \sum_{c \in C} \Ind[\bp[CS]{c}{f}{y} p (\val{c}{f} - \delta_C y_c) \ge \kappa_C] \bp[CS]{c}{f}{y} \Big( p ( \val{f}{c} - \delta_F y_f) - \kappa_F \Big)^+ \right).
\end{align}
Similarly, for all $y \in Y$ and $c \in C$ we have that
\begin{align} 
   	&\lim_{\lambda p \rightarrow 1} \left( \delta_C y_c + \lambda \sum_{f \in F} \Ind[\bp[FS]{c}{f}{y} p (\val{f}{c} - \delta_F y_f) \ge \kappa_F] \Big( \bp[FS]{c}{f}{y} p ( \val{c}{f} - \delta_C y_c) - \kappa_C \Big)^+ \right)\\
   	= &\delta_C y_c + \lambda \sum_{f \in F} \Ind \left[f =  \text{argmax}_{f' \in F: \val{f'}{c} - \delta_F y_{f'} > \kappa_F \land \val{c}{f'} - \delta_C y_c > \kappa_C} \val{c}{f'} \right] \Big( \val{c}{f} - \delta_C y_c - \kappa_C \Big)\\
   	= &\lim_{\lambda p \rightarrow 1} \left( \delta_C y_c + \lambda \sum_{f \in F} \Ind[p (\val{f}{c} - \delta_F y_f) \ge \kappa_F] \bp[CS]{c}{f}{y} \Big( p ( \val{c}{f} - \delta_C y_c) - \kappa_C \Big)^+ \right).
\end{align}
and hence $\lim_{\lambda p \rightarrow 1} |\ut{i}{\strat[FS]{i}{y}} - \ut{i}{\strat[CS]{i}{y}}| = 0$. This concludes the proof for families. For children, the proof is analogous and omitted.
\end{proof}

In order to see why \Cref{proposition:limit_p_lambda} holds, notice that the probability of child $c$ matching with his first choice from $M_c(\strat{}{})$ goes to 1 as $\lambda p \rightarrow 1$. Thus, the contribution of all other families from $M_c(\strat{}{})$ goes to zero. As only these first choices contribute to utilities, the difference between FS and CS again disappears.

\subsection{Low Market Thickness}

As the supply on the family side becomes very small, the differences between CS and FS disappear in equilibrium. 

\begin{proposition}\label{proposition:limit_lambda_small}
   	As $\lambda \rightarrow 0$, $|\ut[FS]{i}{s} - \ut[CS]{i}{s}| \rightarrow 0$ for all $s \in S$, $i \in A$.
\end{proposition}
\begin{proof}{Proof.}
Notice that the unique fixpoints $y^{FS}, y^{CS} \in Y$ of $T_{\strat{}{}}^{FS}(y)$ and $T_{\strat{}{}}^{CS}(y)$ (recall definitions from the proof of \Cref{proposition:limit_kappa_small} define agents' utilities under $\strat{}{}$ in FS and CS, respectively.
For all $y \in Y$ and $f \in F$ it holds that
\begin{align} 
   	&\lim_{\lambda \rightarrow 0} \left( \delta_F y_f + \frac{1}{n} \sum_{c \in C} 	\strat{c}{f} \strat{f}{c} \Big( \bp{c}{f}{\strat{}{}} p ( \val{f}{c} - \delta_F y_f) - \kappa_F \Big) \right)\\
   	= &\delta_F y_f + \frac{1}{n} \sum_{c \in C} \strat{c}{f} \strat{f}{c} \Big( p ( \val{f}{c} - \delta_F y_f) -\kappa_F \Big)\\
   	= &\lim_{\lambda \rightarrow 0} \left( \delta_F y_f + \frac{1}{n} \sum_{c \in C} 	\strat{c}{f} \strat{f}{c} \bp{c}{f}{\strat{}{}} \Big( p ( \val{f}{c} - \delta_F y_f) - \kappa_F \Big) \right),
\end{align}
because $\bp{c}{f}{\strat{}{}} \rightarrow 1$ as $\lambda \rightarrow 0$. Similarly, for all $y \in Y$ and $c \in C$ we have that
\begin{align} 
   	&\lim_{\lambda \rightarrow 0} \left( \delta_C y_c + \lambda \sum_{f \in F} \strat{c}{f} \strat{f}{c} \Big( \bp{c}{f}{\strat{}{}} p ( \val{c}{f} - \delta_C y_c) - \kappa_C \Big) \right)\\
   	= &\delta_C y_c \\
   	= &\lim_{\lambda \rightarrow 0} \left( \delta_C y_c + \lambda \sum_{f \in F} \strat{c}{f} \strat{f}{c} \bp{c}{f}{\strat{}{}} \Big( p ( \val{c}{f} - \delta_C y_c) - \kappa_C \Big) \right),
\end{align}
and hence $\lim_{\lambda \rightarrow 0} |\ut[FS]{i}{\strat{}{}} - \ut[CS]{i}{\strat{}{}}| = 0$ for all $i \in A$.
\end{proof}

The reason why the statement holds is that for very small $\lambda$, families do not have to worry about competition in FS, as it is likely that there is no other family active at any given time step.

\subsection{Patient Agents}

If agents are sufficiently patient, the problem becomes equivalent to a classic stable matching problem. 
To see this, we first need to revisit classical matching markets. The matching market \emph{induced} by $\instance$ is a tuple $(C,F,\succ)$, where $f \succ_c f'$ if and only if $\val{c}{f} > \val{c}{f'}$ and $c \succ_f c'$ if and only if $\val{f}{c} > \val{f}{c'}$. Further, $f \succ_c c$ if and only if $p \val{c}{f} \ge \kappa_C$ and $c \succ_f f$ if and only if $p \val{f}{c} \ge \kappa_F$. 

Let $(C,F,\succ)$ be a matching market where agents have strict preferences.

\begin{definition}
   	A pair of functions $g=(g_C,g_F)$ is called a \emph{pre-matching} if $g_C: C \rightarrow A$ and $g_F: F \rightarrow A$, such that if $g_C(c) \neq c$ then $g_C(c) \in F$ and if $g_F(f) \neq f$ then $g_F(f) \in C$. 
\end{definition}

We say that a pre-matching $g$ \emph{induces} a matching $w$ if the function $w: A \rightarrow A$ defined by $w(i) = g(i)$ is a matching. Consider the following set of equations:
\begin{align}
   	g_C(c) = \max_{\succ_c} \left( \{f \in F \mid c \succeq_f g_F(f)\} \cup \{c\} \right), \quad c \in C,\label{equation:stable_matching_condition}\\
   	g_F(f) = \max_{\succ_f} \left( \{c \in C \mid f \succeq_c g_C(c)\} \cup \{f\}\right), \quad f \in F,
\end{align}
where the maxima are taken with respect to agents' preferences.

\begin{lemma}[\citet{adachi_search_2003}]
   	If a matching $w$ is stable, then the pre-matching $g$ defined by $w$ solves the above equations. If a pre-matching $g$ solves the above equations, then $g$ induces a stable matching $w$.
\end{lemma}

With the above lemma, we can now prove \Cref{lemma:limit_delta}. Assume that $\delta_C = \delta_F =: \delta$.

\begin{lemma}\label{lemma:limit_delta}
   	There exists $\bar \delta \in [0,1)$, such that for all $\delta \in [\bar \delta, 1)$ the set of equilibrium matching correspondences are identical in FS and CS and coincide with the set of stable matchings in the induced marriage market.
\end{lemma}

\begin{proof}{Proof.}
Let $s \in S^{FS}$ and $c \in C$. 
$c$'s utility under $s$ is the unique value $\ut[FS]{c}{s}$ that satisfies
\begin{align}
   	&\ut{c}{s} = \delta \ut{c}{s} + \lambda \sum_{f \in M_c(\strat{}{})} \strat{f}{c} \Big( \bp{c}{f}{s} p ( \val{c}{f} - \delta \ut{c}{s}) - \kappa_C \Big)^+\\
   	\iff & (1-\delta) \ut{c}{s} = \lambda \sum_{f \in M_c(\strat{}{})} \strat{f}{c} \Big( \bp{c}{f}{s} p ( \val{c}{f} - \delta \ut{c}{s}) - \kappa_C \Big)^+.
\end{align}
Therefore, as $\delta \rightarrow 1$ we get
\begin{equation}\label{equation:delta_one_converge_zero}
   	\lambda \sum_{f \in M_c(\strat{}{})} \strat{f}{c} \Big( \bp{c}{f}{s} p ( \val{f}{c} - \delta \ut[FS]{c}{s}) - \kappa_C \Big)^+ \rightarrow 0.
\end{equation}
For the sake of contradiction, assume that $|M_c(\strat{}{})| \ge 2$. Let $f^* = \arg\max_{f \in F:  \strat{f^*}{c} = 1} \val{c}{f}$ and $f \in M_c(\strat{}{}) \setminus \{f^*\}$.
Because of \Cref{equation:delta_one_converge_zero}, it must hold that 
\begin{equation}
   	\bp{c}{f}{s} \Big( \val{c}{f} - \delta \ut{c}{s} - \kappa_C/p \Big)^+ \rightarrow 0.
\end{equation}
However, since $v_c(f^*) > \val{c}{f}$ and $\bp{c}{f^*}{s} > \bp{c}{f}{s}$ we get that 
\begin{equation}
   	\bp{c}{f^*}{s} \Big( \val{c}{f^*} - \delta \ut{c}{s} - \kappa_C/p \Big)^+ \rightarrow \epsilon
\end{equation}
for some $\epsilon > 0$, a contradiction.
Hence, $M_c(\strat{}{}) = \{f^*\}$ if $p \val{c}{f^*} \ge \kappa_C$ or $M_c(\strat{}{}) = \emptyset$ otherwise.
Notice that this is equivalent to the expression of \Cref{equation:stable_matching_condition}. Now that we have established that children will be mutually interested in at most one family, the same can similarly be shown for families, which completes the proof for FS. The proof for CS is analogous and thus omitted.
\end{proof}

\section{Family Preferences}\label{app:fam_pref}

Using data on 1,364 families across the state of Florida who completed a multi-step registration and approval process with the platform examined in Section~\ref{sec:empirical_casestudy}, we provide high-level insights on preference alignment to help understand which parameter regions discussed in Section~\ref{sec:num_eval} best represent reality. It should be noted that, as a snapshot of the population registered on the platform, this may not necessarily be representative of all families looking to adopt in Florida. As these preferences are based on families' answers to a registration questionnaire, they are not binding on how families respond when they are solicited for their interest in specific children, nor do they indicate the family's suitability to care for such a child.

\begin{figure}[ht]
	\centering
	\includegraphics[scale=0.55]{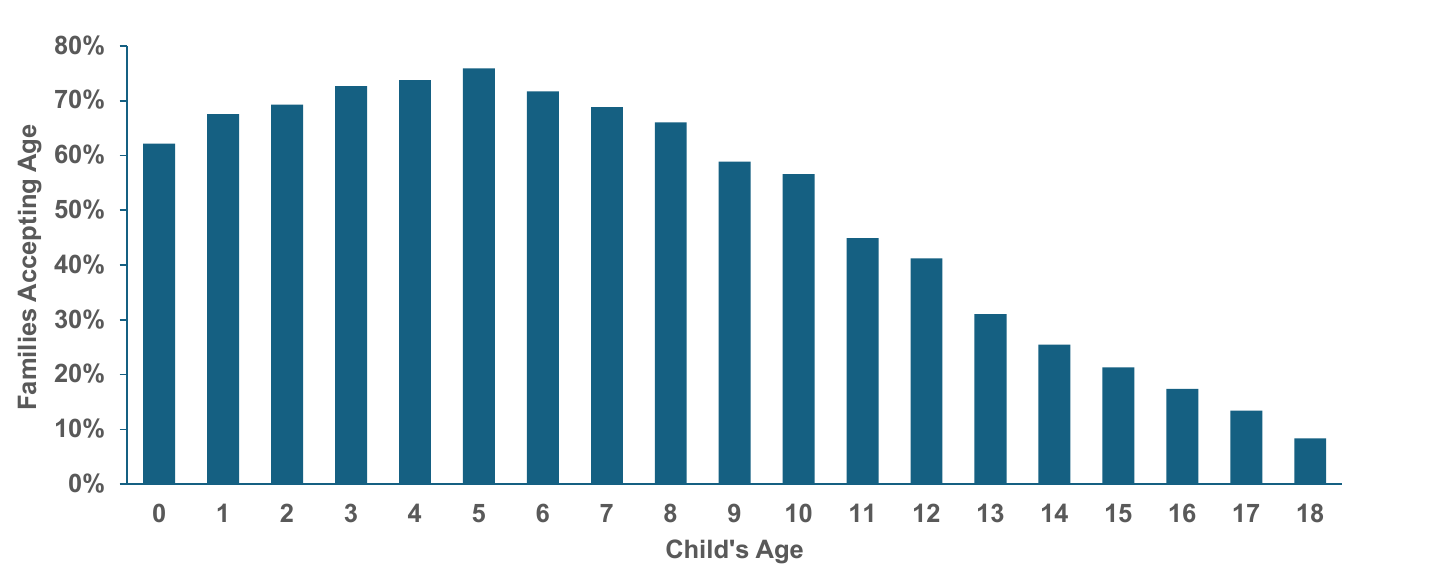}
	\caption{Acceptable child ages for 1,364 families in Florida active on the platform.}
	\label{fig:agepref}
\end{figure}

First, as shown in Figure~\ref{fig:agepref}, age preferences vary significantly. While a mean lower age limit of 1.8 combined with a standard deviation of 3.0 indicates a high willingness among families to adopt younger children, a mean upper age limit of 10.5 with a standard deviation of 4.8 indicates a reluctance to accept older children. The age that families find most acceptable is five years old, which is within the minimum and maximum age range for 75\% of families. 

Concerning the child's gender, almost 70\% of families chose ``no preference,'' with 12.7\% preferring boys and 15.4\% preferring girls. Although not pronounced, this difference in gender preference presents a challenge due to the higher number of boys that become available for adoption (see Table~\ref{tab:summ} in Section~\ref{sec:empirical_casestudy}) and partially explains why male children have worse adoption hazard rate coefficients. We note that --- while outside the scope of what our model considers --- choosing ``no preference'' may also indicate an interest in sibling groups, and the mean maximum number of children that a family expresses interest in is 2.2, with a standard deviation of 1.2. 

 Ethnicity and race offer additional attributes over which families express heterogeneous preferences. While 70\% of families expressed an openness to a child of any ethnicity, 21\% expressed an interest in white children, 16\% in biracial children, and between 7\% and 14\% in each of four other racial categories. Half of families are interested in children of Hispanic or Latino ethnicity, which is indicated separately from race.

\begin{figure}[ht]
	\centering
	\includegraphics[scale=0.55]{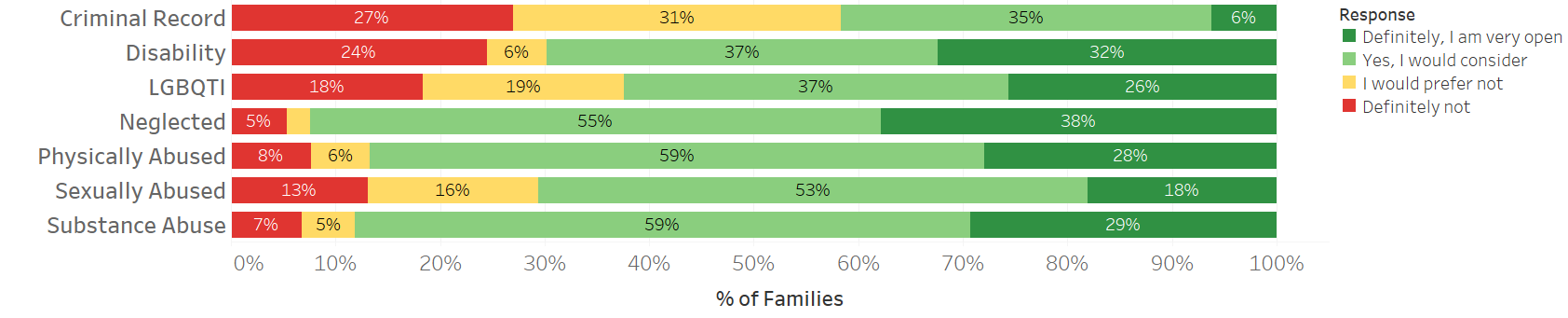}
	\caption{Preferences over child attributes for 1,364 families in Florida active on the platform.}
	\label{fig:fampref}
\end{figure}

Family preferences over seven additional attributes shown in Figure~\ref{fig:fampref} further diminish the plausibility of highly aligned preferences among families on some dimensions. For example, while nearly all families are open to a child impacted by neglect or substance abuse, only 41\% of families would consider or definitely be interested in a child with a criminal record.

\section{Numerical Evaluation: Supplementary Material} \label{app:num_eval_further}

\subsection{Families}\label{app:num_eval_families}

\begin{figure}[ht]
	\centering
	\includegraphics[scale=0.55]{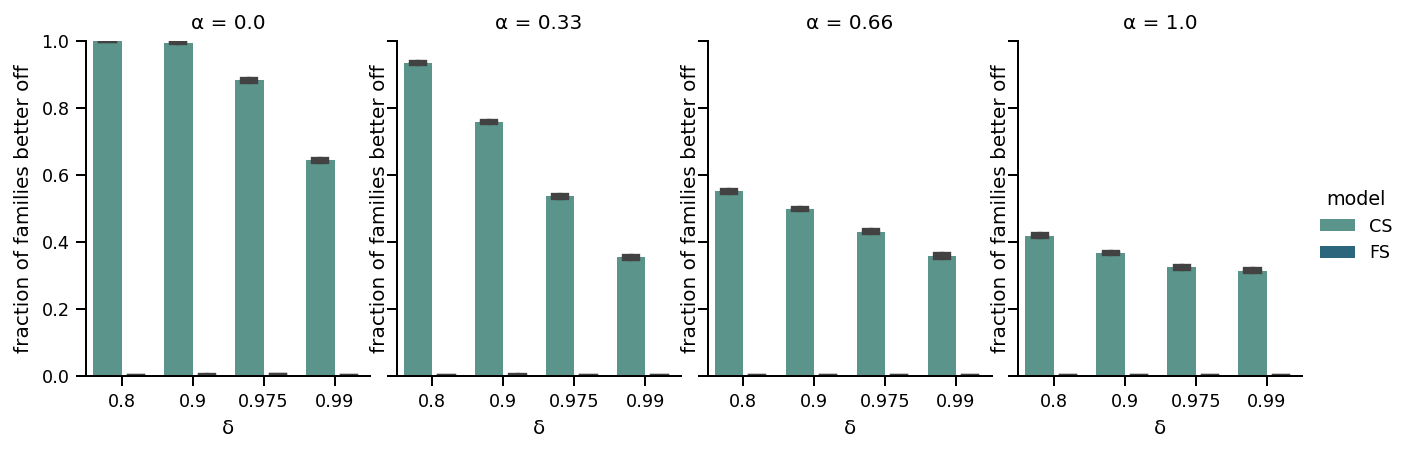}
	\caption{The ratio of families that are on average (strictly) better off in either approach in the family-optimal equilibrium for different combinations of agents' patience and the level of preference correlation.}
	\label{fig:families_better_off_utilities}
\end{figure}

From \Cref{fig:families_better_off_utilities}, we can see that the majority of families achieves a higher utility under CS, and almost no families achieve a higher utility under FS. 

\subsection{Match Probabilities}\label{app:num_eval_match_probabilities}

\begin{figure}[ht]
   	\centering
   	\includegraphics[scale=0.55]{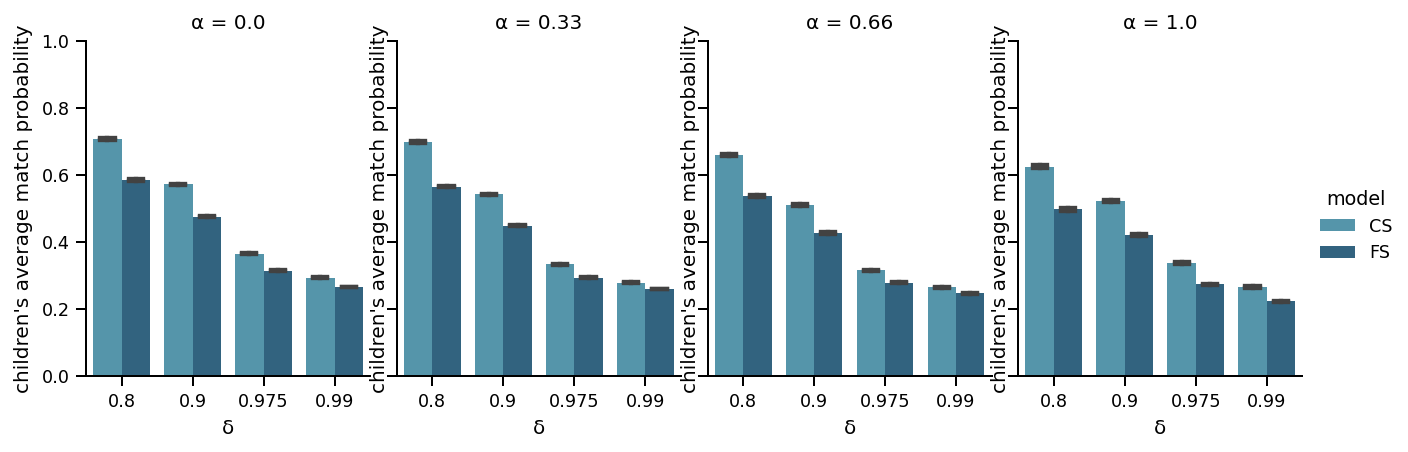}
   	\caption{Children's average match probability (averaged over all children and over all instances) in the family-optimal equilibrium for different combinations of agents' patience and the level of preference correlation.}
   	\label{fig:match_probabilities}
\end{figure}

\Cref{fig:match_probabilities} shows that, on average, children are more likely to get matched in CS than FS at any given time step.
 
\section{Empirical Analysis: Supplementary Material} \label{app:further_casestudy}

\subsection{Data: Further Details}\label{app:further_data_description}

\begin{figure}[t!]
	\begin{center}
	\includegraphics[width=\textwidth]{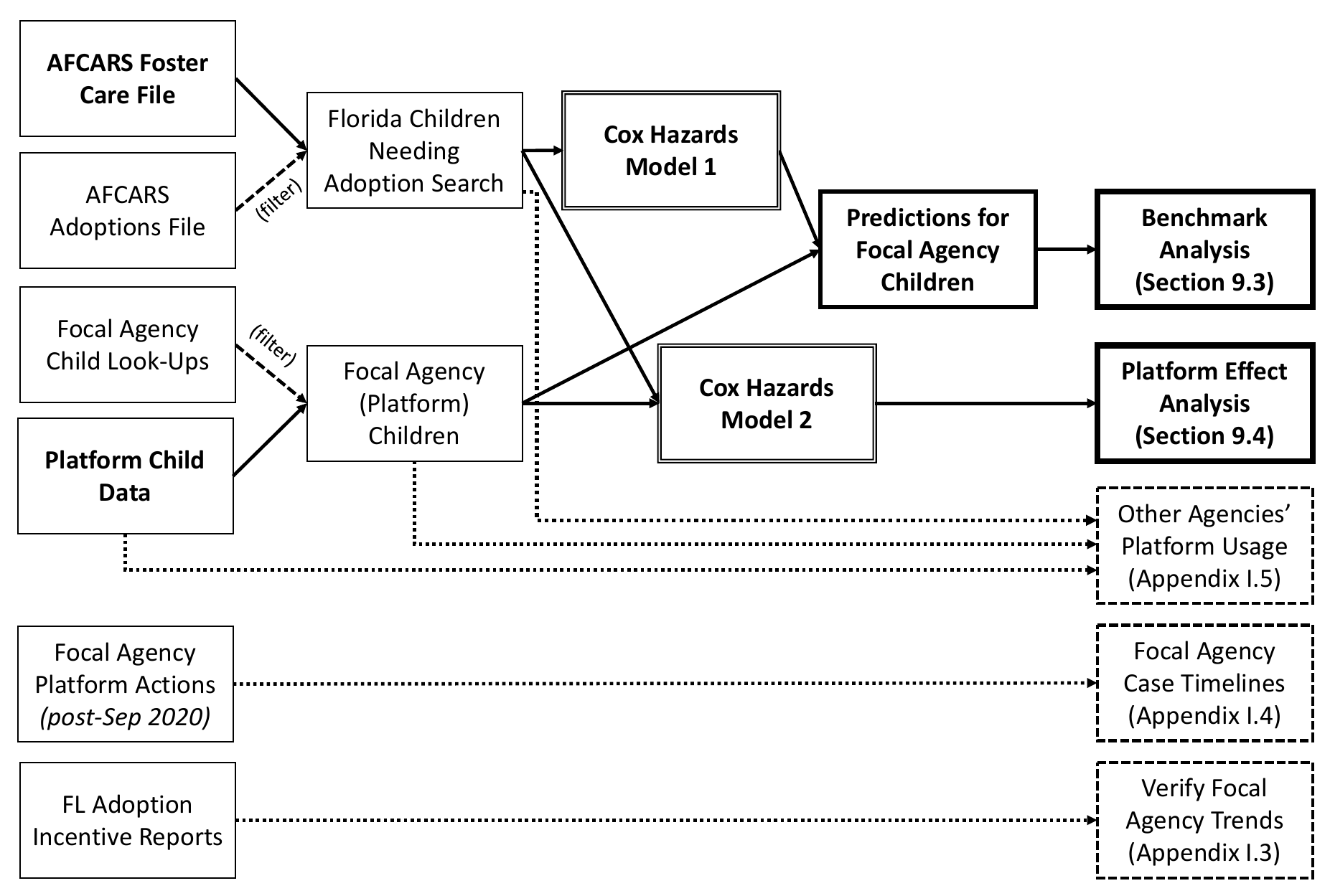}
 \end{center}
	\caption{Data sources used in the analyses of Section~\ref{sec:empirical_casestudy} and online Appendix~\ref{app:further_validation}. 
 }
	\label{fig:data-maintext}
\end{figure}

Our main analysis relies on multiple data sets: the AFCARS Foster Care 6-month File \citep{afcarsFoster}, the AFCARS Adoption File \citep{afcarsAdopt}, and case history data from the platform. 
From 766,527 AFCARS foster care 6-month update records for Florida children, we identified 9,544 children as legally free and clear for adoption with cases starting after October 1, 2014, which is the closest federal fiscal year cut-off for the data in Figure~\ref{fig:fl}. Regrettably, and despite significant efforts, the agency implementing the platform could not extract information from Florida's statewide case management system and assemble its own comparable case history dataset. However, we provide supplementary context in online Appendix~\ref{app:further_validation} that uses circuit-level data in state reports to assess that the platform had roughly average performance for all adoptions compared to other circuits in the state. Filtering children based on case goals and the relationships with adoptive families proved to be the biggest obstacles to using the agency's data. However, the agency was helpful in manually tracking down the outcomes of children who left the platform without an adoptive placement. In Figure~\ref{fig:data-maintext}, we provide an outline of how we combined different data sources to create the dataset used and how it is used in various analyses. 

\begin{figure}[t!]
	\begin{center}
	\includegraphics[width=\textwidth]{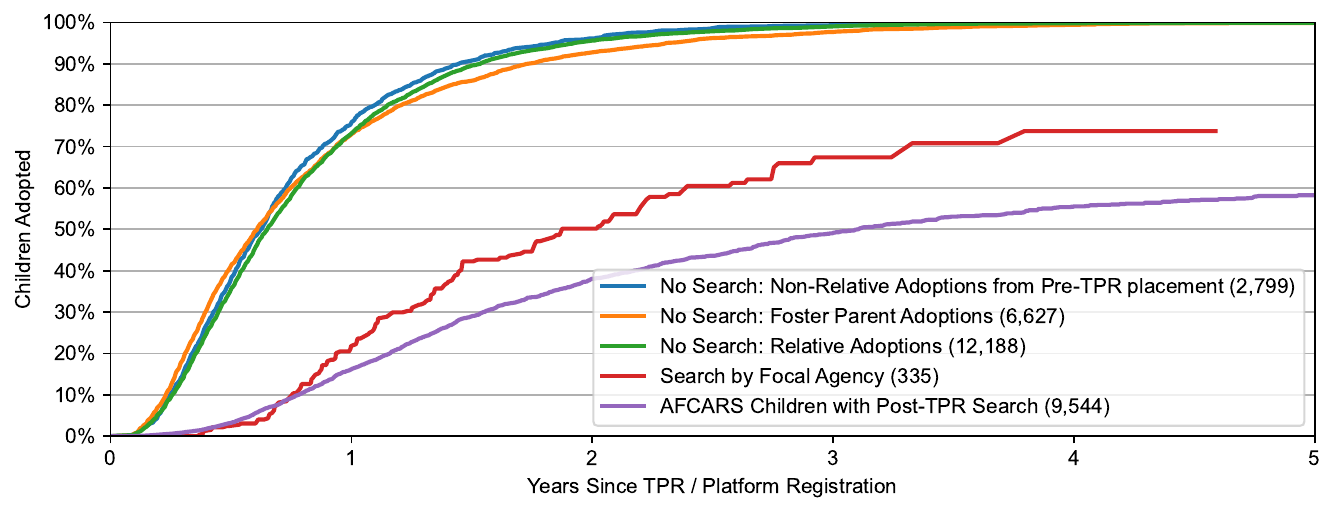}
 \end{center}
	\caption{Adoption outcomes for different populations of children within the AFCARS dataset, with a comparison to outcomes for focal agency children served by the platform. Values are computed as one minus the Kaplan-Meier survival curve, representing cumulative adoption over time.}
	\label{fig:filterAFCARS}
\end{figure}

For our analysis, children's timelines begin with the termination of parental rights (TPR) order and end with the last status update, which could be the adoption finalization date. For AFCARS cases to qualify for this analysis, children had to have a case goal of adoption in their final record or previously had a case goal of adoption with (a) a resulting non-relative adoption, (b) a final case goal of ``emancipation'', or (c) a discharge reason of ``emancipation.'' We excluded children listed in the Adoption File as being adopted by a relative, step-parent, or foster parent. Finally, (d) we excluded any adopted child where the pre-adoptive placement began before the TPR date. Because adoption search services can only be provided after TPR, those children, by definition, did not receive such services. Figure~\ref{fig:filterAFCARS} presents adoption outcomes for these populations within the AFCARS dataset and shows that children who did not require an adoption search reached finalization much more quickly. Because the most recent AFCARS Adoptions File only covers adoptions through September 30, 2021, we excluded case data after that date, as we would not be able to determine whether adoptions resulted from relative, step-parent, or foster placements. Note that this means empirical adoption rates are significantly lower than Kaplan-Meier survival curves, because many children are observed for only a short period.

We also note that the AFCARS dataset may still include some unidentifiable children who did not receive an active search for an adoptive family, as an adoptive placement was already identified at TPR, but the child had not yet been placed with the adoptive family. The inclusion of this group in our benchmark results in a slightly conservative assessment of the platform's performance, as it underestimates the platform's estimated association with faster placements.
Child welfare professionals have also mentioned the possibility that users of the AFCARS reporting system may inadvertently classify adoptions by foster parents as non-relative adoptions. These potential limitations underscore the importance of explicitly identifying children for whom an active search is underway in child welfare data sources to aid performance evaluation.

We received case data from the platform about 335 children in need of adoptive placements who had TPR before the AFCARS cut-off date of October 1, 2021. The platform provided its first matches around July 1, 2018. Because the platform allows caseworkers to register children and indicate the time since TPR using time intervals, the TPR date is estimated as the average of the interval endpoints, then subtracted from the platform case creation date. Of the 335 children, 146 had finalized adoptions through the platform by February 1, 2023. While the platform dataset does not include information about the final outcomes of the remaining children, we have obtained a supplemental ``Agency Case Look-Ups'' file manually collected by the focal agency, showing that an additional 19 of the children registered on the platform found adoptive placements through some other channel, such as serendipitous encounters between caseworkers in the agency's office. Additionally, the supplementary file allowed us to exclude an additional 75 children (beyond the 335 children we consider) who were registered on the platform at one time but achieved permanency through adoptive placement with relatives or foster care parents, guardianship, or reunification. Our data include only activity and case updates through February 1, 2023; children adopted after that date are not counted as adopted in our dataset.

Note that although the platform dataset also contains information for children for whom agencies in other circuits created profiles on the platform, we do not have access to the agency-side case data for those children. Consequently, we cannot identify what happened to children not adopted through the platform. Those children may have been adopted via FS or other channels, may have left the welfare system without adoption (e.g., due to emancipation), or may have only been removed from the platform for organizational reasons, but are still in need of adoption placements. We therefore cannot include them in our main analysis. However, we consider those cases in our supplemental platform use in other circuits found in Appendix~\ref{app:further_otherAg_usage}.

\subsection{Robustness Checks}\label{app:further_benchmark}

\begin{table}[t]
\centering
\footnotesize
\caption{Compiled Benchmark Results for Robustness Checks}
\label{tab:benchmark}
\begin{tabular}{p{4.2cm} c l rrr}
\hline
Model Variant &  Benchmark & Adoptions & 1 Year & 2 Years & 3 Years \\
\hline

\multirow{5}{4.2cm}{Base Model (Section~9.3)}
& \multirow{5}{*}{Model $1$}
& Platform         & 60   & 123  & 143   \\
&                   & Off-Platform & 8    & 15   & 19    \\
&                   & Total        & 68   & 138  & 162   \\
&                   & Benchmark    & 54.0 & 95.3 & 111.8 \\
&                   & \% above Benchmark & 26.0\% & 44.8\% & 44.9\% \\
\hline

\multirow{5}{4.2cm}{Without Conditional Probabilities (Appendix~\ref{app:further_no_varying_cox})}
& \multirow{5}{*}{Model $1$}
& Platform         & 60   & 123  & 143   \\
&                   & Off-Platform & 8    & 15   & 19    \\
&                   & Total        & 68   & 138  & 162   \\
&                   & Benchmark    & 45.1 & 95.2 & 112.5 \\
&                   & \% above Benchmark & 50.9\% & 45.0\% & 44.0\% \\
\hline

\multirow{5}{4.2cm}{Truncated at Sept.\ 30, 2021 (Appendix~\ref{app:truncate})}
& \multirow{5}{*}{Model $1$}
& Platform         &      44 &   98   &    105   \\
&                   & Off-Platform &    7  &   11   &   12    \\
&                   & Total        &    51  &  109    &  117     \\
&                   & Benchmark    &   50.3   &   83.2   &  93.5     \\
&                   & \% above Benchmark &  1.3\%     & 31.1\%     & 25.2\%      \\
\hline

\multirow{5}{4.2cm}{Excluding Off-Platform Adoptions (Appendix~\ref{app:further_excl_offPlatform})}
& \multirow{5}{*}{Model $1$}
& Platform         & 60   & 123  & 143   \\
&                   & Off-Platform & --   & --   & --    \\
&                   & Total        & 60   & 123  & 143   \\
&                   & Benchmark    & 50.8 & 89.1 & 104.4 \\
&                   & \% above Benchmark & 18.0\% & 38.0\% & 37.0\% \\
\hline

\multirow{5}{4.2cm}{More Restrictive `Black' Variable (Appendix~\ref{app:further_excl_black})}
& \multirow{5}{*}{Model $1^b$}
& Platform         &   60   &    123  &   143    \\
&                   & Off-Platform &  8    &  15    &  19     \\
&                   & Total        &   68   &  138    &  162     \\
&                   & Benchmark    &   52.4   &   92.9   &  109.2     \\
&                   & \% above Benchmark &  29.8\%    &  48.6\%    &   48.3\%    \\
\hline
\end{tabular}
\end{table}

\begin{table}[h]
\centering
\footnotesize
\caption{Cox Proportional Hazards Model of Time Until Adoption for Model Variants}
\label{tab:coxRobust}
\begin{tabular}{l d{4.6} d{4.6} d{4.6} d{4.6} d{4.6} d{4.6}}
                       & \multicolumn{1}{c}{Model $3$} & \multicolumn{1}{c}{Model $3^i$} & \multicolumn{1}{c}{Model $2^t$} & \multicolumn{1}{c}{Model $2^x$} & \multicolumn{1}{c}{Model $1^b$} & \multicolumn{1}{c}{Model $2^b$} \\ 
                       Appendix & \multicolumn{1}{c}{\ref{app:further_no_varying_cox}} & \multicolumn{1}{c}{\ref{app:further_no_varying_cox}} & \multicolumn{1}{c}{\ref{app:truncate}} & \multicolumn{1}{c}{\ref{app:further_excl_offPlatform}} & \multicolumn{1}{c}{\ref{app:further_excl_black}} & \multicolumn{1}{c}{\ref{app:further_excl_black}} \\ \hline     

Female                 
& 1.051 & 1.051 & 1.051 & 1.046 & 1.038 & 1.048 \\
& (1.302) & (1.290) & (1.278) & (1.157) & (0.939) & (1.223) \\

Black                  
& 0.663^{\ast\ast\ast} & 0.659^{\ast\ast\ast} & 0.667^{\ast\ast\ast} & 0.667^{\ast\ast\ast} & 0.687^{\ast\ast\ast} & 0.677^{\ast\ast\ast} \\
& (-9.664) & (-9.808) & (-9.455) & (-9.474) & (-8.035) & (-8.541) \\

Hispanic               
& 0.758^{\ast\ast\ast} & 0.756^{\ast\ast\ast} & 0.766^{\ast\ast\ast} & 0.766^{\ast\ast\ast} & 0.762^{\ast\ast\ast} & 0.774^{\ast\ast\ast} \\
& (-4.793) & (-4.835) & (-4.616) & (-4.621) & (-4.657) & (-4.437) \\

Age at TPR (years)     
& 0.903^{\ast\ast\ast} & 0.904^{\ast\ast\ast} & 0.903^{\ast\ast\ast} & 0.905^{\ast\ast\ast} & 0.901^{\ast\ast\ast} & 0.905^{\ast\ast\ast} \\
& (-6.232) & (-6.143) & (-6.147) & (-6.045) & (-6.174) & (-6.098) \\

(Age at TPR)$^2$ 
& 0.998 & 0.998 & 0.998 & 0.998 & 0.998 & 0.998 \\
& (-1.778) & (-1.863) & (-1.834) & (-1.954) & (-1.535) & (-1.868) \\

Disability             
& 0.872^{\ast\ast} & 0.868^{\ast\ast} & 0.869^{\ast\ast} & 0.871^{\ast\ast} & 0.893^{\ast\ast} & 0.871^{\ast\ast} \\
& (-3.256) & (-3.368) & (-3.303) & (-3.276) & (-2.609) & (-3.295) \\

TPR in FY2016 
& 0.967 & 0.972 & 0.958 & 0.953 & 0.984 & 0.961 \\
& (-0.492) & (-0.424) & (-0.633) & (-0.707) & (-0.238) & (-0.594) \\

TPR in FY2017 
& 0.982 & 0.988 & 0.980 & 0.976 & 1.023 & 0.986 \\
& (-0.268) & (-0.184) & (-0.297) & (-0.362) & (0.339) & (-0.203) \\

TPR in FY2018 
& 0.769^{\ast\ast\ast} & 0.768^{\ast\ast\ast} & 0.763^{\ast\ast\ast} & 0.754^{\ast\ast\ast} & 0.798^{\ast\ast} & 0.751^{\ast\ast\ast} \\
& (-3.956) & (-3.973) & (-4.076) & (-4.244) & (-3.371) & (-4.314) \\

TPR in FY2019             
& 0.602^{\ast\ast\ast} & 0.597^{\ast\ast\ast} & 0.590^{\ast\ast\ast} & 0.594^{\ast\ast\ast} & 0.590^{\ast\ast\ast} & 0.598^{\ast\ast\ast} \\
& (-7.279) & (-7.389) & (-7.540) & (-7.474) & (-7.407) & (-7.391) \\

TPR in FY2020          
& 0.332^{\ast\ast\ast} & 0.331^{\ast\ast\ast} & 0.321^{\ast\ast\ast} & 0.324^{\ast\ast\ast} & 0.312^{\ast\ast\ast} & 0.329^{\ast\ast\ast} \\
& (-13.133) & (-13.161) & (-13.251) & (-13.313) & (-13.236) & (-13.224) \\

TPR in FY2021 
& 0.211^{\ast\ast\ast} & 0.212^{\ast\ast\ast} & 0.185^{\ast\ast\ast} & 0.193^{\ast\ast\ast} & 0.175^{\ast\ast\ast} & 0.207^{\ast\ast\ast} \\
& (-11.523) & (-11.508) & (-11.024) & (-11.745) & (-11.102) & (-11.662) \\

Focal Agency               
& 1.695^{\ast\ast\ast} & 1.326^{\ast\ast} & 1.318^{\ast\ast} & 1.426^{\ast\ast\ast} & \multicolumn{1}{c}{-}  & 1.594^{\ast\ast\ast} \\
\emph{(Time-varying: $2^t$, $2^x$, $2^b$)}& (6.415) & (3.420) & (2.871) & (4.046) &  & (5.634) \\          
\hline
Focal Agency Children 
& \multicolumn{1}{c}{335} & \multicolumn{1}{c}{335} & \multicolumn{1}{c}{284} & \multicolumn{1}{c}{316} & \multicolumn{1}{c}{-} & \multicolumn{1}{c}{335} \\

N 
& \multicolumn{1}{c}{9879} & \multicolumn{1}{c}{9879} & \multicolumn{1}{c}{9828} & \multicolumn{1}{c}{9860} & \multicolumn{1}{c}{9544} & \multicolumn{1}{c}{9879} \\

Concordance 
& \multicolumn{1}{c}{0.704}  & \multicolumn{1}{c}{0.703}  & \multicolumn{1}{c}{-}  & \multicolumn{1}{c}{-}  &\multicolumn{1}{c}{0.706} & \multicolumn{1}{c}{-}  \\

Log-likelihood ratio test 
& \multicolumn{1}{c}{1451.129} & \multicolumn{1}{c}{1437.793} & \multicolumn{1}{c}{1431.757} & \multicolumn{1}{c}{1442.261} & \multicolumn{1}{c}{1363.882} & \multicolumn{1}{c}{1428.620} \\ 
& \multicolumn{1}{c}{on 13 d.f.} & \multicolumn{1}{c}{on 13 d.f.} & \multicolumn{1}{c}{on 13 d.f.} & \multicolumn{1}{c}{on 13 d.f.} & \multicolumn{1}{c}{on 12 d.f.} & \multicolumn{1}{c}{on 13 d.f.} \\ \hline

\multicolumn{7}{l}{\textit{Note: all coefficients are exponentiated with z-statistics in parentheses.}}\\
\multicolumn{7}{l}{$^\ast p < 0.05$, $^{\ast\ast} p < 0.01$, $^{\ast\ast\ast} p < 0.001$}\\
\end{tabular}
\end{table}
We now provide additional model variations for our empirical analysis as a robustness check. Table~\ref{tab:benchmark} reports the performance against the statewide AFCARS benchmark for relevant models, and Table~\ref{tab:coxRobust} presents the Cox proportional hazards models of time until adoption for each variant.  

\subsubsection{Analysis without Conditional Adoption Probabilities or Time-Varying Focal Agency Indicator}
\label{app:further_no_varying_cox}

\begin{figure}[t!]
	\begin{center}
	\includegraphics[scale=0.55]{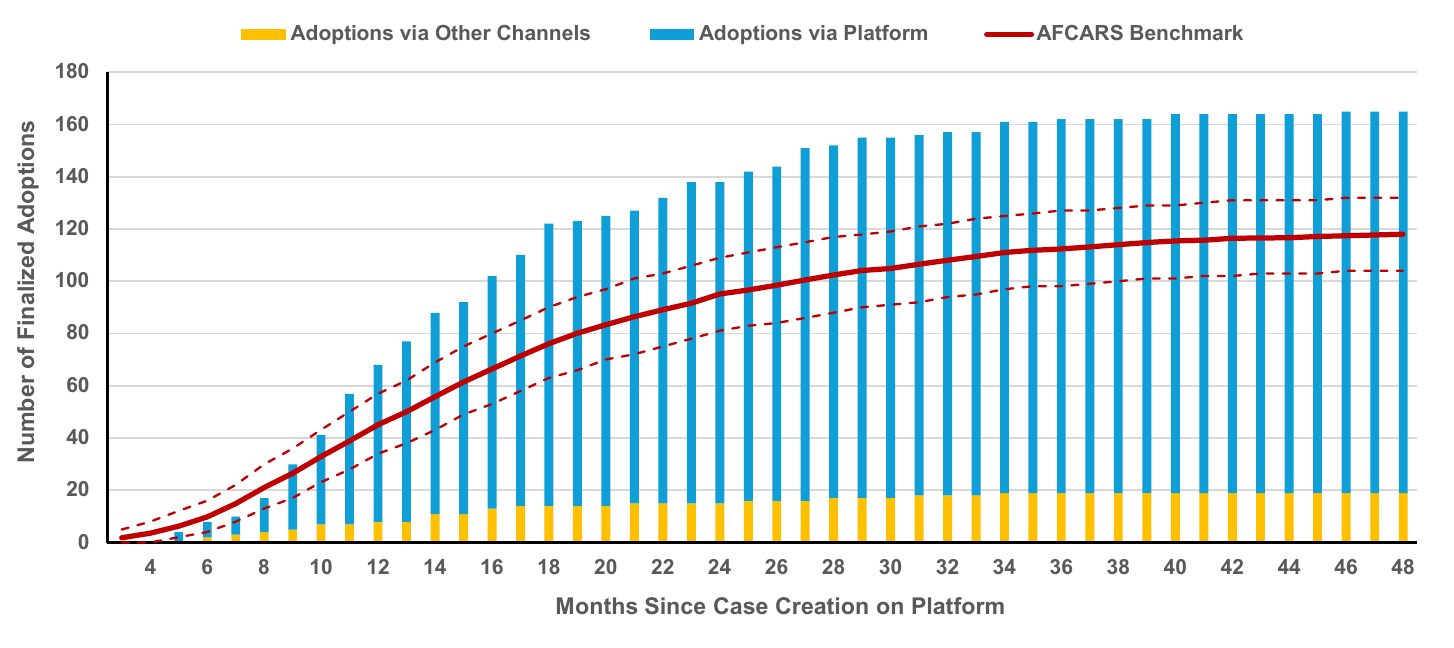}
 \end{center}
	\caption{Actual adoptions by the agency using the platform and other channels compared to Florida AFCARS benchmark model, using a model without conditional adoption probabilities or time-varying focal agency indicator. 
 }
	\label{fig:bench-nocondit}
\end{figure}

While the benchmark approach of Section~\ref{sec:empirical_benchmark} and the time-varying focal indicator approach of Section~\ref{sec:empirical_add_cox} account for elapsed time between TPR and case creation---which may arise because of startup effects or placements identified at TPR that later fail --- an alternative specification instead aligns each child's timeline with the start of adoptive search. Specifically, we let time begin at TPR for children in AFCARS and at platform case creation for children served through the platform. This removes the need to model conditional adoption probabilities based on time elapsed between TPR and case creation.

Table~\ref{tab:benchmark} and Figure~\ref{fig:bench-nocondit} show that the substantive benchmark conclusions remain similar under this alternative specification. The agency records 68 total adoptions within one year, 138 within two years, and 162 within three years, compared with benchmark values of 45.1, 95.2, and 112.5, respectively. Thus, actual adoptions remain well above benchmark levels at all three horizons: by 50.9\% at one year, 45.0\% at two years, and 44.0\% at three years. Relative to the base benchmark model, the main difference is that the benchmark prediction is lower at the one-year mark, which mechanically increases the estimated percentage improvement in the first year, while the two- and three-year comparisons remain very similar.

For the hazard analysis, we also estimate a standard Cox proportional hazards model with a static focal-agency indicator rather than a time-varying indicator. Formally, for each child $i$, we define a binary covariate $\textit{FocalAgency}_i$ that equals 1 if the child is ever registered on the platform before adoption and 0 otherwise. We refer to this specification as Model $3$. The estimates in Table~\ref{tab:coxRobust} again show a strong positive association between platform use and adoption timing: the estimated hazard ratio on the platform indicator is 1.695 ($p<0.001$), implying substantially faster adoption for focal agency children in this specification.

Even an exceedingly conservative variant in which children's timelines start immediately at TPR (or July 1, 2018, for children with TPR before the platform's inception), rather than when the focal agency's CS began, yields a significant positive focal agency effect of $1.326$ ($p<0.002$). 

\subsubsection{Truncating Outcomes on September 30, 2021}
\label{app:truncate}

\begin{figure}[t!]
	\begin{center}
	\includegraphics[scale=0.55]{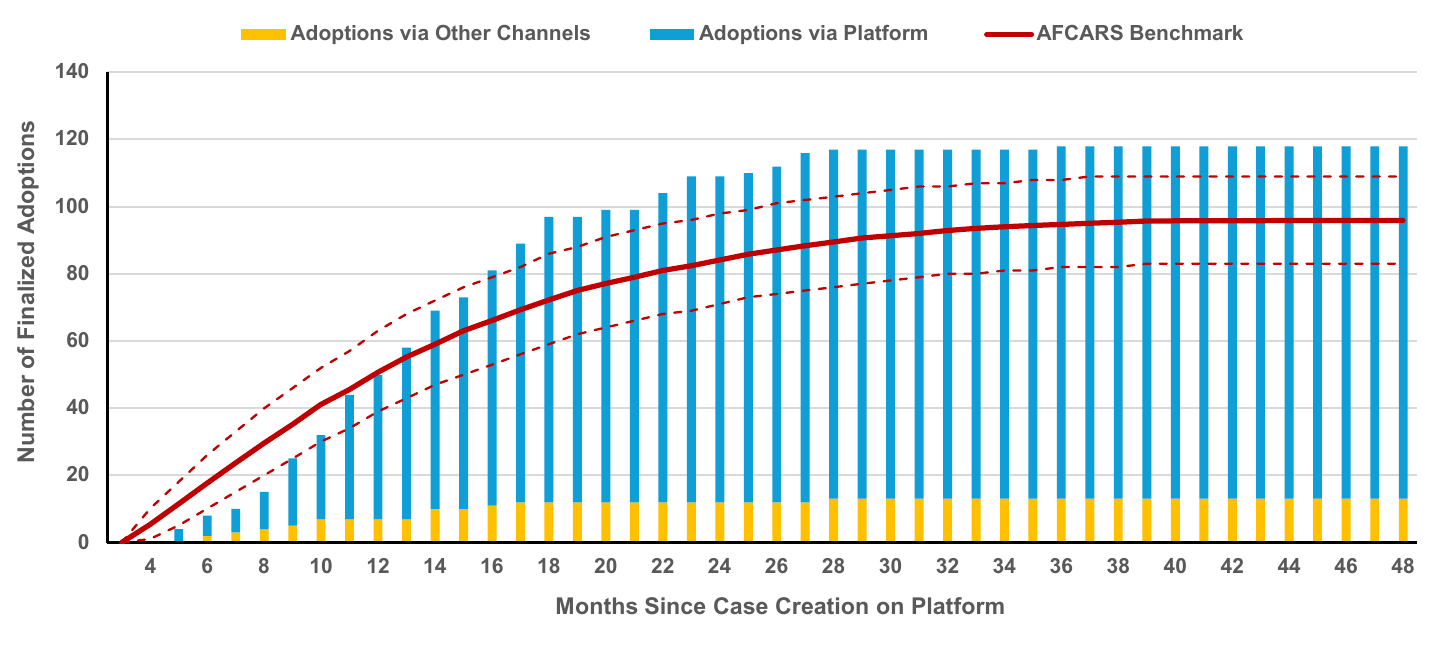}
 \end{center}
	\caption{Actual adoptions by the agency using the platform and other channels compared to Florida AFCARS benchmark model, with outcomes truncated to September 30, 2021. 
 }
	\label{fig:bench-trunc}
\end{figure}

Because the AFCARS data is only available through September 30, 2021, while the focal agency's data extends to February 1, 2023, we present an analysis in which the platform data and outcomes are truncated to September 30, 2021. Excluding children who could not have had an adoption finalized before October 1, 2021, due to the required delay between TPR and adoption, this specification focuses on 284 focal agency children.

Table~\ref{tab:benchmark} shows that, even under this more restrictive sample, the platform continues to outperform the AFCARS benchmark. As displayed in Figure~\ref{fig:bench-trunc}, the agency records 51 total adoptions within one year, 109 within two years, and 117 within three years, compared with benchmark values of 50.3, 83.2, and 93.5, respectively. While performance at the one-year mark is nearly identical to the benchmark (1.3\% higher), the platform continues to generate substantially more adoptions at longer horizons, exceeding the benchmark by 31.1\% at two years and 25.2\% at three years. These results suggest that the platform's advantage does not require the extended observation window in the full sample, but instead reflects sustained improvements in adoption outcomes over time.

For the hazard analysis, we estimate Model $2^t$, which allows the focal agency indicator to vary over time, as in our main Model 2.  The results in Table~\ref{tab:coxRobust} again indicate a positive and statistically significant effect of the focal agency on adoption timing, with a hazard ratio of 1.318 ($p<0.01$). Although this estimate is smaller than in the baseline specification, it continues to imply meaningfully faster adoptions for children served through the platform.

\subsubsection{Excluding Off-Platform Adoptions}
\label{app:further_excl_offPlatform}

\begin{figure}[t!]
	\begin{center}
	\includegraphics[scale=0.55]{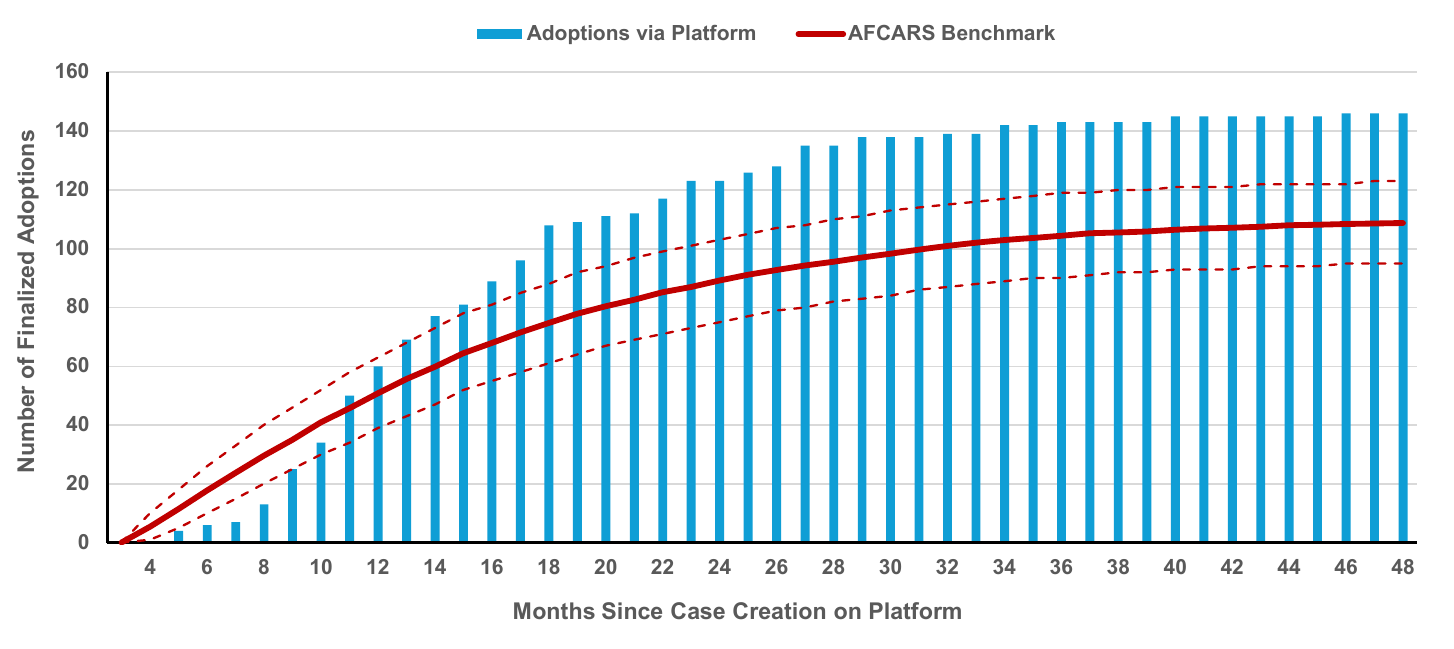}
 \end{center}
	\caption{Actual adoptions by the agency using the platform and other channels compared to Florida AFCARS benchmark model, excluding children adopted via channels other than the platform. 
 }
	\label{fig:bench-excloff}
\end{figure}

We also present results in Figure~\ref{fig:bench-excloff} that exclude children who were adopted through channels other than the platform. With these 19 off-platform adoptions removed, the analysis isolates the outcomes directly attributable to platform-facilitated matches. Table~\ref{tab:benchmark} shows that the platform continues to outperform the statewide benchmark even under this restriction. The agency records 60 adoptions within one year, 123 within two years, and 143 within three years, compared with benchmark values of 50.8, 89.1, and 104.4, respectively. This corresponds to improvements of 18.0\% at one year, 38.0\% at two years, and 37.0\% at three years. Thus, even when excluding all off-platform placements, adoption outcomes remain substantially higher than predicted by the benchmark model across all time horizons.

For the hazards analysis, we estimate Model $2^x$, which allows the focal agency indicator to vary over time while excluding off-platform adoptions from the outcome measure. The results in Table~\ref{tab:coxRobust} continue to show a strong and statistically significant association between platform use and adoption timing, with an estimated hazard ratio of 1.426 ($p<0.001$). This indicates that children served through the platform experience meaningfully faster adoptions even when attention is restricted solely to platform-generated matches.

\subsubsection{A More Exclusive AFCARS Race Variable}
\label{app:further_excl_black}

\begin{figure}[t!]
	\begin{center}
	\includegraphics[scale=0.55]{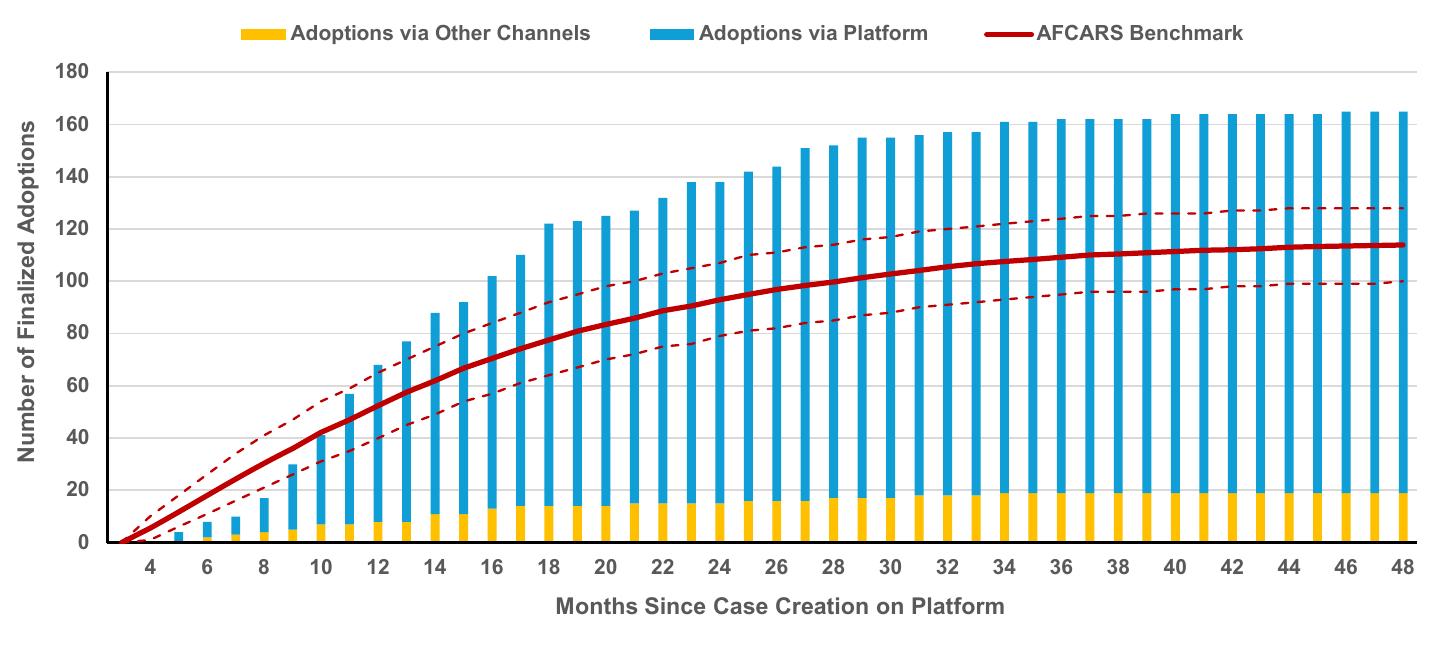}
 \end{center}
	\caption{Actual adoptions by the agency using the platform and other channels compared to Florida AFCARS benchmark model, using a more restrictive definition for the Black variable in the AFCARS data. 
 }
	\label{fig:bench-black}
\end{figure}

We also consider an alternative characterization of the \textit{Black} variable, as the approach in Section~\ref{sec:empirical_casestudy} adopts a conservative definition due to differences in how the AFCARS and platform datasets record race. In AFCARS, children may be assigned to multiple race categories, whereas the platform restricts entries to a single race and includes an ``other'' category that likely captures some multi-racial children.

In this alternative specification, we define the \textit{Black} indicator to include only children identified \emph{exclusively} as Black in the AFCARS data; that is, the Black race category equals 1 and all other race indicators equal 0. Under this definition, the share of children classified as Black in the AFCARS dataset decreases from 36.0\% to 28.8\%.

Table~\ref{tab:coxRobust} reports the corresponding hazards estimates for Model $2^b$. The coefficient on the Black variable increases in magnitude (i.e., the hazard ratio moves further below 1), indicating slower adoption for children identified exclusively as Black. This pattern suggests that multi-racial children tend to experience higher adoption hazards than those identified solely as Black. The focal agency effect remains positive and highly statistically significant, with an estimated hazard ratio of 1.594 ($p<0.001$), slightly smaller than in the baseline specification but still indicating substantially faster adoption for focal agency children.

Figure~\ref{fig:bench-black} and Table~\ref{tab:benchmark} present the corresponding benchmark comparison using Model $1^b$. The agency records 68 total adoptions within one year, 138 within two years, and 162 within three years, compared with benchmark values of 52.4, 92.9, and 109.2, respectively. This corresponds to improvements of 29.8\%, 48.6\%, and 48.3\% across the three horizons. Relative to the baseline specification, the benchmark predictions are modestly lower, which slightly increases the gap between observed and expected adoptions.

Overall, these results indicate that our findings are robust to alternative definitions of race in the AFCARS data. In particular, the platform remains strongly associated with faster adoptions, and the benchmark comparison continues to show substantial improvements over predicted outcomes.

\subsubsection{Fiscal Year Fixed Effects}\label{sec:platform_year}

We now provide a more detailed year-by-year analysis of the focal agency's performance after adopting CS using the platform. This allows us to better observe long-term trends and shocks in Florida stemming from the opioid crisis and pandemic disruptions, as well as possible organizational learning as the focal agency gained experience with the platform. Table~\ref{tab:coxyear} presents the results of this specification with federal fiscal-year-specific indicators. For robustness, we include three variations: Model 4 includes fiscal year fixed effects for both the focal agency and the rest of Florida, Model 5 only includes fiscal year fixed effects for the focal agency, while Model 6 again includes fiscal year fixed effects for both the focal agency and the rest of Florida, but omits TPR year fixed effects. Note that the $p$ values for fiscal year fixed effects in Table~\ref{tab:coxyear} all compare to the respective omitted variable (i.e., the rest of Florida in 2019 for Models 4 and 6, and the non-time varying rest of Florida fixed effect in Model 5). The differences between the focal agency's fixed effect and the respective fiscal year's fixed effect for the rest of Florida, including  $z$ statistics and $p$ values, are separately provided in Table~\ref{tab:pyear}.  

First, Model 4 reveals a clear temporal pattern in the platform’s effect on adoption hazards. While the rest of Florida has seen a stark and continuing decline in adoption hazard rates, the focal agency has resisted this trend after switching to CS. 
The coefficient for the focal agency in the first federal fiscal year following the platform's introduction (FY2019, October 2018--September 2019) is below that of the rest of Florida ($0.630$ vs $0.953$), suggesting that children served through the platform initially experienced a lower adoption hazard than comparable children. However, this decrease was not statistically significant. This early negative effect is consistent with a rollout period in which the agency was still learning how to use the system and integrating it into its existing adoption search processes. In FY2020, the estimated focal agency effect is already positive, but not yet statistically significant. 

However, beginning in FY2021, the focal agency’s effect becomes both large and highly statistically significant. The estimated hazard ratio rises to 2.184 in FY2021 and continues to increase in FY2022 and FY2023. While the latter years lack a direct same-year baseline for the rest of Florida, these coefficients remain significant even relative to the FY2015 statewide baseline, indicating sustained improvements in adoption hazards at the focal agency.

Model 5 shows broadly similar dynamics, but the removal of fiscal-year effects for the rest of Florida shifts common shocks (e.g., most notably the COVID-19 pandemic) into the TPR-year controls. As a result, the negative TPR-year effects in later years are very pronounced, potentially leading to the focal agency’s performance in 2022 and 2023 being overstated due to the absence of comparable rest-of-state benchmarks.

Conversely, Model 6 removes the TPR-year effects, forcing all temporal variation to be absorbed by the within-year covariates. Prior to the end of the AFCARS observation window, this yields the expected pattern: the focal agency exhibits (non-significantly) lower hazards in 2019, then slightly higher hazards in 2020, followed by a statistically significant increase in 2021. In 2022 and 2023, Model 6 shows a decrease in the adoption hazard of the focal agency children compared to the focal agency in 2021.  This is unsurprising given the persistent downward trend in year fixed effects between 2015 and 2021. While no separate data on children in need of adoptive search is available for 2022 and 2023, the downward trend in general adoption finalizations across Florida continued due to aftereffects of the pandemic and the height of the opioid crisis from 2021 to 2023 \citep{hackworth2025florida}, only reaching its low point in 2023 (see Figure \ref{fig:fl} and \citet{fl2019,fl2024}). While the focal agency's performance during these years dropped (non-significantly) below the Florida-wide pre-pandemic performance, it remained significantly above the most recent 2021 statewide covariate.

Taken together, these estimates imply that, after the initial adjustment period, children served by the focal agency experienced adoption rates nearly twice as high as comparable children in other circuits. The results suggest that any potential positive effects of using the platform emerged gradually rather than immediately, likely reflecting a combination of adoption timelines spanning multiple years, organizational learning by agency staff, improved engagement by prospective adoptive families, and the maturation of the platform’s search and recommendation capabilities. The timing of these improvements is also notable given the broader system disruptions during the last few years, which substantially reduced adoption hazards overall, as reflected in the negative year effects in the table. It is also possible that search costs increased substantially during the pandemic, and that the efficiency gains from shifting from FS to CS enabled agencies to maintain more effective adoption search processes when staffing constraints and operational disruptions made traditional search methods more difficult.

\begin{table}[h!]
\centering
\footnotesize
\caption{Cox Proportional Hazards Model of Time until Adoption: \\ Models 4, 5, and 6 with time varying focal agency indicator.}
\label{tab:coxyear}
\begin{tabular}{l  d{4.6} d{4.6} d{4.6} }
                       & \multicolumn{1}{c}{Model 4} & \multicolumn{1}{c}{Model 5} & \multicolumn{1}{c}{Model 6}  \\ \hline

Female                 & 1.052 & 1.053 & 1.048   \\
                       & (1.321) & (1.350) & (1.231)   \\

Black                  & 0.659^{\ast\ast\ast} & 0.661^{\ast\ast\ast} & 0.655^{\ast\ast\ast}  \\
                       & (-9.780) & (-9.743) & (-9.940)   \\

Hispanic               & 0.770^{\ast\ast\ast} & 0.769^{\ast\ast\ast} & 0.763^{\ast\ast\ast}  \\
                      & (-4.524) & (-4.536) & (-4.674)\\

Age at TPR (years)     & 0.902^{\ast\ast\ast} & 0.902^{\ast\ast\ast} & 0.906^{\ast\ast\ast}  \\
                       & (-6.255) & (-6.301) & (-6.034) \\

(Age at TPR)$^2$ & 0.998 & 0.998 & 0.998^{\ast} \\
                       & (-1.781) & (-1.728) & (-1.981)  \\

Disability             & 0.865^{\ast\ast} & 0.868^{\ast\ast} & 0.869^{\ast\ast}  \\
                       & (-3.444) & (-3.366) & (-3.331)\\

TPR in FY2016 
& 0.988 & 0.954 & \multicolumn{1}{c}{-}     \\
                      & (-0.132) & (-0.696) &      \\

TPR in FY2017 
& 1.115 & 0.977 & \multicolumn{1}{c}{-}    \\
                      & (0.905) & (-0.346) &    \\

TPR in FY2018 
& 0.960 & 0.769^{\ast\ast\ast} & \multicolumn{1}{c}{-}      \\
     & (-0.260) & (-3.973) &        \\

TPR in FY2019             & 0.793 & 0.582^{\ast\ast\ast} & \multicolumn{1}{c}{-}    \\
                      & (-1.168) & (-7.759) &   \\

TPR in FY2020          & 0.469^{\ast\ast} & 0.305^{\ast\ast\ast} & \multicolumn{1}{c}{-}    \\
                       & (-3.186) & (-13.958) &   \\

TPR in FY2021 
 & 0.286^{\ast\ast\ast} & 0.171^{\ast\ast\ast} & \multicolumn{1}{c}{-}     \\
                       & (-4.302) & (-12.375) &        \\

Rest of FL in 2015 & 1.433 & \multicolumn{1}{c}{-}  & 1.710^{\ast\ast\ast} \\
                   & (1.621) &   & (3.794) \\

Rest of FL in 2016 & 1.138 & \multicolumn{1}{c}{-}  & 1.266^{\ast\ast} \\
                   & (0.790) &   & (2.900) \\

Rest of FL in 2017 & 1.270^{\ast} & \multicolumn{1}{c}{-}  & 1.370^{\ast\ast\ast} \\
                   & (2.010) &   & (4.705) \\

Rest of FL in 2018 & 1.066 & \multicolumn{1}{c}{-}  & 1.158^{\ast} \\
                   & (0.759) &   & (2.249) \\

Rest of FL in 2020 & 0.963 & \multicolumn{1}{c}{-}  & 0.808^{\ast\ast} \\
                   & (-0.464) &   & (-3.379) \\

Rest of FL in 2021 & 0.768^{\ast} & \multicolumn{1}{c}{-}  & 0.446^{\ast\ast\ast} \\
                   & (-2.263) &   & (-12.089) \\

Focal agency in 2019 & 0.667 & 0.669 & 0.678 \\
                     & (-1.816) & (-1.819) & (-1.741) \\

Focal agency in 2020 & 1.165 & 1.266 & 1.011 \\
                     & (0.909) & (1.500) & (0.068) \\

Focal agency in 2021 & 2.209^{\ast\ast\ast} & 2.646^{\ast\ast\ast} & 1.395^{\ast} \\
                     & (4.690) & (6.973) & (2.334) \\

Focal agency in 2022 & 2.073^{\ast\ast} & 2.739^{\ast\ast\ast} & 0.858 \\
                      & (3.242) & (5.554) & (-0.870) \\

Focal agency in 2023 & 2.590^{\ast\ast} & 3.653^{\ast\ast\ast} & 0.910 \\
                     & (2.780) & (4.277) & (-0.322) \\

\hline
Focal agency children & \multicolumn{1}{c}{335} & \multicolumn{1}{c}{335} & \multicolumn{1}{c}{335}  \\
N & \multicolumn{1}{c}{9879} & \multicolumn{1}{c}{9879} & \multicolumn{1}{c}{9879} \\
Concordance &  \multicolumn{1}{c}{-} &  \multicolumn{1}{c}{-} & \multicolumn{1}{c}{-}  \\

Log-likelihood ratio test 
& \multicolumn{1}{c}{1511.142}  
& \multicolumn{1}{c}{1498.127} 
& \multicolumn{1}{c}{1426.904} \\ 

 & \multicolumn{1}{c}{on 23 d.f.}  
 & \multicolumn{1}{c}{on 17 d.f.} 
 & \multicolumn{1}{c}{on 17 d.f.} \\ \hline

\multicolumn{4}{l}{\textit{Note: all coefficients are exponentiated with z-statistics in parentheses.}}\\
\multicolumn{4}{l}{$^\ast p < 0.05$, $^{\ast\ast} p < 0.01$, $^{\ast\ast\ast} p < 0.001$}\\

\end{tabular}
\end{table}

\begin{table}[h!]
\centering
\footnotesize
\caption{Ratio of exponentiated coefficients for the Focal Agency and the Rest of Florida in different fiscal years.}
\label{tab:pyear}
\begin{tabular}{l  d{4.6} d{4.6} }
                       & \multicolumn{1}{c}{Model 4} & \multicolumn{1}{c}{Model 6}  \\ \hline

Focal agency in 2020 /  Rest of FL in 2020                     & 1.208 & 1.252 \\
                       & (1.177) & (1.395)   \\

Focal agency in 2021 /  Rest of FL in 2021                     & 2.876^{\ast\ast\ast}   & 3.13^{\ast\ast\ast}  \\
                       & (7.335) & (7.957)   \\    

Focal agency in 2022 /  Rest of FL in 2020                     & 2.151^{\ast\ast\ast}   & 1.062 \\
                       & (3.79) & (0.347)   \\  
Focal agency in 2022 /  Rest of FL in 2021                     & 2.700^{\ast\ast\ast}   & 1.925^{\ast\ast\ast}  \\
                       & (5.375) & (3.716)   \\  
 
Focal agency in 2023 /  Rest of FL in 2020                     & 2.688^{\ast\ast\ast}   & 1.126  \\
                       & (3.052) & ( 0.404)   \\  
Focal agency in 2023 /  Rest of FL in 2021                     & 3.374^{\ast\ast\ast}   & 2.041^{\ast}  \\
                       & (3.947) & (2.423)   \\  
\hline
\multicolumn{3}{l}{\textit{Note: We display the ratio of the exponentiated coefficients with z-statistics in parentheses.}}\\
\multicolumn{3}{l}{$^\ast p < 0.05$, $^{\ast\ast} p < 0.01$, $^{\ast\ast\ast} p < 0.001$}\\
\end{tabular}
\end{table}

\subsection{Validation from State Reports}\label{app:further_validation}

To validate our empirical analysis --- especially to understand how the circuit that implemented the platform compares to statewide averages before and after implementation --- we use the ``Adoption Incentive'' annual reports published by the Florida Department of Children and Families (\citeyear{fl2019}, \citeyear{fl2024}). The analysis using AFCARS data implicitly assumes that the agency is representative of statewide patterns; if the agency already outperformed statewide averages (controlling for the population of children) before implementing the platform, gains cannot be attributed to the platform. 

The Adoption Incentive reports provide annual statistics for how the 20 circuits in Florida perform on various measures. We note that Florida provides data on 19 entities; while most are individual circuits, some circuits are split or combined. Of the provided statistics, adoption success is best measured by the ``number of children who were eligible for adoption on 7/1 who were adopted by 6/30,'' which is displayed in Figure~\ref{fig:fl} in the main paper. In this case, eligibility refers to children for whom a termination of parental rights order has been granted. We refer to this metric as the \emph{adoption clearance rate}. Using the case timelines of the children on the platform, approximately 45\% of the children eligible for adoption every year on July 1 belonged to the set of 335 children who required search services through the platform. The remaining children likely already had a path to adoption identified through a foster parent or relative, and would be expected to have a faster path to adoption. 

\begin{table}[h]
\centering
\caption{Mean annual adoption clearance rate for the circuit that implemented a caseworker-driven search platform compared to state averages. The clearance rate refers to the percentage of children eligible for adoption on July 1 of a year whose adoption is finalized by June 30 of the following year.}
\label{tab:incentivereport}
\begin{tabular}{lrr}
                                                   & Before                                  & After                                   \\ \hline
\multicolumn{1}{|l|}{Comparison Period}            & \multicolumn{1}{r|}{7/1/2014-6/30/2018} & \multicolumn{1}{r|}{7/1/2019-6/30/2024}  \\ \hline
\multicolumn{1}{|l|}{Focal Agency Circuit Annual Mean \% Adopted}          & \multicolumn{1}{r|}{57\%}               & \multicolumn{1}{r|}{58\%}               \\ \hline
\multicolumn{1}{|l|}{Statewide Annual Mean \% Adopted}        & \multicolumn{1}{r|}{55\%}               & \multicolumn{1}{r|}{51\%}               \\ \hline
\multicolumn{1}{|l|}{Circuit-Statewide Ratio Mean} & \multicolumn{1}{r|}{103\%}              & \multicolumn{1}{r|}{115\%}              \\ \hline
\multicolumn{1}{|l|}{Mean Rank (of 19 circuits)}            & \multicolumn{1}{r|}{9.25}               & \multicolumn{1}{r|}{5.60}               \\ \hline
\end{tabular}
\end{table}

In Table~\ref{tab:incentivereport}, we compare the circuit's average adoption case clearance rate performance against the statewide average for the four years before implementing the platform and the five years after implementation. Because the monthly case creation peaked after July 1, 2018, as the platform's usage gradually ramped up into fall 2018, we disregard the annual report for July 1, 2018, to June 30, 2019, as a transition period. Thus, we compare the mean across annual statistics from July 1, 2014, until June 30, 2018, against July 1, 2019, until June 30, 2024. In the four years before implementation, the agency's performance was only 3\% higher than the statewide average, corresponding to an average ranking among all circuits of 9.25 out of 19. Thus, we expect the statewide AFCARS case data used for benchmarking to accurately reflect the agency's performance without the platform. 

Considering the averages over the five years since implementation, the circuit has seen a slight increase in its own performance and outperformed statewide averages. We note that the statewide average for the percentage of eligible children adopted decreased compared to before 2018, which could reflect higher acuity in children's needs or increased difficulties in casework and judicial processes from the COVID-19 pandemic. Considering the five years of data after implementation, the mean ratio for the circuit's case clearance rate compared to the statewide rate increased to 114.6\%. While it is difficult to directly link the percentage of eligible children adopted over a one-year time frame to the outcomes explored in Section~\ref{sec:empirical_casestudy}, this data indicates the focal agency has improved its performance in relation to the state as a whole and is consistent with our previous analysis using benchmarks from statewide AFCARS data.

\subsection{Insights from User Activity Logs}\label{app:further_audit}

Beginning in September 2020, the platform introduced enhanced functionality to track user actions associated with specific children or sibling sets. This tracking capability initially emerged from user-interface debugging tools, but platform managers later recognized its potential for process analytics and have since expanded its scope. These data allow us to observe on-platform activity and reconstruct the realized search process for a subset of children in our sample from the focal agency. Complete activity timelines based on user actions are only available for cases created after the expanded logging capability was implemented. This yields a subsample of 96 children out of the 335 children analyzed in Section~\ref{app:further_data_description} whose cases were created after September 1, 2020. Actions are tracked at the case level, as these 96 children are grouped into 58 \emph{cases} to account for sibling sets. Recorded actions extend to the end of 2024; some of the children who were not identified as adopted by the February 2023 cut-off for the Section~\ref{app:further_data_description} analysis were ultimately adopted. 

Important actions include:
\begin{description}
    \item[Case Creation.] A case begins with a case creation action, which corresponds to the completion of a simple questionnaire by the child's caseworker, providing basic details about the child that the caseworker should already have available.  After creating the case, the user completes an assessment questionnaire about the child that is used to generate recommendations. 
    \item[View Matches.] To begin the search, the child's caseworker clicks on a specific feature on the platform to view potential families for the child, either as an ordered list or in map view. For some children, this step may be delayed as the agency waits for opportunities with foster families or relatives to unfold, only to be unsuccessful. Families are shown along with a compatibility percentage score based on a proprietary algorithm that uses questionnaire data. Unfortunately, the agency did not track which families the caseworker viewed. 
    \item[Matched.] Once the caseworker has identified a family that is a good fit and willing to adopt the child,  they initiate the process towards an adoptive placement by updating the child's status to ``matched.'' A child can only be matched with one family at a time. Typically, the family will meet the child after the child has been matched with the family in preparation for a placement with the family.
    \item[Placed.] Once the child is living in the home with a family that intends to adopt it, the status changes to placement.
    \item[Adopted.] The final event is the legal finalization of the adoption. 
\end{description}

\begin{figure}[t!]
	\begin{center}
	\includegraphics[width=\textwidth]{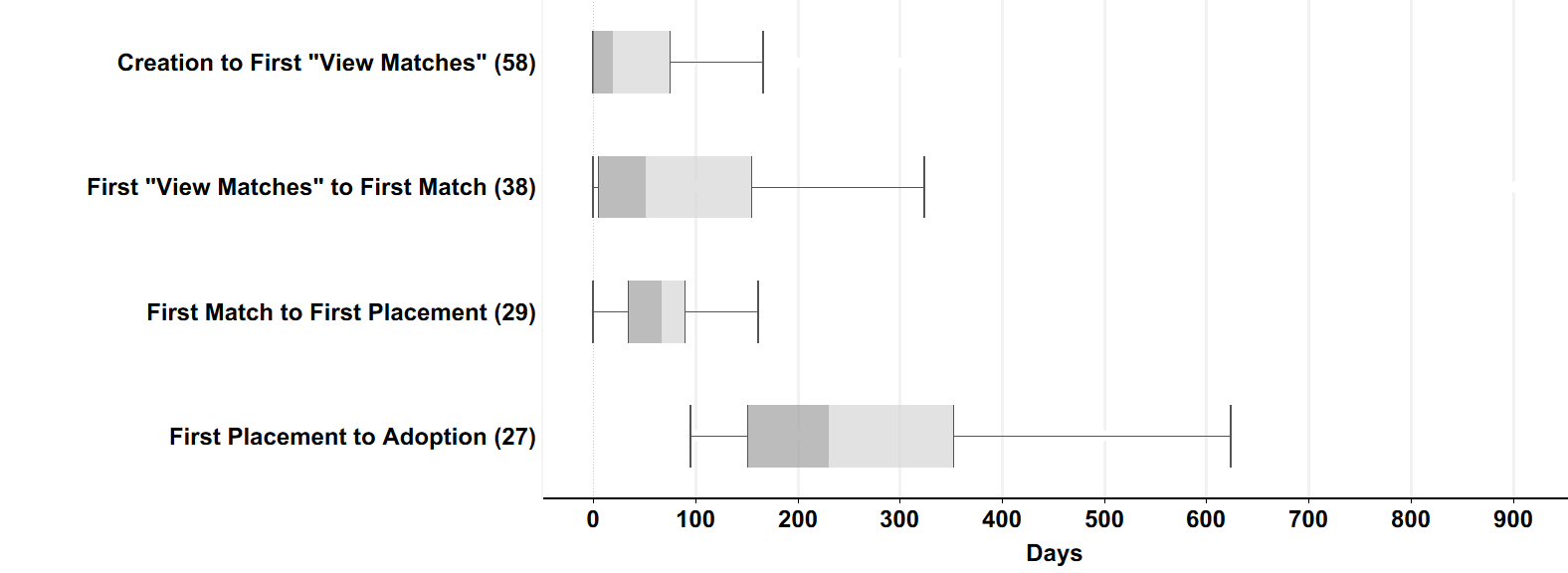}
 \end{center}
	\caption{Time between select events in caseworker activity data and number of sibling cases (out of 58) who progressed to each step. Boxes denote the interquartile range, and whiskers extend to the most extreme observations within 1.5 IQRs of the quartiles.}
	\label{fig:audit}
\end{figure}

Figure~\ref{fig:audit} shows the time between case creation, the first ``View Matches'' action, placement, and adoption for the 58 cases identified above. In over half of the cases, caseworkers used the ``View Matches'' action within 20 days; a manager shared that outliers for delays in the View Matches or Match stages are likely cases in which placements with relatives (i.e., outside the platform) were being explored that ultimately failed after several months, or even years. The median time between viewing matches and selecting a match is 52 days, and the median 67 days for the time between matching and a placement. Finalized adoptions occurred for 27 of the 58 cases, with a median time from first placement to adoption of 230 days. For finalized adoptions, the mean time from case creation to adoption was 432 days, with a range of approximately six months to three years.

\subsection{Platform Usage by Other Agencies}\label{app:further_otherAg_usage}

\begin{figure}[t!]
	\centering
	\includegraphics[width=\textwidth]{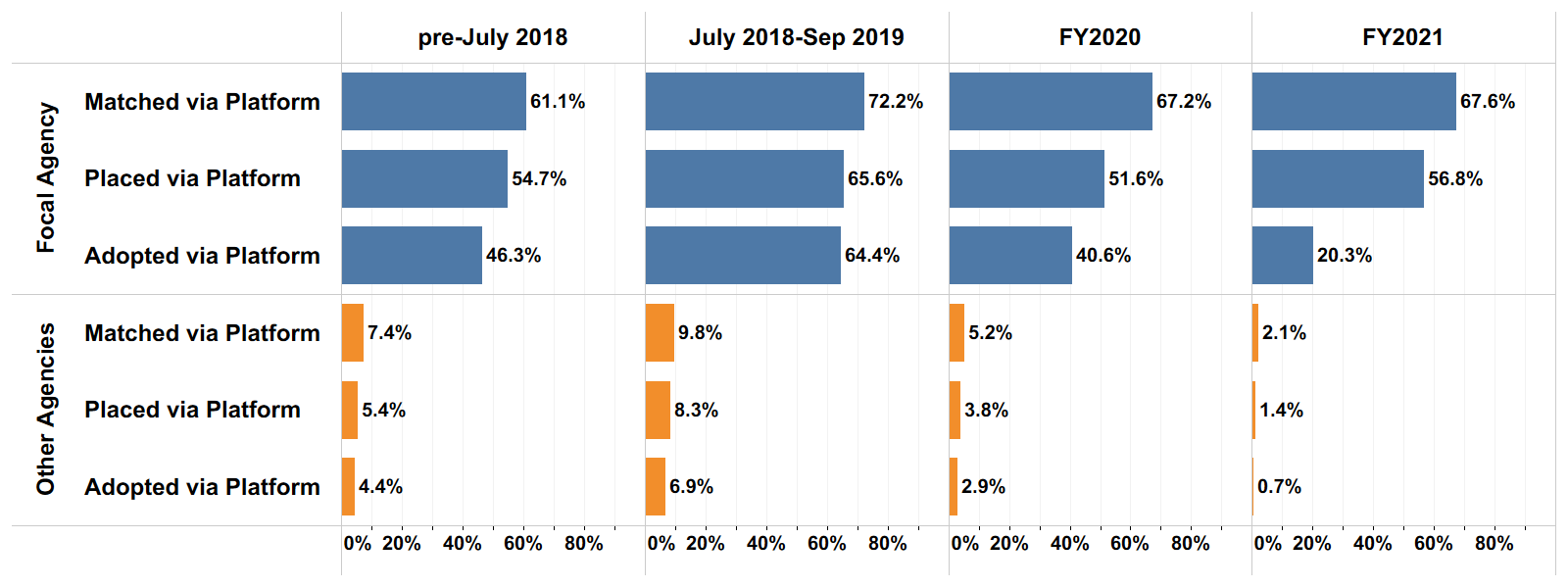}
	\caption{Milestone frequency as a ratio of all events occurring in the platform's dataset for children in Florida to the number of all children in need of adoptive search. Columns represent the child's TPR order date, with ``pre-July 2018'' referring to TPR orders that occurred before the platform began operation. The chart encompasses a total of  7,919 children statewide with active searches between July 1, 2018, and September 30, 2021. }
	\label{fig:otheragencies}
\end{figure}

To assess platform usage by other agencies across Florida, we combine the AFCARS dataset with the platform child data for platform events in other circuits through February 1, 2023. We emphasize that the focal agency was the only agency in Florida to use the platform to switch its primary search method from an FS to CS approach. All other agencies at most used it as a secondary strategy, employed opportunistically to supplement their existing FS approaches. This analysis characterizes the differences in platform usage prevalence at the focal agency relative to the rest of Florida. 

The expanded platform dataset includes the same data for all children registered on the platform and served by any agency in Florida.
Agency and platform managers have also confirmed in interviews that 100\% of children served by the focal agency were registered on the platform. Taken together, we can infer the frequency of platform milestones for children served by other agencies as a percentage of the total statewide population  --- excluding the focal agency --- of children needing adoptive search. Specifically, we measure the ratio of platform events for children in other circuits from the platform dataset to the total population of children in need of adoptive search services from the AFCARS dataset with TPR during the respective year and, in the case of children with TPR pre-July 2018, still in the child welfare system as of July 1, 2018. For example, in FY2020, according to the AFCARS dataset, TPR occurred for 1,345 children in Florida in need of an adoptive search, while  64 children with TPR in FY2020 were listed at the focal agency. Consequently, 1,281 children required adoptive search in circuits other than the focal agency's circuit. Conversely, other agencies matched 66 children who had TPR in FY2020 with families during the observed time period. Thus, 5.2\% of children in need of an adoption search outside of the focal agency received a match through the platform. We similarly compute ratios for the number of children placed and adopted via the platform.

Displaying these ratios for children based on when the TPR order occurred, 
Figure~\ref{fig:otheragencies} highlights both limited statewide usage of the platform and low downstream engagement among agencies other than the focal agency.
Progression towards adoption remained limited, as most agencies decided to remain focused on their original FS approach. Consequently, we see matches for only between 2 and 10\%  of the population outside the focal agency, and adoption finalization of 7\% or less through the platform. In contrast, the focal agency focused on CS and not only registered all children, but also converted a much larger share of the cases it created into matches, placements, and adoptions. Children in some TPR time intervals had adoption rates above 60\%, with later years showing lower observed adoption rates due to the February 2023 cut-off of the platform dataset. 

Taken together, these patterns suggest variation between the focal agency and other agencies in how the platform was used and integrated into existing workflows. These results reflect platform managers' view that the other agencies' implementations were fundamentally different from the focal agency's: families did not need to register with the platform, the agencies continued focusing on other FS-style search practices, and, at most, registered a subset of the children in need of an adoptive search opportunistically. By contrast, the focal agency represents a setting in which the platform was consistently applied across cases, providing a clearer view of outcomes under a sustained CS approach.
\end{document}